\documentclass[aps,prd,oneside,twocolumn,superscriptaddress,floatfix,nofootinbib,showpacs]{revtex4-2}
\usepackage{amsmath}

\usepackage{bm}
\usepackage{xcolor}
\usepackage{hyperref}
\usepackage{url}
\usepackage{graphicx}
\usepackage[large]{subfigure}
\usepackage{amssymb}
\usepackage[amssymb]{SIunits}

\usepackage{natbib}
\usepackage{multirow}
\usepackage{amsmath}

\usepackage{cleveref}

\usepackage{simpler-wick}

\usepackage{verbatim}

\usepackage{color}
\usepackage{amsmath}

\usepackage{bm}

\usepackage{hyperref}
\usepackage{url}
\usepackage{graphicx}
\usepackage[large]{subfigure}
\usepackage{amssymb}
\usepackage[amssymb]{SIunits}

\usepackage{natbib}
\usepackage{multirow}
\usepackage{amsmath}

\usepackage{cleveref}

\usepackage{simpler-wick}

\usepackage{booktabs} 

\newcommand{\axioncamb}{\texttt{axionCAMB}}

\newcommand{\ev}{{\rm eV}}

\newcommand{\apjs}{Astrophys. J. Suppl. Ser.}
\newcommand{\aap}{Astron. Astrophys.}

\newcommand{\jcap}{J. Cosmol. Astropart. Phys.}
\newcommand{\mnras}{Mon. Not. R. Astron. Soc.}
\newcommand{\physrep}{Phys. Rep.}

\renewcommand\prd{{\rm PhRvD}}
\renewcommand\prl{{\rm PhRvL}}
\newcommand\pasp{{\rm PASP}}

\newcommand{\healpix}[1]{\texttt{HEALPix} #1}
\newcommand{\emcee}[1]{\texttt{emcee} #1}

\begin{document}

\title{Detecting ultralight axions from multifrequency observations of neutral hydrogen intensity mapping}

\author{Qianshuo Liu}

\author{Chang Feng}
\email{changfeng@ustc.edu.cn (corresponding author)}
\affiliation{CAS Key Laboratory for Research in Galaxies and Cosmology, Department of Astronomy, University of Science and Technology of China, Hefei, Anhui 230026, China}
\affiliation{School of Astronomy and Space Science, University of Science and Technology of China, Hefei 230026, China}

\author{Filipe B. Abdalla}
\affiliation{CAS Key Laboratory for Research in Galaxies and Cosmology, Department of Astronomy, University of Science and Technology of China, Hefei, Anhui 230026, China}
\affiliation{School of Astronomy and Space Science, University of Science and Technology of China, Hefei 230026, China}

\begin{abstract}
We investigate the effects of ultralight axions (ULAs) on the differential brightness temperature fluctuations of neutral hydrogen at the post-reionization stage. Unique structure suppression features under the influence of ULAs are studied from angular correlations of the neutral hydrogen signals observed at different frequencies. Moreover, ULAs can behave like dark energy or dark matter at different mass regimes, implying that the fraction of the ULA energy density would be correlated with the sum of neutrino masses and the parameters of the dark
energy equation of state for dark matter and dark energy like ULAs, respectively. We have explored the parameter space for the ULA physics and possible degeneracies among dark energy, neutrinos, and ULAs using the theoretical angular correlation functions expected from future intensity mapping surveys. 
\end{abstract}

\maketitle

\section{Introduction}
Gravitational effects of dark matter (DM) have been detected from observations at different wavelengths but its microscale physics remains unknown. It is still unclear whether dark matter particles are made of cold dark matter (CDM), neutrinos, or other exotic particles, such as axions which are hypothetical particles introduced to solve the strong CP problem in particle physics~\cite {1977PhRvL..38.1440P, 1978PhRvL..40..223W} and are believed to be so light that the de Broglie wavelength can be as long as the astrophysical scale.  One interesting example is a particular mass around $10^{-22}$ {\ev}, which leads to a halo scale wavelength and can form condensates to solve the dark matter halo cuspy problem. This is the so-called —``fuzzy dark matter (FDM)'' scenario~\cite{2000PhRvL..85.1158H}. However, the large-scale structure is more influenced by the much longer wavelengths of the ultralight axions (ULAs) which have typical masses of approximately $10^{-33}$ {\ev} (corresponding to the Hubble scale $H_0$ today) to $10^{-22}$ \ev (corresponding to the fuzzy dark matter regime today) ~\cite{2010PhRvD..81l3530A}. 
On the other hand, neutrinos behave like CDM at large scales but cannot cluster within free streaming scales~\cite{PhysRevLett.45.1980, 2006PhR...429..307L}. Therefore, in neutrino scenarios, there is mass-dependent structure suppression of matter fluctuations due to massive neutrinos. Theoretical calculations show that neutrinos can introduce approximately 6\% suppression on small scales~\cite{PhysRevLett.80.5255}.

The axions are expected to be frozen at early times, behaving like dark energy (DE) with a constant equation of state; at later times, they are expected to begin oscillations so their averaged equation of state is zero, behaving like dark matter~\cite{2010PhRvD..82j3528M}. Therefore, the axion mechanism can describe both the dark sectors~\cite{2006PhLB..642..192A}, corresponding to axion masses at $m_a>10^{-27}$ {\ev} and $m_a<10^{-27}$ \ev, respectively~\cite{2006PhLB..642..192A}. Different from neutrinos, the free streaming scales of axions are extremely large, but the axion condensates have a finite sound speed that has a critical scale, beyond which there is also structure suppression, comparable with the neutrino free streaming scales~\cite{2012PhRvD..85j3514M}. The axions are not produced thermally and move at non-relativistic speeds despite the low masses. Therefore, the axions do not cluster below a certain length scale due to the wave effects. Because of these features, there might be interesting degeneracies among CDM, neutrino, dark energy, and axion signatures according to cosmological and astrophysical observations. A strong degeneracy between the dark energy equation of state and the sum of neutrino masses has been unveiled from observations~\cite{PhysRevLett.95.221301}.

Axion signatures exist at different levels of fluctuation correlation functions, mainly two-point correlation functions, or angular power spectra~\cite{2013PhRvD..87l1701M}. Other statistical quantities such as the peak counts~\cite{2018A&A...619A..38P} and higher-order correlations~\cite{2010ApJ...712..992B, 2023JCAP...06..023R} might also be nontrivially affected by ULAs. The ULA signatures were searched for from different observations including the cosmic microwave background (CMB), galaxy redshift surveys, and Lyman-$\alpha$. Different datasets such as the WMAP, SDSS, and Lyman-$\alpha$ datasets were analyzed to place constraints on the ULA models~\cite{2006PhLB..642..192A,2015PhRvD..91j3512H, 2017PhRvD..96l3514K, 2022JCAP...01..049L, 2021PhRvL.126g1302R}. Furthermore, the axion energy density was constrained to less than 1\% of the critical energy density~\cite{2015PhRvD..91j3512H} within $10^{-32}$ {\ev} to $10^{-26}$ {\ev} using both the CMB temperature and polarization power spectrum measurements from Planck~\cite{2014A&A...571A..15P}, Atacama cosmology telescope~\cite{2014JCAP...04..014D} and south pole telescope~\cite{2011ApJ...743...28K} in conjunction with the matter power spectrum from the WiggleZ Dark Energy Survey~\cite{2010MNRAS.406..803B}.  

The CMB power spectra are mostly affected by modifications of the global evolution history caused by the entire energy density fraction contributed by ULAs. The matter power spectra are more sensitive to ULA masses and other microscopic properties. The CMB lensing power spectrum was first adopted to constrain the ULA models in the following analysis ~\cite{2018MNRAS.476.3063H}. Although no evidence of ULA signatures has been detected, the CMB lensing power spectra from the next-generation will improve the ULA constraining power by a factor of a few, compared to the primary fluctuations. Recent analysis also demonstrated that ULA, as a subcomponent of DM, can offer a possible solution to alleviate the $\sigma_8$ tension~\cite{2023JCAP...06..023R}. 

Axions also have profound small-scale implications and can leave unique suppression imprints on cosmic shear measurements of weak lensing in galaxies. In such a small-scale regime, the fuzzy dark matter models might be more dominant, and the FDM constraints were obtained from the cosmic shear measurements of the Dark Energy Survey~\cite{2022MNRAS.515.5646D}.

Moreover, as perturbations, axions can not only affect large-scale structures but also collapse and form bound objects such axion halos~\cite{1983PhRvL..50..925I}, axion minclusters~\cite{2024arXiv240218221E} and boson stars~\cite{1986PhRvL..57.2485C,2024FrASS..1146820B}. Future high-resolution intensity mapping (IM) experiments may be able to resolve these sources, so we can study them with better sensitivity.

The 21 cm line is an ideal tracer of large-scale structures in both spatial and temporal coordinates at much smaller scales than the CMB. Tomographic intensity mapping of the neutral hydrogen (HI) can thus yield interesting constraints on the ULA models, forming a sensitive probe for ULA physics. Recent forecasts have shown that the axion energy density fraction can be constrained to a sub-percent level by future IM surveys~\cite{2021MNRAS.500.3162B}.

In this work, we study the ULA signatures of the HI power spectra measured from future IM surveys at different wavelengths and further investigate the possible origins of ULA-DE and ULA-neutrino degeneracies.

\section{Theoretical models of differential brightness temperature fluctuations}
\begin{figure}
	\includegraphics[width=1.0\linewidth]{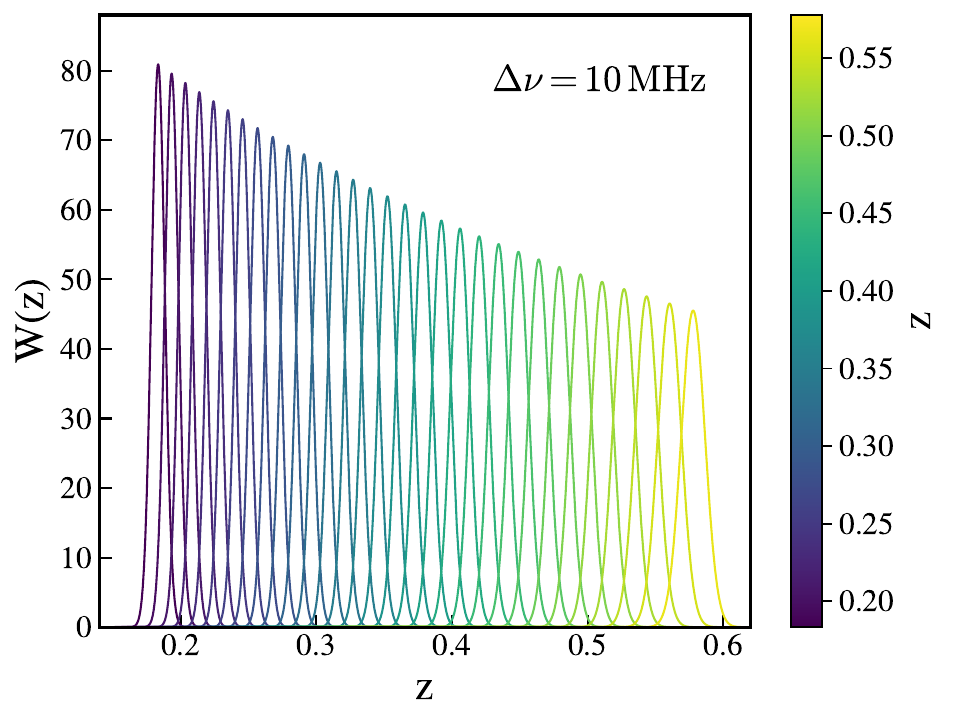}
	\caption{Window functions for different frequency channels considered in this work. The frequency range is 900 to 1200 MHz, and $\Delta\nu$ = 10 MHz. } 
	\label{window}
\end{figure}

Different components contribute to the 21 cm brightness temperature fluctuations $\delta T_b$, mainly baryon density fluctuations $\delta_b$~\cite{2006PhR...433..181F}. This work only assumes $\delta T_b\sim\delta_b$ and extracts the ULA modified perturbations using the public code \axioncamb~\cite{2021MNRAS.500.3162B}. The differential brightness temperature fluctuations can be measured at different frequencies $\{\nu \}$, and the cross-frequency power spectrum of the differential brightness temperatures measured at a frequency-pair $\nu_1\nu_2$ is
\begin{equation}
C^{\nu_1\nu_2}_{\ell}=4\pi\int d\ln k \mathcal{P}(k)\Delta^{\nu_1}_{\ell}(k)\Delta^{\nu_2}_{\ell}(k),\label{exact}
\end{equation}
where $\ell$ is the multiple, $k$ denotes spatial and temporal coordinates, $\mathcal{P}(k)$ is the power spectrum of primordial curvature perturbations, and $\Delta^{\nu}_{\ell}(k)$ is the transfer function expressed as 
\begin{equation}
\Delta^{\nu}_{\ell}(k)=\int dz T_b(z)W_{\nu}(z)\mathcal{S}(k,z).
\end{equation}
Here, $z$ is the redshift, $\chi$ is the comoving distance, and $j_{\ell}$ is the spherical Bessel function. The mean brightness temperature $T_b$  is ~\cite{2006PhR...433..181F} 
\begin{equation}
{T}_b(z) =0.188  h \Omega_{\mathrm{HI}}(z) \frac{(1+z)^2}{E(z)} \mathrm{~K},
\end{equation}
where $\Omega_{\mathrm{HI}}$ is the fraction of neutral hydrogen and is assumed to be constant that can be derived from the observational constraint $\Omega_{\mathrm{HI}}b_{\rm HI}=6.2\times 10^{-4}$~\cite{2013MNRAS.434L..46S}. $E(z)$ is the reduced Hubble constant $H(z)/H_0$ and $H_0$ is the Hubble constant today with $H_0=100 h\,\rm{km}\, {\rm s}^{-1}\,{\rm Mpc}^{-1}$. $H(z)$ is the Hubble constant at redshift $z$. The galaxy bias $b_{\rm HI}$ is assumed to be unity in this work. In this work, we only consider a source term $\mathcal{S}(k,z)$ contributed by the baryon density fluctuations, i.e., $\mathcal{S}(k,z)=\delta_b(k,z)j_{\ell}(k\chi)$, and consider two types of window functions for the computation of the power spectrum, i.e., a top-hat window and a Gaussian window. The top-hat window is used to simulate an intensity mapping survey with a top-hat frequency band, and the Gaussian window
\begin{equation}
W_{\nu}(z)=\frac{1}{\sqrt{2\pi}\Delta z_{\nu}}\exp\Big[-\frac{1}{2}\Big(\frac{z-z_{\nu}}{\Delta z_{\nu}}\Big)^2\Big]
\end{equation}
is used to perform several computational tests because it suffers from fewer computational issues, such as aliasing. Here, the observed frequency is $\nu=\nu_0/(1+z)$, where $\nu_0$ is the rest-frame 21 cm frequency, so $\Delta z$ can be directly determined at the central frequency of each band once a bandwidth $\Delta \nu$ is given. The narrow band window functions from 900 to 1200 MHz with a bandwidth $\Delta\nu$ = 10 MHz are displayed in Fig. \ref{window}.

\begin{figure*}
    \centering   
      \includegraphics[width=8cm,height=7cm]{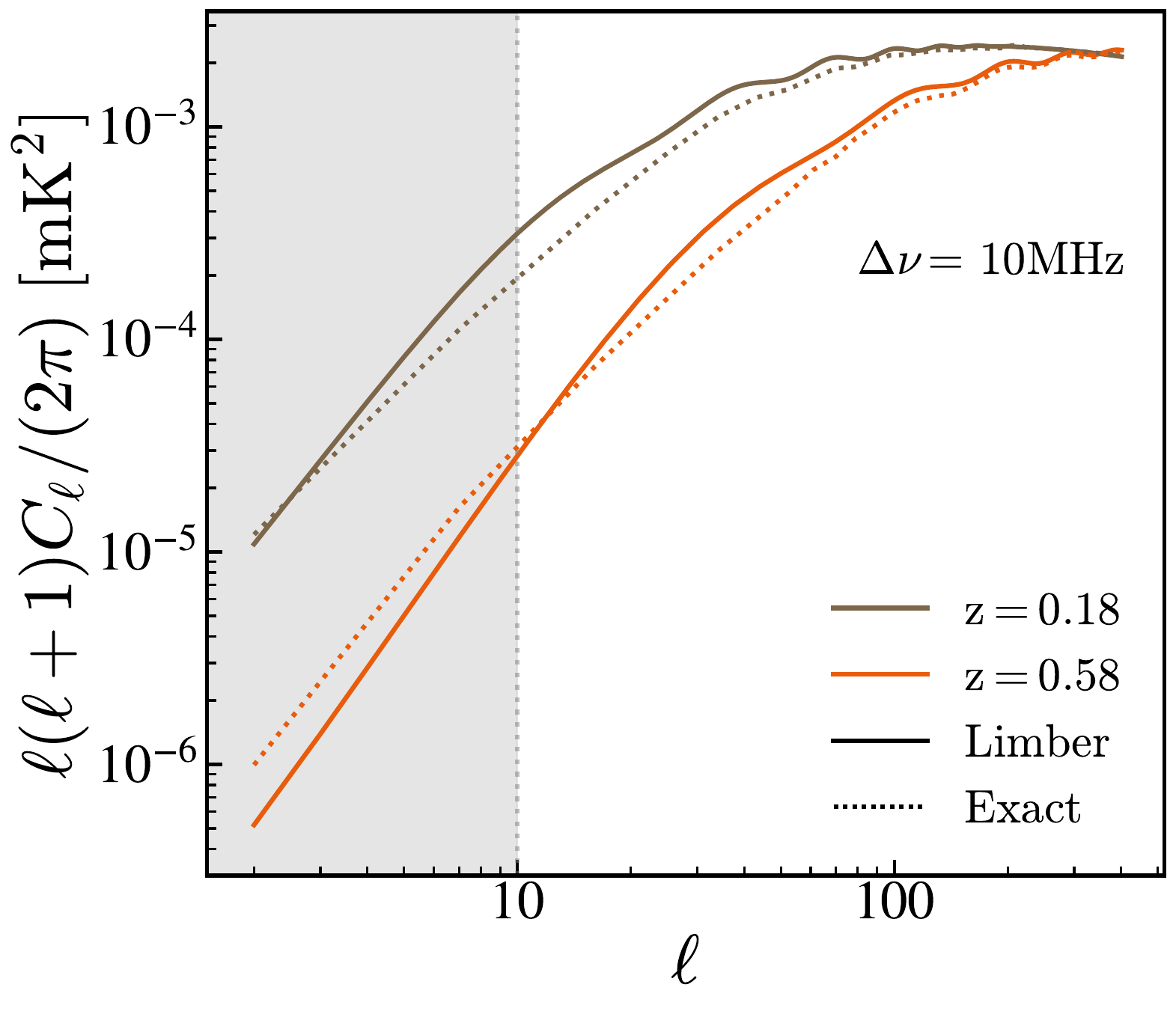 }
       \includegraphics[width=8cm,height=7cm]{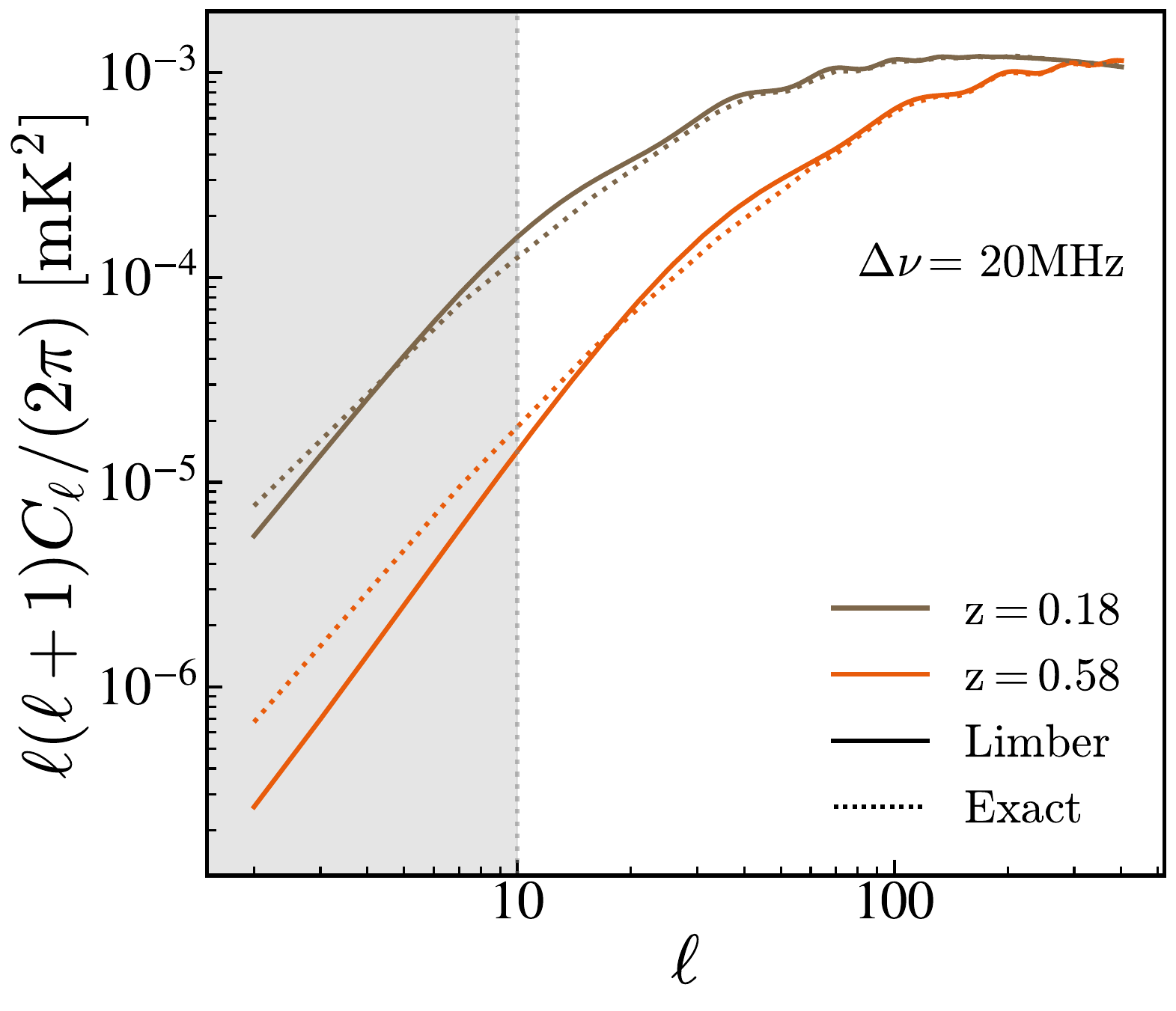}
    \includegraphics[width=8cm,height=7cm]{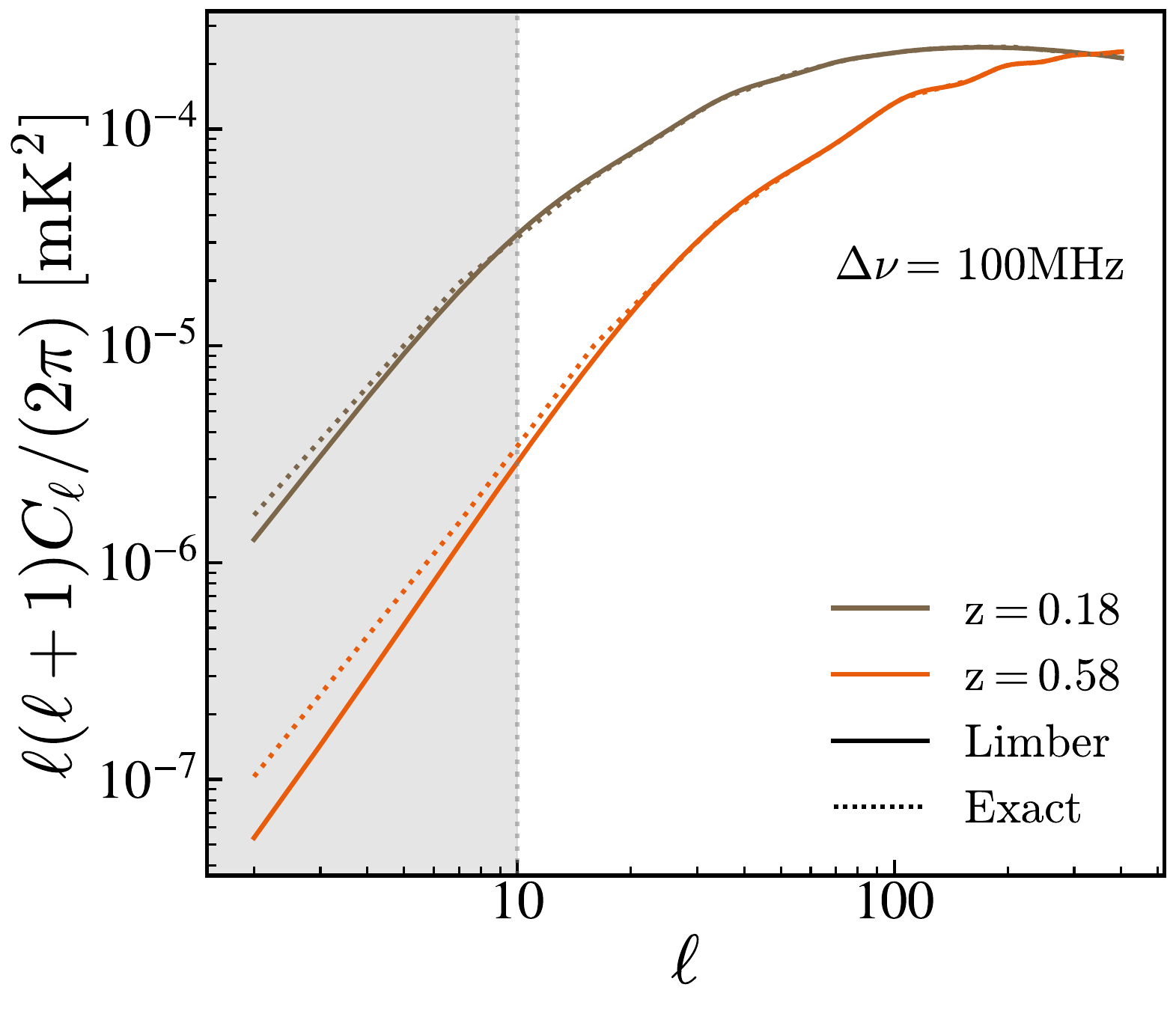}
       \includegraphics[width=8cm,height=7cm]{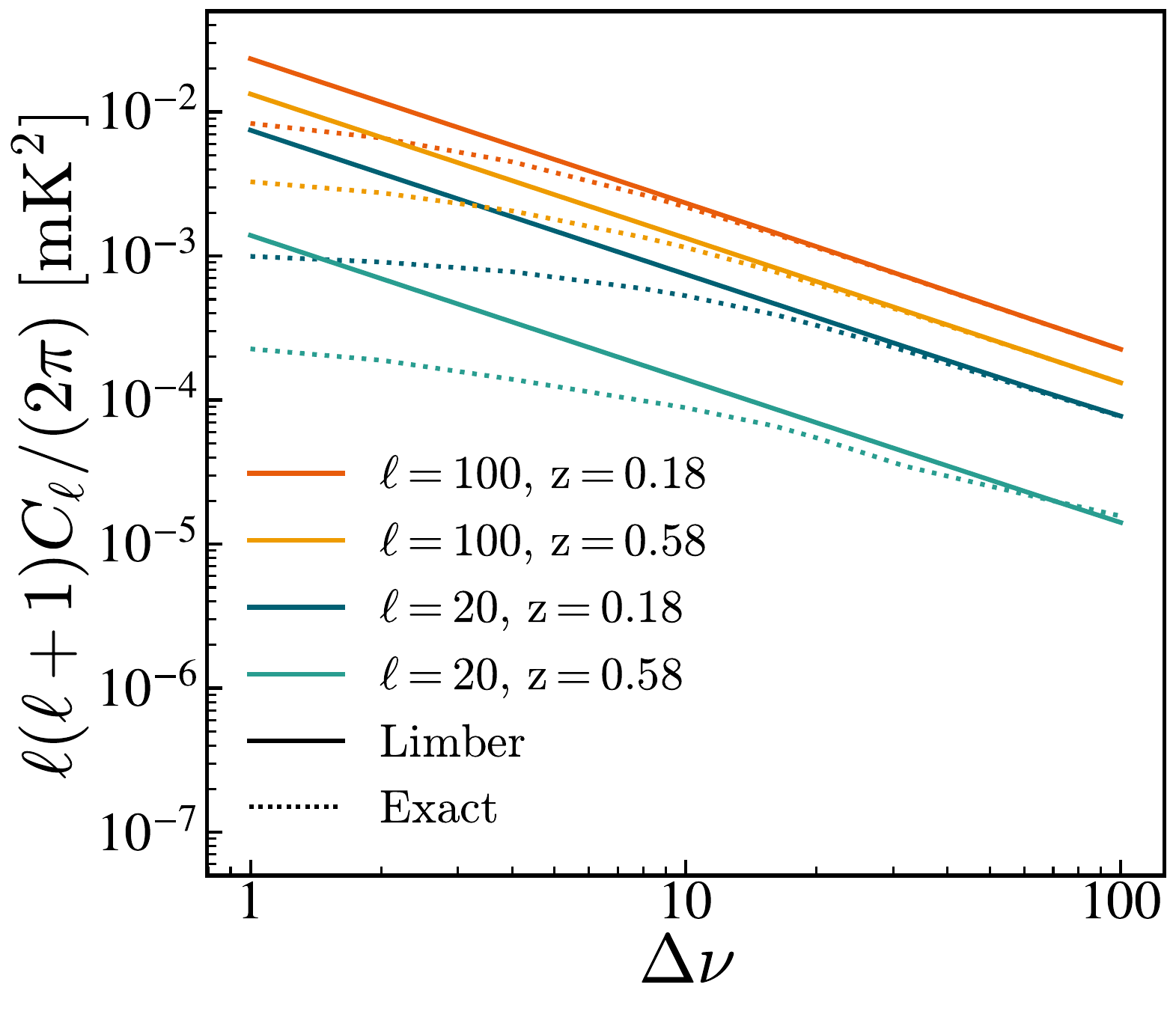}
    \caption {Comparisons of the exact calculation and the Limber approximation at redshifts of $z$ = 0.18 and $z$ = 0.58, respectively. The Limber approximation is very efficient and also consistent with the exact calculations. The deviations of the HI power spectra as a function of the bandwidth $\Delta \nu$ in the last subplot indicate that tomographic IM surveys with larger bandwidths can effectively avoid biases in the power spectrum calculations.  } 
    \label{exactandlimber} 
\end{figure*}

\begin{figure*}   
   \includegraphics[width=1.0\linewidth]{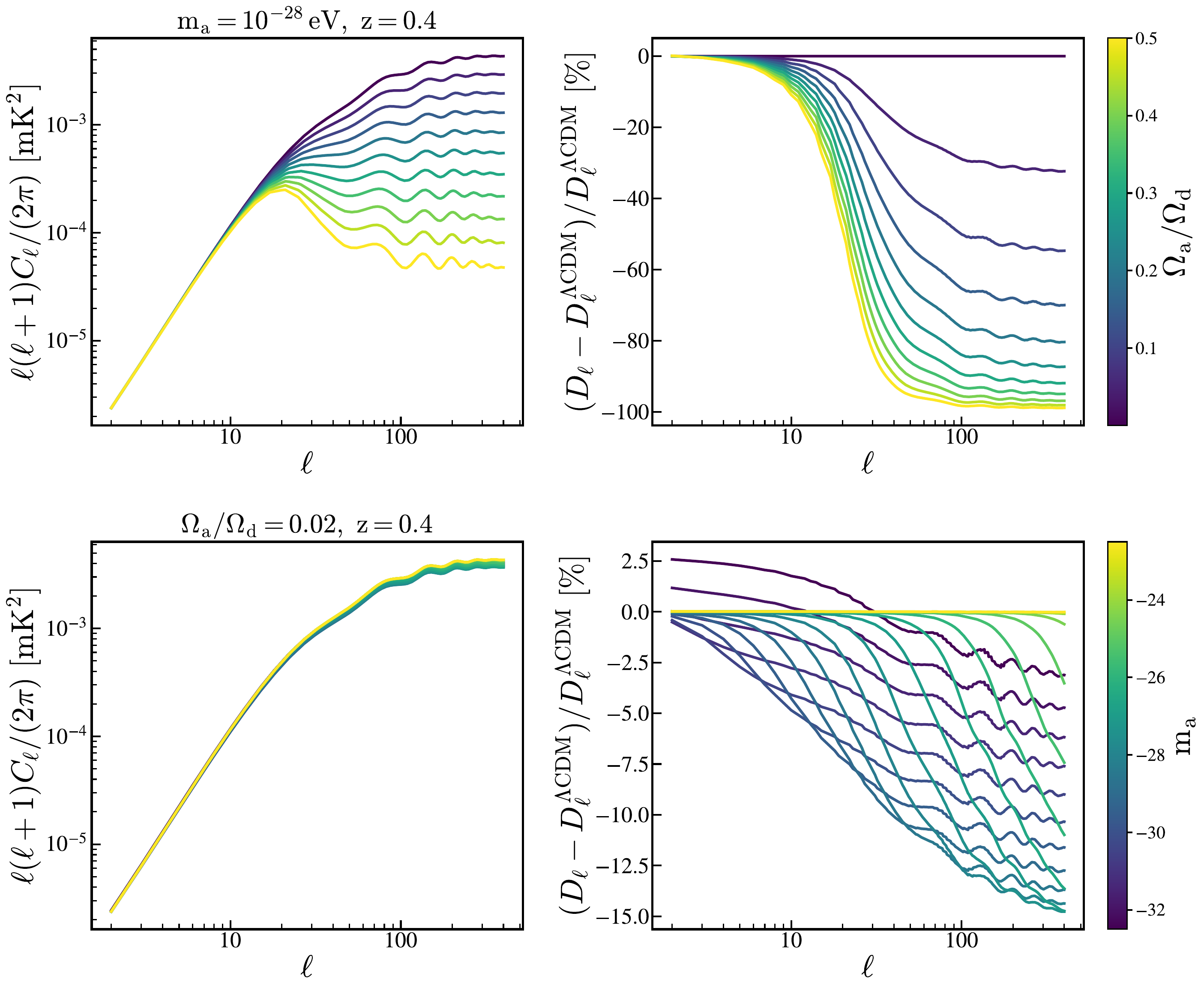}
    \caption{ULAs impact on HI angular power spectra with respect to energy density fractions $\Omega_a$ and masses $m_a$ while fixing the sum of neutrino masses at $\Sigma m_{\nu} = 0.06\, \ev$. To compare with the $\Lambda \mathrm{CDM}$ scenario, we fix the fraction ratio of ULAs to total dark matter energy density $\Omega_a/\Omega_d = 0$, the sum of neutrino masses $\Sigma m_{\nu} = 0.06 \, \ev$, and vary the total energy density fraction of dark matter $\Omega_d=\Omega_c+\Omega_{\nu}+\Omega_a$ accordingly while the fractions of ULA energy density $\Omega_a$ vary. Here, $\Omega_c$ and $\Omega_{\nu}$ are the energy density fractions of cold dark matter and neutrinos, respectively. The structure suppression of ULAs is sensitive to the ULA fraction. Moreover, ULAs leave unique mass signatures on spatial clustering properties. In this plot, the convention is defined as $D_{\ell}=\ell(\ell+1)/(2\pi)C_{\ell}$.} 
    \label{axioneffectsanother} 
\end{figure*}

\begin{figure*}   
   \includegraphics[width=1.0\linewidth]{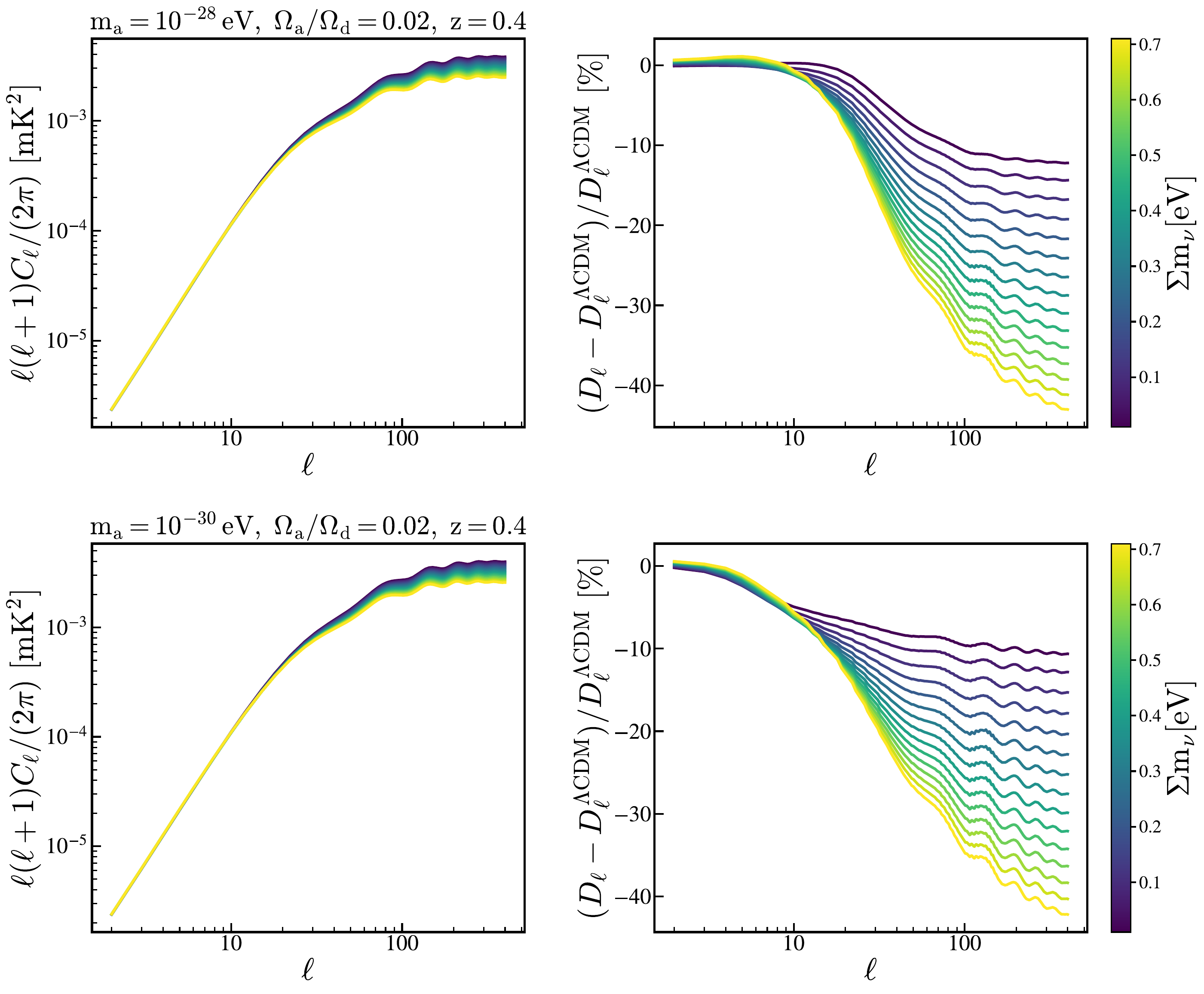}
    \caption{The impact of neutrino mass on HI angular power spectra at a fixed fraction ratio of ULAs to total dark matter energy density $\Omega_a/\Omega_d = 0.02$ and two different ULA masses, i.e., $m_a=10^{-28}\,\ev$ and $m_a=10^{-30}\,\ev$, in the first and second rows, respectively. To compare with the $\Lambda \mathrm{CDM}$ scenario, we fix $\Omega_a/\Omega_d = 0$, $\Sigma m_{\nu} = 0.06 \, \ev$, and vary the total energy density fraction of dark matter $\Omega_d=\Omega_c+\Omega_{\nu}+\Omega_a$ accordingly while varying the energy density fraction of neutrinos $\Omega_{\nu}$. The structure suppression due to neutrino masses is different from that of ULAs, making both suppression effects distinguishable in future IM surveys. } 
    \label{neutrinoeffectsanother1} 
\end{figure*}

The perturbation variables such as the baryon density contrast $\delta_b$ are generated by \axioncamb\ on discrete grids from $10^{-5}<k<5\,h{\rm Mpc}^{-1}$ and $0<z<1.2$, respectively, and are further interpolated for the power spectrum calculations as described by Eq. (\ref{exact}). 

In this work, we adopt the best-fit Planck cosmological parameters~\cite{2020A&A...641A...6P} and ULA parameters from~\cite{2021MNRAS.500.3162B}. Specifically, the fiducial parameters are \{$w_0, \Omega_a/\Omega_d, \sum m_{\nu}, h, \Omega_b, \Omega_d$\}=\{-1, 0.02, 0.2, 0.69, 0.04667, 0.25142\}.

There are a few simplifications made for the power spectrum calculations. The matter power spectrum only accounts for linear clustering, i.e., two-halo contribution. This simplification is largely justified in~\cite{2021MNRAS.500.3162B} where it is found that the non-linear scales probed by near-future IM experiments can mildly affect the constraints on the ULAs. Furthermore, we will mostly avoid the non-linear regime in our analysis given our choices of angular-scale ranges explored by the near future IM experiments. Nevertheless, non-linear effects can become more constraining once the next-generation CMB datasets such as CMB-S4~\cite{2019arXiv190704473A} are combined. In addition, we neglect redshift-space distortions (RSDs) and assume that the HI bias is constant, i.e., $b_{\rm HI}=1$, which is different from the halo-model formalism and it predicts that the HI bias can be enhanced at a higher ULA mass regime. Accurately accounting for the RSDs and bias is beyond the scope of this work, mainly because dedicated numerical simulations are required to model ULA effects. In principle, the halo model~\cite{2002PhR...372....1C} could be implemented to better address these issues but it could still introduce significant modeling uncertainties complicating the studies of this work.

The exact expression in Eq. (\ref{exact}) can be reduced to the approximated form, i.e., the Limber approximation, using the orthogonal relation $\int k^2dk j_{\ell}(k\chi)j_{\ell}(k\chi')=\pi/(2\chi^2)\delta(\chi-\chi')$ where $\delta$ denotes a delta function. Thus, the power spectrum with the Limber approximation~\cite{1953ApJ...117..134L} is reduced to
\begin{equation}
C^{\nu_1\nu_2}_{\ell}=\int\frac{d\chi}{\chi^2}T_b^2(z)W_{\nu_1}(z)W_{\nu_2}(z)\mathcal{P}(k)\mathcal{S}^2(k,z).\label{limber}
\end{equation}

However, the Limber approximation may not be sufficiently accurate for narrow tomographic bands, as argued in the literature~\cite{2020JCAP...05..010F}. We compare the power spectrum calculations of the exact formalism with those of the Limber approximation to determine whether the latter can still yield consistent results for multipole ranges at the scales of interest in this work. Comparisons between the exact calculation and the Limber approximation are shown for two representative frequencies in Fig. \ref{exactandlimber} at redshifts of $z$ = 0.18 and $z$ = 0.58, respectively. 

The Limber approximation agrees with the exact calculations when the bandwidth $\Delta \nu$ is sufficiently large. However, it will deviate from the exact calculations when the bandwidth $\Delta \nu$ is too narrow. This difference lies in the fact that a narrow tomographic bin violates the orthogonality relation $\int k^2dk j_{\ell}(k\chi)j_{\ell}(k\chi')=\pi/(2\chi^2)\delta(\chi-\chi')$ based on which the exact formalism is reduced to the Limber. Furthermore, we investigate how the deviations change as a function of the bandwidth $\Delta \nu$ in Fig. \ref{exactandlimber}.  

The validations in Fig. \ref{exactandlimber} indicate that the power spectrum calculation with the Limber approximation is not only quite satisfactory but also more efficient than the exact calculations~\cite{2008PhRvD..78l3506L}. Therefore, we adopt the power spectrum calculation with the Limber approximation as our baseline formalism.

We investigate the fractional differences of the HI power spectra with different ULA masses while varying the fraction ratio of ULAs to total dark matter $\Omega_a/\Omega_d$. The results in the top panel of Fig. \ref{axioneffectsanother} show the fractional differences at a ULA mass of $m_a=10^{-28}$ {\ev} in the dark-energy-like (DE-like) mass regime, implying that the structure suppression is sensitive to the ULA fractions.

We fix the fraction ratio of ULAs to total dark matter to be $\Omega_a/\Omega_d$ = 0.02 and study how the power spectra are affected by the ULA masses in the bottom panel of Fig. \ref{axioneffectsanother}. Although they are DE-like at low mass regimes, ULAs can still enhance the matter power spectra at very large scales. This feature is consistent with the result in~\cite{2021MNRAS.500.3162B}. 

The bottom panel of Fig. \ref{neutrinoeffectsanother1} shows the power spectrum calculations with the sum of neutrino masses $\Sigma m_{\nu}$ varying. Different masses show a generic trend of structure suppression~\cite{2012arXiv1212.6154L} which is different from that of ULAs. We also compare the HI angular power spectrum with different ULA masses in Fig. \ref{neutrinoeffectsanother1} and the neutrino-induced structure suppression is further modified. This indicates that some level of degeneracy between these parameters should be expected given the similar structure suppression.

\section{Experimental configurations of this work}
We only consider a single-dish IM experiment with $N_b$ narrow bands with a bandwidth $\Delta\nu$ = 10 MHz since we mainly focus on the large-scale modes of the HI fluctuations. The cross-power spectra among different frequencies are neglected because of the large bandwidth and the frequency channels are almost decorrelated. 

In this work, we only consider a minimum experimental setting for a particular IM survey in Table \ref{expset} and defer detailed instrumental and systematic investigations to future work. 

\begin{table}[h]
\centering
\caption{Experiment configuration of this work. Only a single-dish-like IM experiment is assumed for this work for simplicity, and the impact of different experimental designs on the parameters is deferred to future work. }
\begin{tabular}{lc}
\toprule

\midrule 
Frequency range (MHz) & [900, 1200]  \\
Redshift range & [0.18, 0.58]  \\
Beam resolution $\theta_{\mathrm{FWHM}}$ (arcmin) & 30 \\
Sky coverage $f_{\rm sky}$ & $1/4$  \\
Channel bandwidth $\delta \nu$ (MHz) & 10  \\
Number of channels $N_{\mathrm{bin}}$ & 31  \\
$\mathrm{White \,\,noise} \,\, w^{-1}$ (mK) & $  2.08\times 10^{-4}$  \\
\bottomrule 
\end{tabular}
\label{expset}
\end{table}

\begin{figure}   
      \includegraphics[width=8cm,height=7cm]{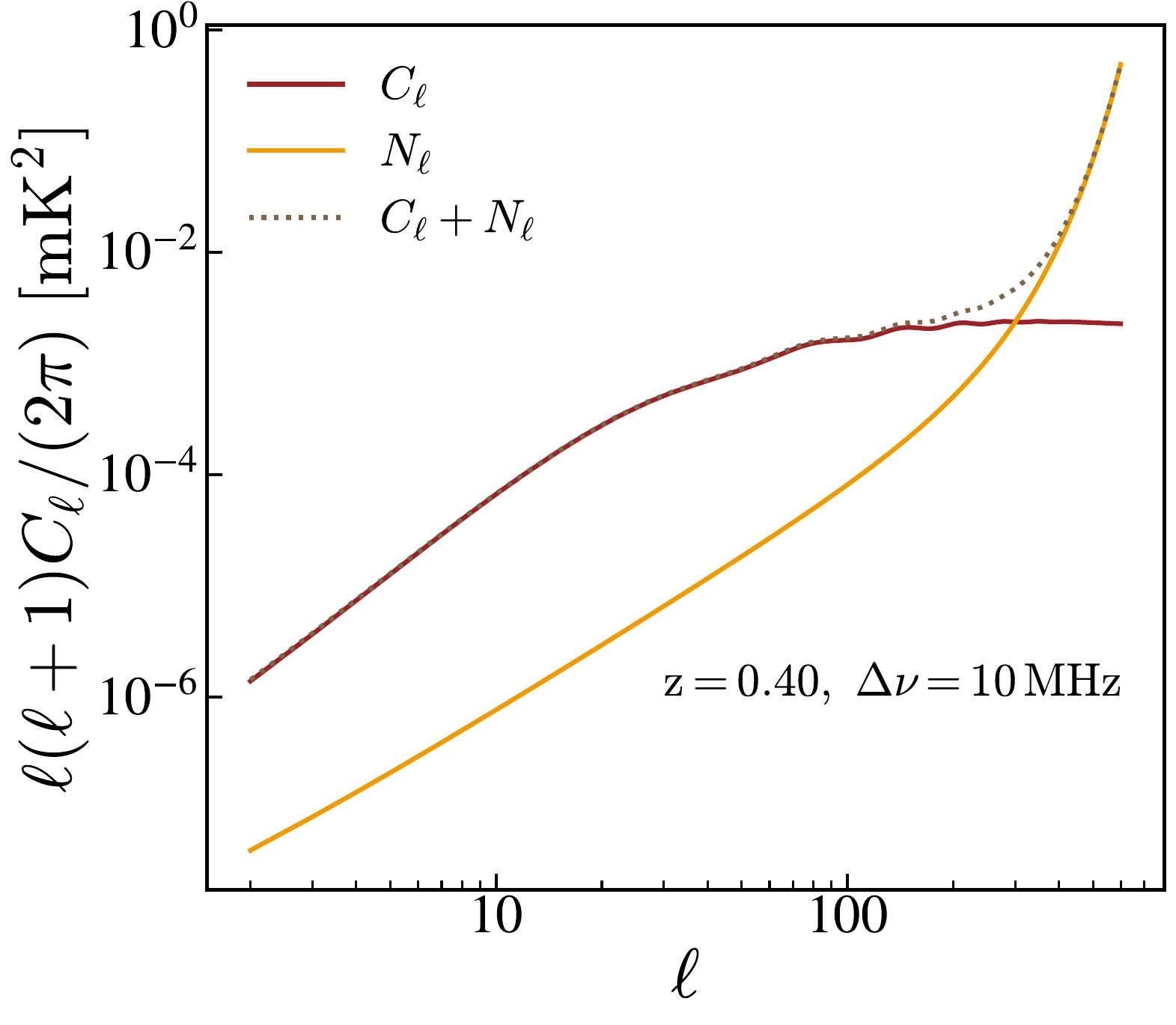}
    \caption{A representative plot of the power spectra of the HI signal and thermal noise at $z=0.4$. We only assume a white noise model with a low noise level and assume a frequency-independent Gaussian beam profile for the IM experiment. The beam-deconvolved noise component is shown as a yellow solid line. } 
    \label{clnlcompare} 
\end{figure}

\begin{figure}
\includegraphics[width=8cm,height=8cm]{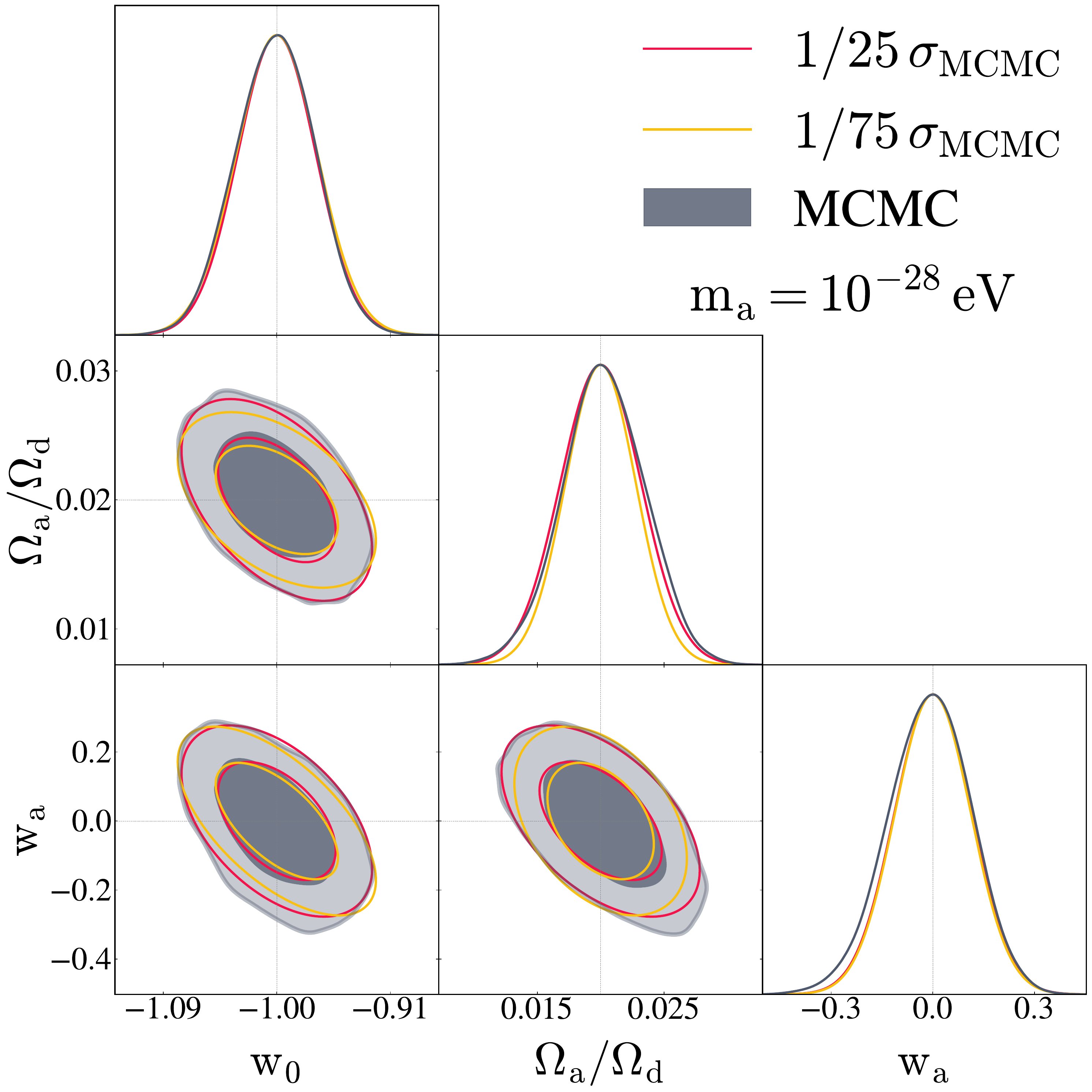}
\caption{Crosschecks of derivative step sizes used in the FM analyses. Two stepsizes in both the red and yellow curves are tested in this work, reaching consistent contours.} 
\label{comparedifferfisherstep}
\end{figure}

\begin{figure}   
    \includegraphics[width=8cm,height=8cm]{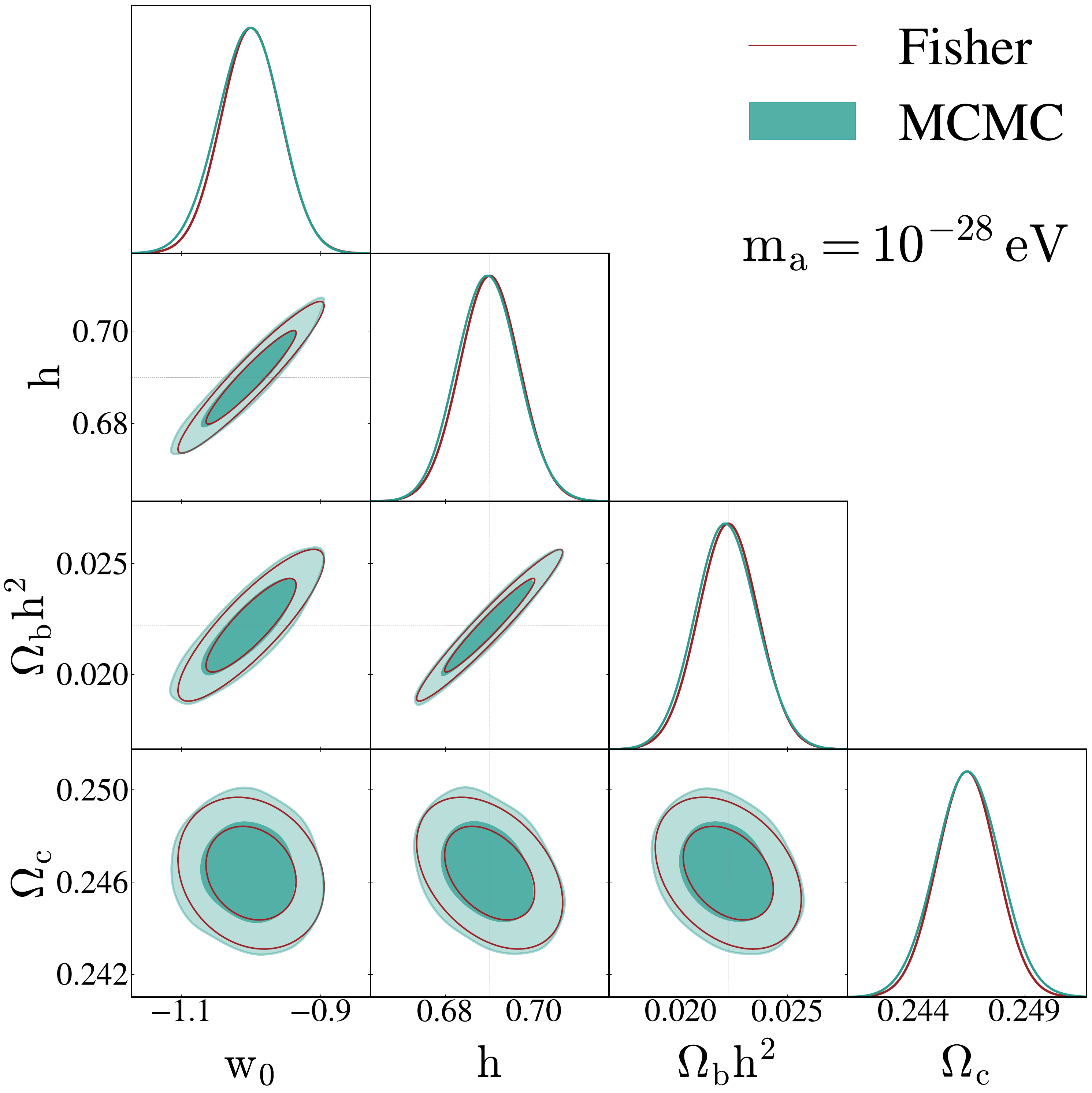} 
    \caption{Consistency check for the posterior distribution functions from both the Fisher matrix formalism and the MCMC method. The $\Lambda$CDM model is considered for this test and both methods can yield consistent results.} 
    \label{lcdm} 
\end{figure}

\begin{figure}   
      \includegraphics[width=8cm,height=8cm]{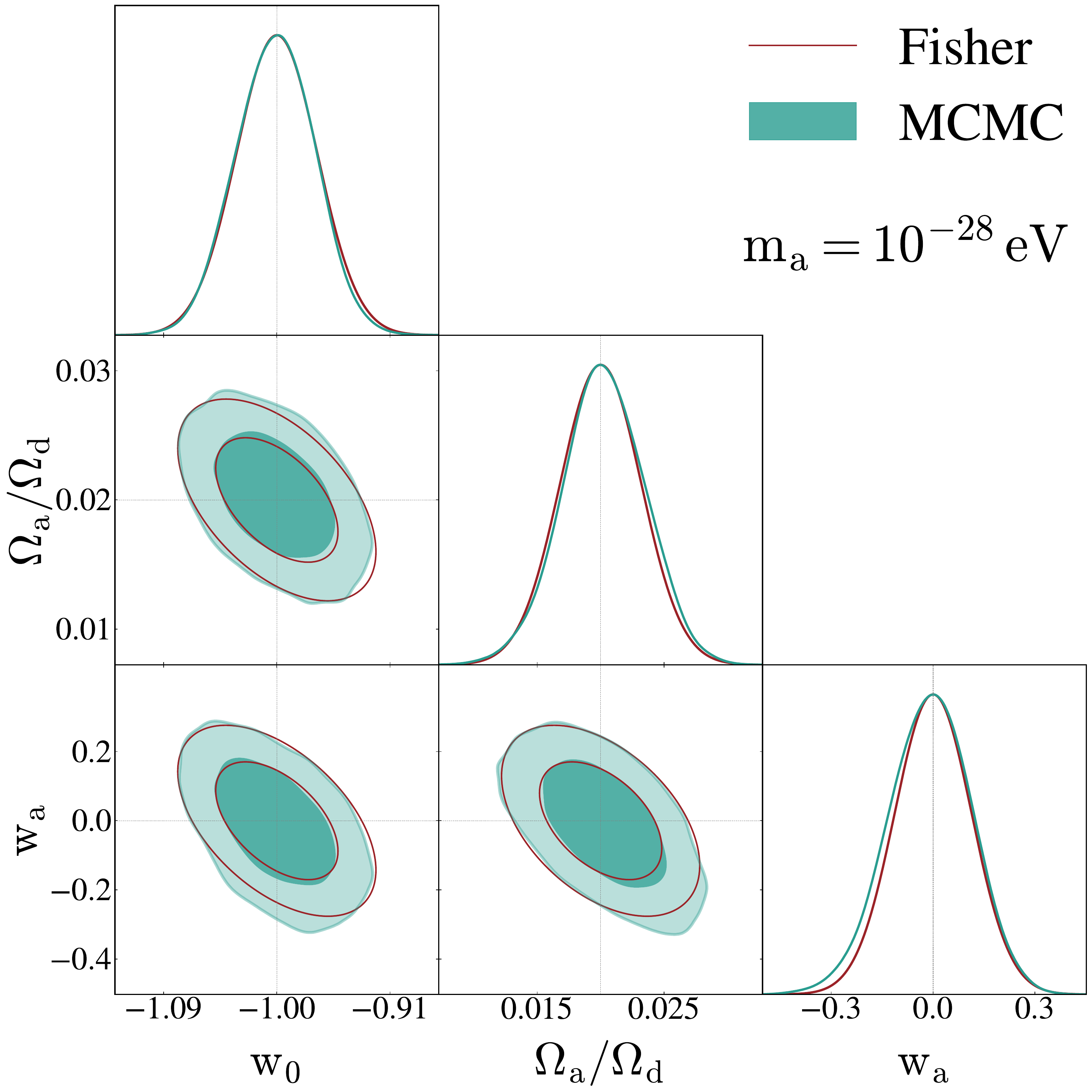}
    \caption{Consistency check for the minimum model with varying fraction ratios of ULAs to total dark matter, i.e., $\Omega_a/\Omega_d$ as the new parameter at a fixed fraction of the total dark matter $\Omega_d$ and a ULA mass $m_a = 10^{-28}\,\ev$ for a cosmic-variance-limited IM experiment.} 
    \label{cplmodel} 
\end{figure}

\begin{figure}         
\includegraphics[width=8cm,height=8cm]{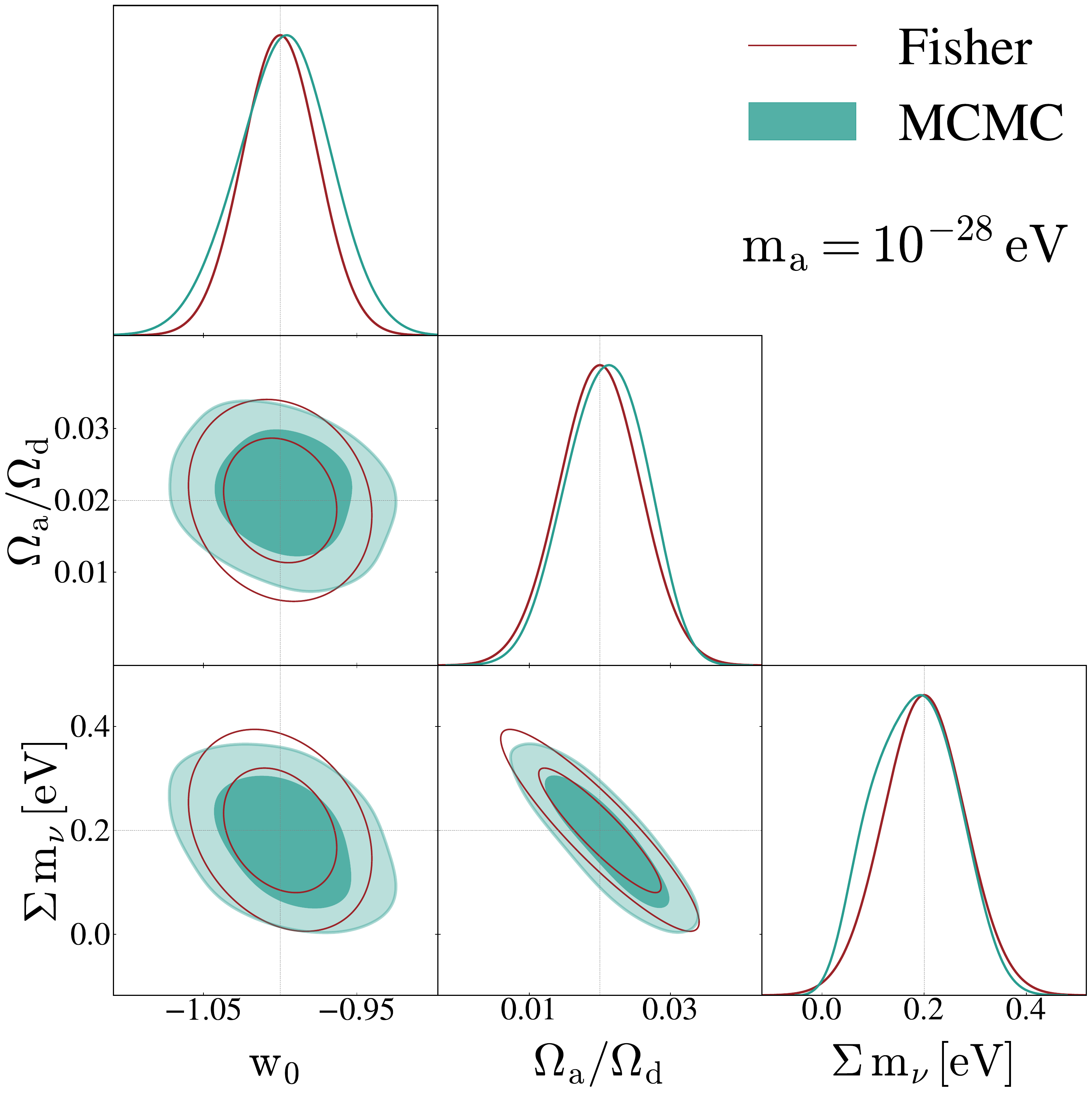}
    \caption{Constraints on the model ${\rm DE}+\sum m_{\nu}+{\rm ULAs}$ at a ULA mass $m_a = 10^{-28}$ {\ev} for a cosmic-variance-limited IM experiment. The input values $\{w_0, \Omega_a/\Omega_d, \Sigma m_{\nu}[\ev]\}=\{-1.0, 0.02, 0.2\}$ are all correctly recovered from the analyses.} 
    \label{neutrinomodel} 
\end{figure}

\begin{figure}
\includegraphics[width=8cm,height=8cm]{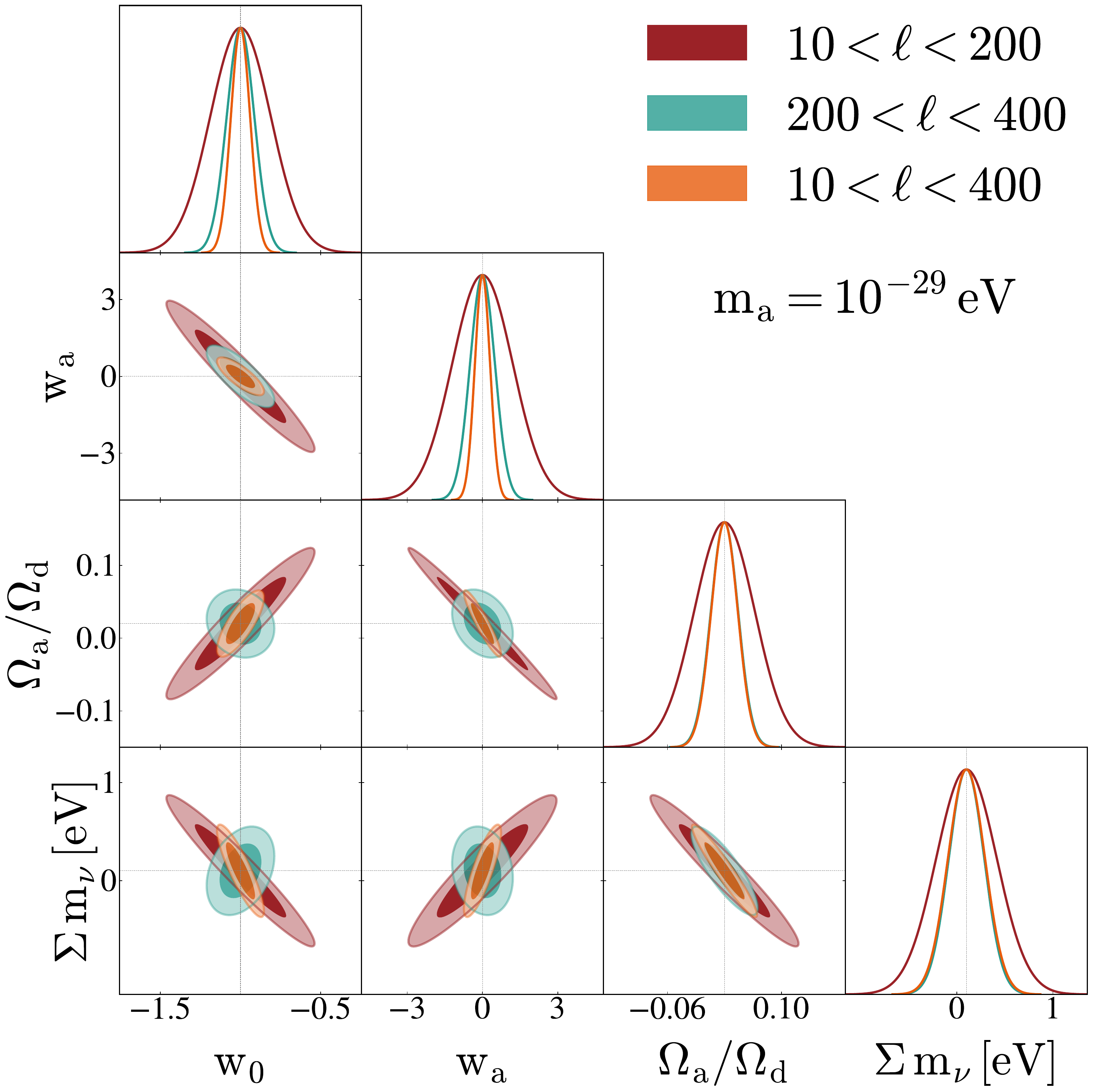}
\caption{Constraints on the model ${\rm CPL}+\sum m_{\nu}+{\rm ULAs}$ with $\ell$ splits. The full $\ell$ range $10<\ell<400$ is split into two ranges and different $\ell$ ranges show different parameter correlations. Tight constraints on the ULA fraction and neutrino mass can be achieved from the full $\ell$ range. } 
\label{elltest}
\end{figure}

\begin{figure}
\includegraphics[width=8cm,height=8cm]{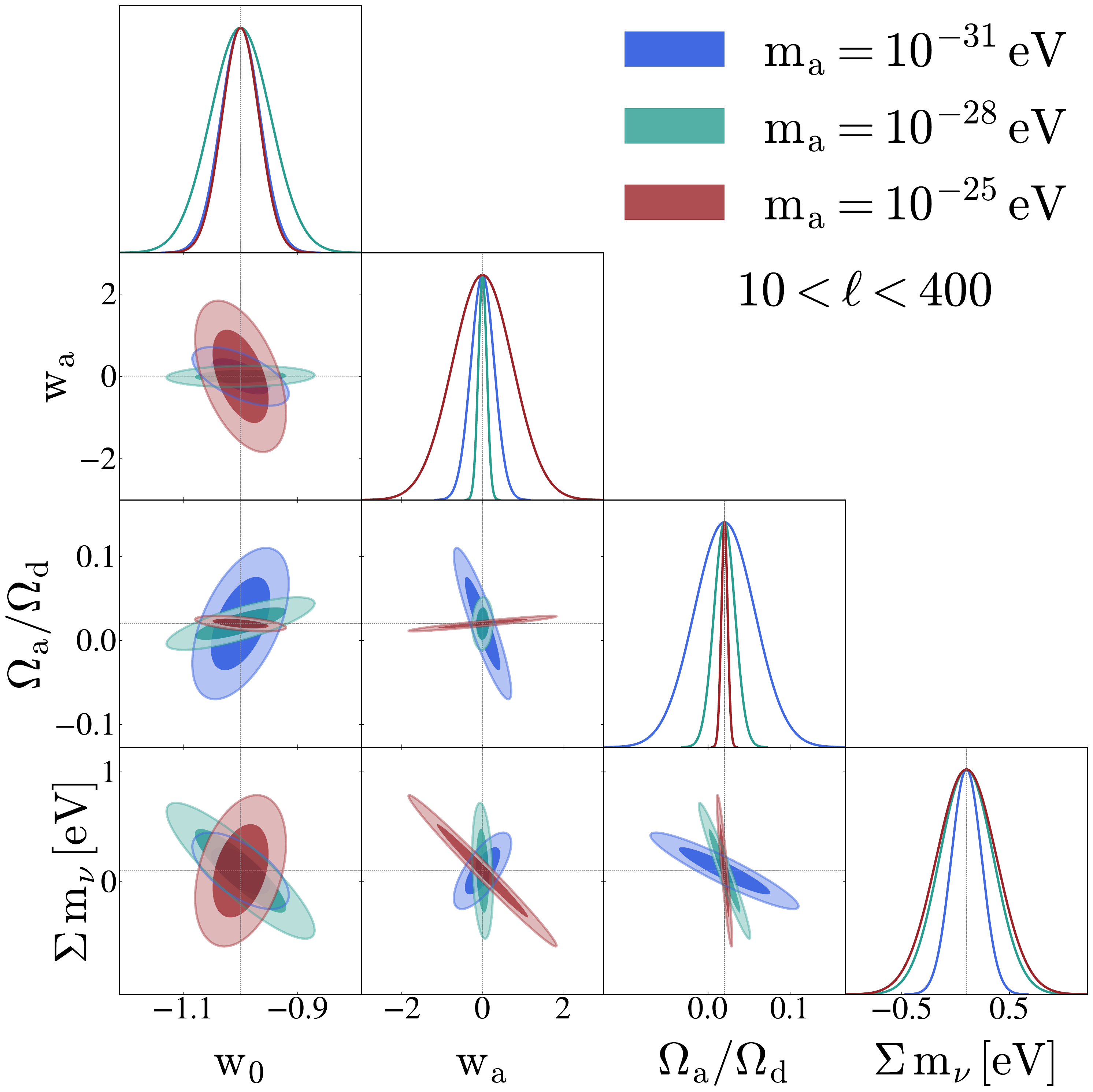}
\caption{Constraints on the model ${\rm CPL}+\sum m_{\nu}+{\rm ULAs}$ with varying ULA masses. Due to the scale-dependent spatial correlations, the ULAs and neutrinos manifest different correlation patterns at different ULA masses which determine either the DE-like or DM-like nature of the ULAs. } 
\label{masstest}
\end{figure}

\begin{table*}
\centering
\caption{Different tests performed in this work. The goal of these tests is to investigate particular correlations among parameters. The first column is a list of the test models, the second column shows the parameter sets for different models, and the third column lists the constraints imposed on different energy density fractions. A spatially flat universe is assumed in this work, i.e., $\Omega_k$ = 0. In the following table, $\Omega_i$ refers to a specific particle species and $i\in\{a,b, c, \nu,\Lambda\}$ = \{ULAs, baryons, cold dark matter, neutrinos, dark energy\}.}
\begin{tabular}{c|c|c|c}
\hline
&{\rm Models}&{\rm Parameter\,\, Set}&{\rm Constraints}\\\hline
1&Flat-wCDM & \{$w_0,h,\Omega_b,\Omega_c$\} & $\Omega_c + \Omega_b + \Omega_{\Lambda} = 1$  \\
2&${\rm CPL}+{\rm ULAs}$ & \{$w_0,w_a,h,\Omega_a/\Omega_d$\} & $\Omega_c + \Omega_a + \Omega_b + \Omega_{\Lambda} = 1$  \\
3&${\rm DE}+\sum m_{\nu}+{\rm ULAs}$ & \{$w_0,\Omega_a/\Omega_d,\sum m_{\nu}$\} & $\Omega_c + \Omega_a + \Omega_b + \Omega_{\nu} + \Omega_{\Lambda} = 1$\\
4&${\rm CPL}+\sum m_{\nu}+{\rm ULAs}$ & \{$w_0,w_a,\Omega_a/\Omega_d,\sum m_{\nu}$\} & $\Omega_c + \Omega_a + \Omega_b + \Omega_{\nu} + \Omega_{\Lambda} = 1$ \\
5&${\rm DE}+\sum m_{\nu}+{\rm ULAs}$+Flat-wCDM & \{$w_0,\Omega_a/\Omega_d,\sum m_{\nu},h,\Omega_b,\Omega_d$\} & $\Omega_c + \Omega_a + \Omega_b + \Omega_{\nu} + \Omega_{\Lambda} = 1$ \\
\hline
\end{tabular}
\label{mcmctests}
\end{table*}

The thermal noise is assumed to be white noise, which can be modeled as $N_{\ell} = w^{-2}b_{\ell}^{-2}$~\cite{2015ApJ...803...21B,2022MNRAS.510.1495X}. Here, $w^{-1}$ is the instrumental noise property, and $b_{\ell}$ is the beam transfer function. A representative plot of the beam-deconvolved noise power spectrum is shown in Fig. \ref{clnlcompare}, where we consider a frequency-independent Gaussian beam with full-width-at-half-maximum (FWHM) $30$ arcmin and the beam transfer function is $b_{\ell}=\exp{[-\ell^2\theta_{\rm FWHM}^2/(16\ln2)]}$. Additionally, we consider a low-noise IM experiment and determine the noise property $w^{-1}$, assuming that the noise power spectrum is at the same level as the signal at $\ell=300$. We do not consider noise correlations at different frequencies. A detailed investigation of correlated noise among different frequencies, such as $1/f$ noise~\cite{2002A&A...391.1185S,2018MNRAS.478.2416H}, is beyond the scope of this work.

Therefore, the mock power spectrum is constructed as
\begin{equation}
C^{\nu_1\nu_2, \rm obs}_{\ell}=C^{\nu_1\nu_2, \rm HI}_{\ell}+N^{\nu_1\nu_2}_{\ell}
\end{equation}
at each frequency pair. We assume perfect foreground removal for this work and do not add foreground residual terms to the mock power spectra. Different foreground removal algorithms have been proposed in the literature~\cite{2003ApJS..148...97B,2009A&A...493..835D,2014PhRvD..90b3018L, 2014PhRvD..90b3019L, 2023arXiv230814777F} and a $10^{-5}$ to $10^{-6}$ level of foreground removal could be achievable in future IM surveys, making the foreground residual much smaller than the thermal noise. Thus, we neglect the impact of the foreground residuals on ULA physics.

We infer the ULA properties from the log-likelihood function 
\begin{equation}
-2\ln \mathcal{L}(\textbf{P})=\displaystyle\sum_{\ell} \sum_{\{\nu_1\nu_2\}}\Big[\frac{C^{\nu_1\nu_2, \rm obs}_{\ell}-C^{\nu_1\nu_2, \rm model}_{\ell}(\textbf{P})}{\Delta C^{\nu_1\nu_2}_{\ell}}\Big]^2,\label{loglike}
\end{equation}where the band power errors are theoretically predicted by the Knox formula
\begin{equation}
(\Delta C_{\ell})^2=\frac{2}{(2\ell+1)f_{\rm sky}\Delta\ell}[C^{\nu_1\nu_2, \rm HI}_{\ell}+N^{\nu_1\nu_2}_{\ell}]^2,\label{knox}
\end{equation}
with all the parameters fixed. The theoretical models described as $C^{\nu_1\nu_2, \rm model}_{\ell}(\textbf{P})$ are calculated for a specific parameter set $\textbf{P}$.

We note that Eq. (\ref{loglike}) is only an approximation of the real likelihood and assumes a Gaussianity of the error around the estimate of the power spectrum. This will break down at large scales, which are not considered in this work, making this a perfectly valid approximation. These power spectrum errors in Eq. (\ref{knox}) are essentially the diagonal elements of the full covariance, and the off-diagonal elements should have a negligible impact on the parameter inference due to the broad bandwidth $\Delta \nu$~\cite{2023arXiv230814777F}.  In this work, only the autopower spectra with $\nu_1=\nu_2$ are considered for the analysis. 

\begin{figure*}
\includegraphics[width=8cm,height=8cm]{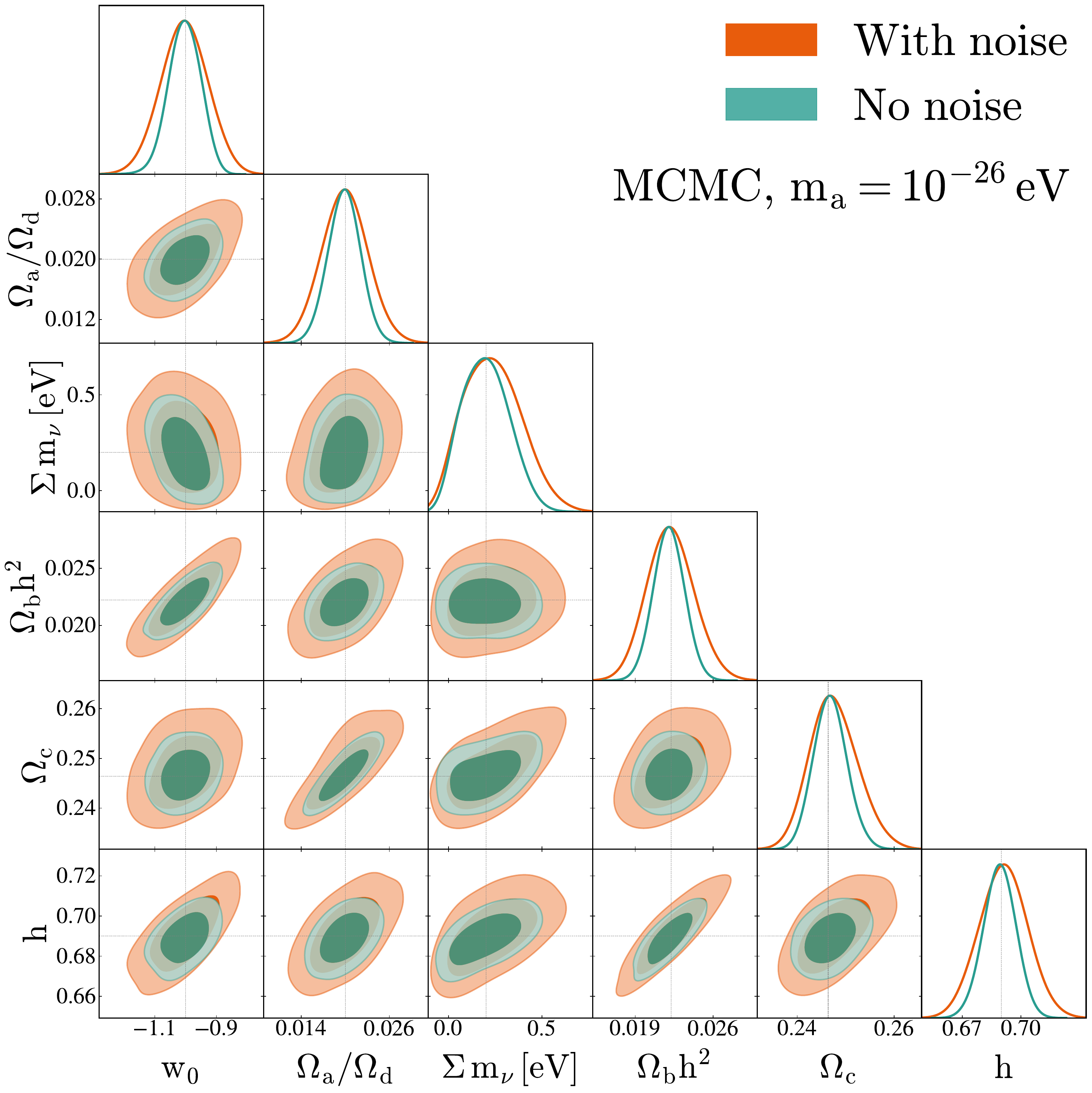}
\includegraphics[width=8cm,height=8cm]{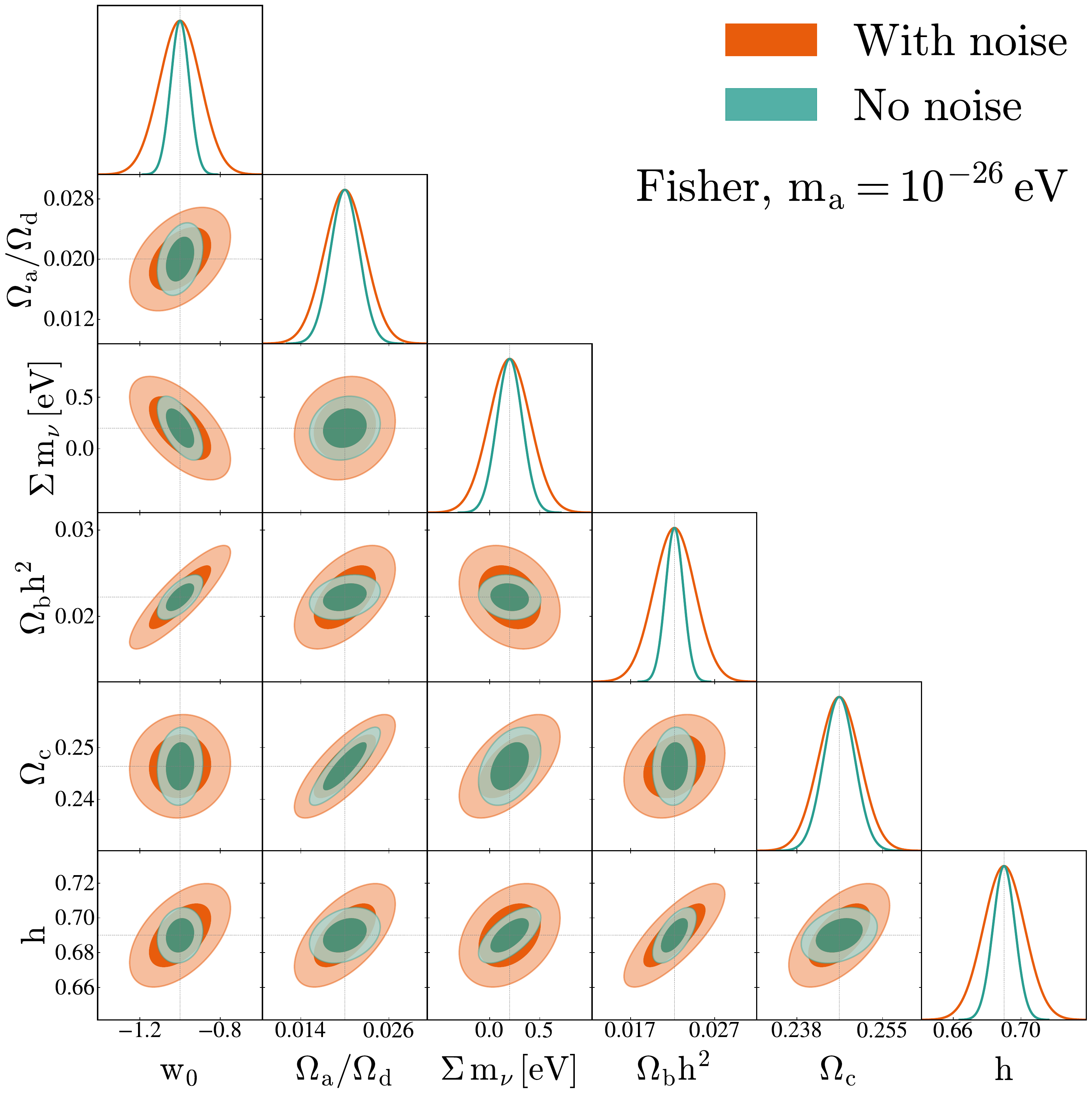}
\caption{Comparison of the MCMC and FM results with and without noise. Both the contours of the MCMC (left) and the FM (right) analyses show consistent degeneracy directions among the different parameters. Parameter constraints are weakened as expected by the instrumental noise. } 
\label{comparenoise1}
\end{figure*}

\section{Validations for both Fisher Matrix formalism and MCMC analyses}

In light of the results in Fig. \ref{axioneffectsanother}, ULAs have distinct suppression patterns from neutrinos, implying that these unique spatial correlations of both ULAs and neutrinos could be detectable in future IM experiments. 

The Fisher Matrix (FM) formalism is very efficient and has sufficient precision to explore the parameter space if it is approximately Gaussian distributed. However, the FM will break down if the parameter space is non-Gaussian. In such a scenario, extended FM methods have been proposed and can perturbatively approximate non-Gaussian contributions. One representative method is the Derivative Approximation for LIkelihoods (DALI) which employs Taylor expansions of the likelihood functions~\cite{2014MNRAS.441.1831S}.

The FM implicitly assumes that the posterior space is Gaussian and may yield biased parameter estimations. 
In this work, we adopt both the FM formalism and Markov chain Monte Carlo (MCMC) analyses to crosscheck constraints on ULA physics. A brute-force calculation of the FM formalism is implemented from the definition, i.e., 
\begin{equation}
    F_{ij}=-\Big\langle\frac{\partial^2\ln \mathcal{L}}{\partial p_i\partial p_j}\Big\rangle 
\end{equation}and a numerical differencing is applied to the derivative calculations as described in Eq. (\ref{nder}). 
\begin{eqnarray}
    \frac{\partial^2 f(x,y)}{\partial x\partial y}
    &=& \frac{1}{4\Delta x \Delta y}\Big[f(x + \Delta x, y + \Delta y) \nonumber\\&+& f(x - \Delta x, y - \Delta y) -f(x + \Delta x, y - \Delta y) \nonumber\\&-& f(x - \Delta x, y + \Delta y)\Big],
\label{nder}
\end{eqnarray}
Here, $p_i$ denotes a specific parameter and $f(x,y)=\ln \mathcal{L}(\theta_i,\theta_j)$. We have carefully chosen the parameter steps to ensure satisfactory convergence.

We test different step sizes chosen for the FM analysis and investigate the impact on parameter constraints in Fig. \ref{comparedifferfisherstep}. The choice of the derivative step is subtle in that the FM results are not completely independent of the step sizes and there are certain limitations of the FM analyses which only encapsulate the second moments of the PDFs with higher moments omitted. Although the steps are less motivated than would be desirable, Fig.\ref{comparedifferfisherstep} shows that there remains good agreement with the MCMC results~\cite{2006JCAP...10..013P}. In this work, we test both a Flat-wCDM Universe with a constant equation of state $w=w_0$ and a Chevallier-Polarski-Linder (CPL) parameterization with $w=w_0+w_az/(1+z)$~\cite{2001IJMPD..10..213C,2003PhRvL..90i1301L}.

Fig. \ref{lcdm} shows the results for the Flat-wCDM model. We use both the FM and MCMC calculations to obtain the two-dimensional posterior distribution functions (PDFs) for parameter pairs among the Flat-wCDM parameter set $\{w_0, h, \Omega_bh^2, \Omega_c\}$. The overlapping contours indicate satisfactory consistency between the FM and MCMC results.

We test a minimum model with ULAs only and consider only three parameters $\{w_0,w_a,\Omega_a/\Omega_d\}$ for a given ULA mass $m_a=10^{-28}$ {\ev} in Fig. \ref{cplmodel}. In the DE-like mass regime, the ULA's energy density fraction is degenerate with the DE's equation of state. Moreover, both the FM and MCMC calculations show consistent results and can recover the input values without biases.

\section{Constraints on ULA properties}
The ULA can describe both the DE and the DM at different mass ranges. In this section, we test ${\rm DE}+\sum m_{\nu}+{\rm ULAs}$ model listed in Table \ref{mcmctests} to study the degeneracies among the DE, neutrinos, and ULAs. The FM and MCMC results are shown in Fig.\ref{neutrinomodel}. The same degeneracy as discussed in~\cite{PhysRevLett.95.221301} is obtained for the parameter pair $w_0\mbox{-}\Sigma m_{\nu}$ at $m_a=10^{-28}$ {\ev}. From the PDFs, we obtain $\sigma(w_0)= 0.03$, $\sigma(\Omega_a/\Omega_d)=0.005$, and $\sigma(\sum m_{\nu})=0.07{\ev}$ . The constraints on the ULA fractions and neutrino masses are quite promising.

Different angular scales may result in different constraining powers on ULA physics. To understand which angular scales contribute the most signal-to-noise ratios (SNRs), we split the $\ell$ range into two regimes $10<\ell<200$ and $200<\ell<400$ and test the model ${\rm CPL}+\sum m_{\nu}+{\rm ULAs}$. The results are presented in Fig.\ref{elltest}. The clustering information at $\ell>200$ has the most constraining power on the parameters and the regime $\ell<200$ is less constraining but both regimes show different correlation directions. Taking advantage of the small-scale information that can effectively break the degeneracy, the full range constraint is thus much improved compared with each of the $\ell$ splits. This can be explained by the result in Fig. \ref{axioneffectsanother} where spatial correlation patterns of ULAs vary at different scales. These characteristics may help distinguish ULAs from neutrinos as DM components. We also run the MCMC to check the FM results and find consistency between the two methods.

In addition, we test the influence of the ULA mass on the model ${\rm CPL}+\sum m_{\nu}+{\rm ULAs}$, as shown in Fig. \ref{masstest}. One interesting finding from this test is that the correlation of $w_0\mbox{-}\Sigma m_{\nu}$ gradually evolved from anti-correlation to correlation as the ULA mass increased from the DE regime to the DM regime. This implies that detailed properties such as ULA masses can be inferred from the power spectrum measurements of future IM surveys.

In the previous sections, we tested the constraints on ULAs and neutrinos with the $\Lambda {\rm CDM}$ parameters fixed. There might also be interesting correlations between the ULAs and the $\Lambda {\rm CDM}$ parameters. We repeat the FM and MCMC calculations for an extended model ${\rm DE}+\sum m_{\nu}+{\rm ULAs}$+Flat-wCDM at a ULA mass of $m_a=10^{-26}$ \ev. All the PDFs from the two methods are shown in Fig. \ref{comparenoise1}, where a comparison with and without noise contribution is also made. As shown in Fig. \ref{axioneffectsanother}, the CDM fraction must be increased to compensate for the clustering power deficit caused by the ULAs in the lower mass regime. This implies that the CDM energy density fraction should increase as the ULA increases. The contour of $\Omega_c\mbox{-}\Omega_a/\Omega_d$ well reflects this correlation.

\begin{figure}
\includegraphics[width=8cm,height=7cm]{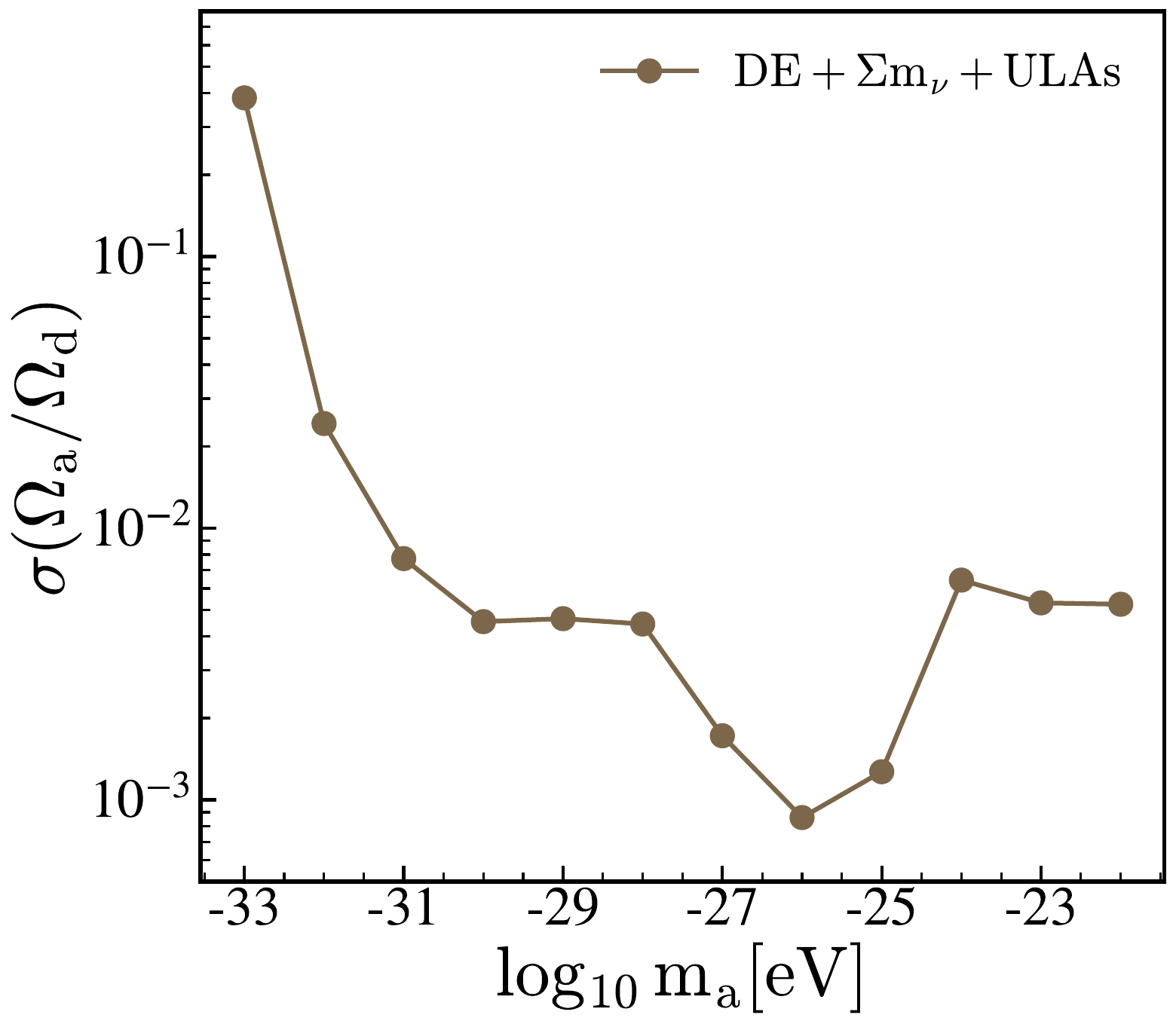}
\caption{Constraints on ULA fractions as a function of mass. The marginalized $1\sigma$ error on $\sigma(\Omega_a/\Omega_d)$ is estimated from the model ${\rm DE}+\sum m_{\nu}+{\rm ULAs}$. Future intensity mapping experiments can achieve subpercent-level measurements of the ULA energy density fraction, demonstrating a substantial improvement over next-generation CMB measurements.} 
\label{axionfractioninf}
\end{figure}

As demonstrated in Fig. \ref{axionfractioninf}, future IM surveys will achieve sub-percent level constraints on the fraction ratio of ULAs to total dark matter and can shed light on the properties of ULAs. This result is consistent with a recent study using the intensity mapping~\cite{2021MNRAS.500.3162B}. We note that only the linear clustering is considered in this work and non-linear clustering would further constrain the models. 

\section{Conclusions}

In this work, we studied the HI signatures from the post-reionization stage of the Universe. Specifically, we calculated the HI power spectra in the frequency range from 900-1200 MHz using both the exact formula and the Limber approximation. Consistency checks between the two methods have been performed, and the Limber approximation can reach satisfactory precision for modeling the power spectra of differential brightness temperature fluctuations. 

We theoretically studied the structure suppression caused by either neutrinos or ULAs and found that ULAs manifest distinct suppression patterns from neutrinos. This unique feature may be used to distinguish ULAs from neutrinos in future tomographic IM datasets. On the other hand, this indicates that there may be degeneracy between the two species.

We sample the posterior distribution functions of different parameters using both the Fisher Matrix (FM) formalism and Markov chain Monte Carlo analyses. The subtle effect of the parameter step size on the numerical differencing of the FM formalism has been carefully investigated. We find good consistency between the FM and the MCMC results.

ULAs can behave like dark energy and dark matter at different mass regimes. At the low mass regime $m_a<10^{-27}$ {\ev}, the ULAs are DE-like since they are not oscillating and have an equation of state $\sim -1$ which is similar to a cosmological constant. In the high mass regime, the ULAs are DM-like and constitute a DM subspecies, becoming degenerate with the CDM and neutrinos. 

To study the degeneracies among the CDM, neutrinos, and ULAs, we perform both the FM and the MCMC calculations to explore the parameter spaces for the Falt-wCDM model, ULA-only model, and neutrino-only model as well as extended models, finding great consistency among different parameters from different models. In the near future, tomographic intensity mapping observations will be a powerful probe of ultralight axions.\\

\acknowledgments
We are grateful for the helpful discussions with Gilbert Holder, Alessandro Marins and Pablo Motta. This work is supported by the starting grant of USTC. We acknowledge the use of the \healpix~\cite{2005ApJ...622..759G} and \emcee~\cite{2013PASP..125..306F} packages.

\bibliography{21cm}

\begin{thebibliography}{54}%
\makeatletter
\providecommand \@ifxundefined [1]{%
 \@ifx{#1\undefined}
}%
\providecommand \@ifnum [1]{%
 \ifnum #1\expandafter \@firstoftwo
 \else \expandafter \@secondoftwo
 \fi
}%
\providecommand \@ifx [1]{%
 \ifx #1\expandafter \@firstoftwo
 \else \expandafter \@secondoftwo
 \fi
}%
\providecommand \natexlab [1]{#1}%
\providecommand \enquote  [1]{``#1''}%
\providecommand \bibnamefont  [1]{#1}%
\providecommand \bibfnamefont [1]{#1}%
\providecommand \citenamefont [1]{#1}%
\providecommand \href@noop [0]{\@secondoftwo}%
\providecommand \href [0]{\begingroup \@sanitize@url \@href}%
\providecommand \@href[1]{\@@startlink{#1}\@@href}%
\providecommand \@@href[1]{\endgroup#1\@@endlink}%
\providecommand \@sanitize@url [0]{\catcode `\\12\catcode `\$12\catcode
  `\&12\catcode `\#12\catcode `\^12\catcode `\_12\catcode `\%12\relax}%
\providecommand \@@startlink[1]{}%
\providecommand \@@endlink[0]{}%
\providecommand \url  [0]{\begingroup\@sanitize@url \@url }%
\providecommand \@url [1]{\endgroup\@href {#1}{\urlprefix }}%
\providecommand \urlprefix  [0]{URL }%
\providecommand \Eprint [0]{\href }%
\providecommand \doibase [0]{https://doi.org/}%
\providecommand \selectlanguage [0]{\@gobble}%
\providecommand \bibinfo  [0]{\@secondoftwo}%
\providecommand \bibfield  [0]{\@secondoftwo}%
\providecommand \translation [1]{[#1]}%
\providecommand \BibitemOpen [0]{}%
\providecommand \bibitemStop [0]{}%
\providecommand \bibitemNoStop [0]{.\EOS\space}%
\providecommand \EOS [0]{\spacefactor3000\relax}%
\providecommand \BibitemShut  [1]{\csname bibitem#1\endcsname}%
\let\auto@bib@innerbib\@empty
\bibitem [{\citenamefont {{Peccei}}\ and\ \citenamefont
  {{Quinn}}(1977)}]{1977PhRvL..38.1440P}%
  \BibitemOpen
  \bibfield  {author} {\bibinfo {author} {\bibfnamefont {R.~D.}\ \bibnamefont
  {{Peccei}}}\ and\ \bibinfo {author} {\bibfnamefont {H.~R.}\ \bibnamefont
  {{Quinn}}},\ }\bibfield  {title} {\bibinfo {title} {{CP conservation in the
  presence of pseudoparticles}},\ }\href
  {https://doi.org/10.1103/PhysRevLett.38.1440} {\bibfield  {journal} {\bibinfo
   {journal} {\prl}\ }\textbf {\bibinfo {volume} {38}},\ \bibinfo {pages}
  {1440} (\bibinfo {year} {1977})}\BibitemShut {NoStop}%
\bibitem [{\citenamefont {{Weinberg}}(1978)}]{1978PhRvL..40..223W}%
  \BibitemOpen
  \bibfield  {author} {\bibinfo {author} {\bibfnamefont {S.}~\bibnamefont
  {{Weinberg}}},\ }\bibfield  {title} {\bibinfo {title} {{A new light
  boson?}},\ }\href {https://doi.org/10.1103/PhysRevLett.40.223} {\bibfield
  {journal} {\bibinfo  {journal} {\prl}\ }\textbf {\bibinfo {volume} {40}},\
  \bibinfo {pages} {223} (\bibinfo {year} {1978})}\BibitemShut {NoStop}%
\bibitem [{\citenamefont {{Hu}}\ \emph {et~al.}(2000)\citenamefont {{Hu}},
  \citenamefont {{Barkana}},\ and\ \citenamefont
  {{Gruzinov}}}]{2000PhRvL..85.1158H}%
  \BibitemOpen
  \bibfield  {author} {\bibinfo {author} {\bibfnamefont {W.}~\bibnamefont
  {{Hu}}}, \bibinfo {author} {\bibfnamefont {R.}~\bibnamefont {{Barkana}}},\
  and\ \bibinfo {author} {\bibfnamefont {A.}~\bibnamefont {{Gruzinov}}},\
  }\bibfield  {title} {\bibinfo {title} {{Fuzzy Cold Dark Matter: The Wave
  Properties of Ultralight Particles}},\ }\href
  {https://doi.org/10.1103/PhysRevLett.85.1158} {\bibfield  {journal} {\bibinfo
   {journal} {\prl}\ }\textbf {\bibinfo {volume} {85}},\ \bibinfo {pages}
  {1158} (\bibinfo {year} {2000})},\ \Eprint
  {https://arxiv.org/abs/astro-ph/0003365} {arXiv:astro-ph/0003365 [astro-ph]}
  \BibitemShut {NoStop}%
\bibitem [{\citenamefont {{Arvanitaki}}\ \emph {et~al.}(2010)\citenamefont
  {{Arvanitaki}}, \citenamefont {{Dimopoulos}}, \citenamefont {{Dubovsky}},
  \citenamefont {{Kaloper}},\ and\ \citenamefont
  {{March-Russell}}}]{2010PhRvD..81l3530A}%
  \BibitemOpen
  \bibfield  {author} {\bibinfo {author} {\bibfnamefont {A.}~\bibnamefont
  {{Arvanitaki}}}, \bibinfo {author} {\bibfnamefont {S.}~\bibnamefont
  {{Dimopoulos}}}, \bibinfo {author} {\bibfnamefont {S.}~\bibnamefont
  {{Dubovsky}}}, \bibinfo {author} {\bibfnamefont {N.}~\bibnamefont
  {{Kaloper}}},\ and\ \bibinfo {author} {\bibfnamefont {J.}~\bibnamefont
  {{March-Russell}}},\ }\bibfield  {title} {\bibinfo {title} {{String
  axiverse}},\ }\href {https://doi.org/10.1103/PhysRevD.81.123530} {\bibfield
  {journal} {\bibinfo  {journal} {\prd}\ }\textbf {\bibinfo {volume} {81}},\
  \bibinfo {eid} {123530} (\bibinfo {year} {2010})},\ \Eprint
  {https://arxiv.org/abs/0905.4720} {arXiv:0905.4720 [hep-th]} \BibitemShut
  {NoStop}%
\bibitem [{\citenamefont {Bond}\ \emph {et~al.}(1980)\citenamefont {Bond},
  \citenamefont {Efstathiou},\ and\ \citenamefont
  {Silk}}]{PhysRevLett.45.1980}%
  \BibitemOpen
  \bibfield  {author} {\bibinfo {author} {\bibfnamefont {J.~R.}\ \bibnamefont
  {Bond}}, \bibinfo {author} {\bibfnamefont {G.}~\bibnamefont {Efstathiou}},\
  and\ \bibinfo {author} {\bibfnamefont {J.}~\bibnamefont {Silk}},\ }\bibfield
  {title} {\bibinfo {title} {Massive neutrinos and the large-scale structure of
  the universe},\ }\href {https://doi.org/10.1103/PhysRevLett.45.1980}
  {\bibfield  {journal} {\bibinfo  {journal} {Phys. Rev. Lett.}\ }\textbf
  {\bibinfo {volume} {45}},\ \bibinfo {pages} {1980} (\bibinfo {year}
  {1980})}\BibitemShut {NoStop}%
\bibitem [{\citenamefont {{Lesgourgues}}\ and\ \citenamefont
  {{Pastor}}(2006)}]{2006PhR...429..307L}%
  \BibitemOpen
  \bibfield  {author} {\bibinfo {author} {\bibfnamefont {J.}~\bibnamefont
  {{Lesgourgues}}}\ and\ \bibinfo {author} {\bibfnamefont {S.}~\bibnamefont
  {{Pastor}}},\ }\bibfield  {title} {\bibinfo {title} {{Massive neutrinos and
  cosmology}},\ }\href {https://doi.org/10.1016/j.physrep.2006.04.001}
  {\bibfield  {journal} {\bibinfo  {journal} {\physrep}\ }\textbf {\bibinfo
  {volume} {429}},\ \bibinfo {pages} {307} (\bibinfo {year} {2006})},\ \Eprint
  {https://arxiv.org/abs/astro-ph/0603494} {arXiv:astro-ph/0603494 [astro-ph]}
  \BibitemShut {NoStop}%
\bibitem [{\citenamefont {Hu}\ \emph {et~al.}(1998)\citenamefont {Hu},
  \citenamefont {Eisenstein},\ and\ \citenamefont
  {Tegmark}}]{PhysRevLett.80.5255}%
  \BibitemOpen
  \bibfield  {author} {\bibinfo {author} {\bibfnamefont {W.}~\bibnamefont
  {Hu}}, \bibinfo {author} {\bibfnamefont {D.~J.}\ \bibnamefont {Eisenstein}},\
  and\ \bibinfo {author} {\bibfnamefont {M.}~\bibnamefont {Tegmark}},\
  }\bibfield  {title} {\bibinfo {title} {Weighing neutrinos with galaxy
  surveys},\ }\href {https://doi.org/10.1103/PhysRevLett.80.5255} {\bibfield
  {journal} {\bibinfo  {journal} {Phys. Rev. Lett.}\ }\textbf {\bibinfo
  {volume} {80}},\ \bibinfo {pages} {5255} (\bibinfo {year}
  {1998})}\BibitemShut {NoStop}%
\bibitem [{\citenamefont {{Marsh}}\ and\ \citenamefont
  {{Ferreira}}(2010)}]{2010PhRvD..82j3528M}%
  \BibitemOpen
  \bibfield  {author} {\bibinfo {author} {\bibfnamefont {D.~J.~E.}\
  \bibnamefont {{Marsh}}}\ and\ \bibinfo {author} {\bibfnamefont {P.~G.}\
  \bibnamefont {{Ferreira}}},\ }\bibfield  {title} {\bibinfo {title}
  {{Ultralight scalar fields and the growth of structure in the Universe}},\
  }\href {https://doi.org/10.1103/PhysRevD.82.103528} {\bibfield  {journal}
  {\bibinfo  {journal} {\prd}\ }\textbf {\bibinfo {volume} {82}},\ \bibinfo
  {eid} {103528} (\bibinfo {year} {2010})},\ \Eprint
  {https://arxiv.org/abs/1009.3501} {arXiv:1009.3501 [hep-ph]} \BibitemShut
  {NoStop}%
\bibitem [{\citenamefont {{Amendola}}\ and\ \citenamefont
  {{Barbieri}}(2006)}]{2006PhLB..642..192A}%
  \BibitemOpen
  \bibfield  {author} {\bibinfo {author} {\bibfnamefont {L.}~\bibnamefont
  {{Amendola}}}\ and\ \bibinfo {author} {\bibfnamefont {R.}~\bibnamefont
  {{Barbieri}}},\ }\bibfield  {title} {\bibinfo {title} {{Dark matter from an
  ultra-light pseudo-Goldsone-boson}},\ }\href
  {https://doi.org/10.1016/j.physletb.2006.08.069} {\bibfield  {journal}
  {\bibinfo  {journal} {Physics Letters B}\ }\textbf {\bibinfo {volume}
  {642}},\ \bibinfo {pages} {192} (\bibinfo {year} {2006})},\ \Eprint
  {https://arxiv.org/abs/hep-ph/0509257} {arXiv:hep-ph/0509257 [hep-ph]}
  \BibitemShut {NoStop}%
\bibitem [{\citenamefont {{Marsh}}\ \emph {et~al.}(2012)\citenamefont
  {{Marsh}}, \citenamefont {{Macaulay}}, \citenamefont {{Trebitsch}},\ and\
  \citenamefont {{Ferreira}}}]{2012PhRvD..85j3514M}%
  \BibitemOpen
  \bibfield  {author} {\bibinfo {author} {\bibfnamefont {D.~J.~E.}\
  \bibnamefont {{Marsh}}}, \bibinfo {author} {\bibfnamefont {E.}~\bibnamefont
  {{Macaulay}}}, \bibinfo {author} {\bibfnamefont {M.}~\bibnamefont
  {{Trebitsch}}},\ and\ \bibinfo {author} {\bibfnamefont {P.~G.}\ \bibnamefont
  {{Ferreira}}},\ }\bibfield  {title} {\bibinfo {title} {{Ultralight axions:
  Degeneracies with massive neutrinos and forecasts for future cosmological
  observations}},\ }\href {https://doi.org/10.1103/PhysRevD.85.103514}
  {\bibfield  {journal} {\bibinfo  {journal} {\prd}\ }\textbf {\bibinfo
  {volume} {85}},\ \bibinfo {eid} {103514} (\bibinfo {year} {2012})},\ \Eprint
  {https://arxiv.org/abs/1110.0502} {arXiv:1110.0502 [astro-ph.CO]}
  \BibitemShut {NoStop}%
\bibitem [{\citenamefont {Hannestad}(2005)}]{PhysRevLett.95.221301}%
  \BibitemOpen
  \bibfield  {author} {\bibinfo {author} {\bibfnamefont {S.}~\bibnamefont
  {Hannestad}},\ }\bibfield  {title} {\bibinfo {title} {Neutrino masses and the
  dark energy equation of state:relaxing the cosmological neutrino mass
  bound},\ }\href {https://doi.org/10.1103/PhysRevLett.95.221301} {\bibfield
  {journal} {\bibinfo  {journal} {Phys. Rev. Lett.}\ }\textbf {\bibinfo
  {volume} {95}},\ \bibinfo {pages} {221301} (\bibinfo {year}
  {2005})}\BibitemShut {NoStop}%
\bibitem [{\citenamefont {{Marsh}}\ \emph {et~al.}(2013)\citenamefont
  {{Marsh}}, \citenamefont {{Grin}}, \citenamefont {{Hlozek}},\ and\
  \citenamefont {{Ferreira}}}]{2013PhRvD..87l1701M}%
  \BibitemOpen
  \bibfield  {author} {\bibinfo {author} {\bibfnamefont {D.~J.~E.}\
  \bibnamefont {{Marsh}}}, \bibinfo {author} {\bibfnamefont {D.}~\bibnamefont
  {{Grin}}}, \bibinfo {author} {\bibfnamefont {R.}~\bibnamefont {{Hlozek}}},\
  and\ \bibinfo {author} {\bibfnamefont {P.~G.}\ \bibnamefont {{Ferreira}}},\
  }\bibfield  {title} {\bibinfo {title} {{Axiverse cosmology and the energy
  scale of inflation}},\ }\href {https://doi.org/10.1103/PhysRevD.87.121701}
  {\bibfield  {journal} {\bibinfo  {journal} {\prd}\ }\textbf {\bibinfo
  {volume} {87}},\ \bibinfo {eid} {121701} (\bibinfo {year} {2013})},\ \Eprint
  {https://arxiv.org/abs/1303.3008} {arXiv:1303.3008 [astro-ph.CO]}
  \BibitemShut {NoStop}%
\bibitem [{\citenamefont {{Peel}}\ \emph {et~al.}(2018)\citenamefont {{Peel}},
  \citenamefont {{Pettorino}}, \citenamefont {{Giocoli}}, \citenamefont
  {{Starck}},\ and\ \citenamefont {{Baldi}}}]{2018A&A...619A..38P}%
  \BibitemOpen
  \bibfield  {author} {\bibinfo {author} {\bibfnamefont {A.}~\bibnamefont
  {{Peel}}}, \bibinfo {author} {\bibfnamefont {V.}~\bibnamefont {{Pettorino}}},
  \bibinfo {author} {\bibfnamefont {C.}~\bibnamefont {{Giocoli}}}, \bibinfo
  {author} {\bibfnamefont {J.-L.}\ \bibnamefont {{Starck}}},\ and\ \bibinfo
  {author} {\bibfnamefont {M.}~\bibnamefont {{Baldi}}},\ }\bibfield  {title}
  {\bibinfo {title} {{Breaking degeneracies in modified gravity with higher
  (than 2nd) order weak-lensing statistics}},\ }\href
  {https://doi.org/10.1051/0004-6361/201833481} {\bibfield  {journal} {\bibinfo
   {journal} {\aap}\ }\textbf {\bibinfo {volume} {619}},\ \bibinfo {eid} {A38}
  (\bibinfo {year} {2018})},\ \Eprint {https://arxiv.org/abs/1805.05146}
  {arXiv:1805.05146 [astro-ph.CO]} \BibitemShut {NoStop}%
\bibitem [{\citenamefont {{Berg{\'e}}}\ \emph {et~al.}(2010)\citenamefont
  {{Berg{\'e}}}, \citenamefont {{Amara}},\ and\ \citenamefont
  {{R{\'e}fr{\'e}gier}}}]{2010ApJ...712..992B}%
  \BibitemOpen
  \bibfield  {author} {\bibinfo {author} {\bibfnamefont {J.}~\bibnamefont
  {{Berg{\'e}}}}, \bibinfo {author} {\bibfnamefont {A.}~\bibnamefont
  {{Amara}}},\ and\ \bibinfo {author} {\bibfnamefont {A.}~\bibnamefont
  {{R{\'e}fr{\'e}gier}}},\ }\bibfield  {title} {\bibinfo {title} {{Optimal
  Capture of Non-Gaussianity in Weak-Lensing Surveys: Power Spectrum,
  Bispectrum, and Halo Counts}},\ }\href
  {https://doi.org/10.1088/0004-637X/712/2/992} {\bibfield  {journal} {\bibinfo
   {journal} {\apj}\ }\textbf {\bibinfo {volume} {712}},\ \bibinfo {pages}
  {992} (\bibinfo {year} {2010})},\ \Eprint {https://arxiv.org/abs/0909.0529}
  {arXiv:0909.0529 [astro-ph.CO]} \BibitemShut {NoStop}%
\bibitem [{\citenamefont {{Rogers}}\ \emph {et~al.}(2023)\citenamefont
  {{Rogers}}, \citenamefont {{Hlo{\v{z}}ek}}, \citenamefont {{Lagu{\"e}}},
  \citenamefont {{Ivanov}}, \citenamefont {{Philcox}}, \citenamefont
  {{Cabass}}, \citenamefont {{Akitsu}},\ and\ \citenamefont
  {{Marsh}}}]{2023JCAP...06..023R}%
  \BibitemOpen
  \bibfield  {author} {\bibinfo {author} {\bibfnamefont {K.~K.}\ \bibnamefont
  {{Rogers}}}, \bibinfo {author} {\bibfnamefont {R.}~\bibnamefont
  {{Hlo{\v{z}}ek}}}, \bibinfo {author} {\bibfnamefont {A.}~\bibnamefont
  {{Lagu{\"e}}}}, \bibinfo {author} {\bibfnamefont {M.~M.}\ \bibnamefont
  {{Ivanov}}}, \bibinfo {author} {\bibfnamefont {O.~H.~E.}\ \bibnamefont
  {{Philcox}}}, \bibinfo {author} {\bibfnamefont {G.}~\bibnamefont {{Cabass}}},
  \bibinfo {author} {\bibfnamefont {K.}~\bibnamefont {{Akitsu}}},\ and\
  \bibinfo {author} {\bibfnamefont {D.~J.~E.}\ \bibnamefont {{Marsh}}},\
  }\bibfield  {title} {\bibinfo {title} {{Ultra-light axions and the S $_{8}$
  tension: joint constraints from the cosmic microwave background and galaxy
  clustering}},\ }\href {https://doi.org/10.1088/1475-7516/2023/06/023}
  {\bibfield  {journal} {\bibinfo  {journal} {\jcap}\ }\textbf {\bibinfo
  {volume} {2023}},\ \bibinfo {eid} {023} (\bibinfo {year} {2023})},\ \Eprint
  {https://arxiv.org/abs/2301.08361} {arXiv:2301.08361 [astro-ph.CO]}
  \BibitemShut {NoStop}%
\bibitem [{\citenamefont {{Hlozek}}\ \emph {et~al.}(2015)\citenamefont
  {{Hlozek}}, \citenamefont {{Grin}}, \citenamefont {{Marsh}},\ and\
  \citenamefont {{Ferreira}}}]{2015PhRvD..91j3512H}%
  \BibitemOpen
  \bibfield  {author} {\bibinfo {author} {\bibfnamefont {R.}~\bibnamefont
  {{Hlozek}}}, \bibinfo {author} {\bibfnamefont {D.}~\bibnamefont {{Grin}}},
  \bibinfo {author} {\bibfnamefont {D.~J.~E.}\ \bibnamefont {{Marsh}}},\ and\
  \bibinfo {author} {\bibfnamefont {P.~G.}\ \bibnamefont {{Ferreira}}},\
  }\bibfield  {title} {\bibinfo {title} {{A search for ultralight axions using
  precision cosmological data}},\ }\href
  {https://doi.org/10.1103/PhysRevD.91.103512} {\bibfield  {journal} {\bibinfo
  {journal} {\prd}\ }\textbf {\bibinfo {volume} {91}},\ \bibinfo {eid} {103512}
  (\bibinfo {year} {2015})},\ \Eprint {https://arxiv.org/abs/1410.2896}
  {arXiv:1410.2896 [astro-ph.CO]} \BibitemShut {NoStop}%
\bibitem [{\citenamefont {{Kobayashi}}\ \emph {et~al.}(2017)\citenamefont
  {{Kobayashi}}, \citenamefont {{Murgia}}, \citenamefont {{De Simone}},
  \citenamefont {{Ir{\v{s}}i{\v{c}}}},\ and\ \citenamefont
  {{Viel}}}]{2017PhRvD..96l3514K}%
  \BibitemOpen
  \bibfield  {author} {\bibinfo {author} {\bibfnamefont {T.}~\bibnamefont
  {{Kobayashi}}}, \bibinfo {author} {\bibfnamefont {R.}~\bibnamefont
  {{Murgia}}}, \bibinfo {author} {\bibfnamefont {A.}~\bibnamefont {{De
  Simone}}}, \bibinfo {author} {\bibfnamefont {V.}~\bibnamefont
  {{Ir{\v{s}}i{\v{c}}}}},\ and\ \bibinfo {author} {\bibfnamefont
  {M.}~\bibnamefont {{Viel}}},\ }\bibfield  {title} {\bibinfo {title}
  {{Lyman-{\ensuremath{\alpha}} constraints on ultralight scalar dark matter:
  Implications for the early and late universe}},\ }\href
  {https://doi.org/10.1103/PhysRevD.96.123514} {\bibfield  {journal} {\bibinfo
  {journal} {\prd}\ }\textbf {\bibinfo {volume} {96}},\ \bibinfo {eid} {123514}
  (\bibinfo {year} {2017})},\ \Eprint {https://arxiv.org/abs/1708.00015}
  {arXiv:1708.00015 [astro-ph.CO]} \BibitemShut {NoStop}%
\bibitem [{\citenamefont {{Lagu{\"e}}}\ \emph {et~al.}(2022)\citenamefont
  {{Lagu{\"e}}}, \citenamefont {{Bond}}, \citenamefont {{Hlo{\v{z}}ek}},
  \citenamefont {{Rogers}}, \citenamefont {{Marsh}},\ and\ \citenamefont
  {{Grin}}}]{2022JCAP...01..049L}%
  \BibitemOpen
  \bibfield  {author} {\bibinfo {author} {\bibfnamefont {A.}~\bibnamefont
  {{Lagu{\"e}}}}, \bibinfo {author} {\bibfnamefont {J.~R.}\ \bibnamefont
  {{Bond}}}, \bibinfo {author} {\bibfnamefont {R.}~\bibnamefont
  {{Hlo{\v{z}}ek}}}, \bibinfo {author} {\bibfnamefont {K.~K.}\ \bibnamefont
  {{Rogers}}}, \bibinfo {author} {\bibfnamefont {D.~J.~E.}\ \bibnamefont
  {{Marsh}}},\ and\ \bibinfo {author} {\bibfnamefont {D.}~\bibnamefont
  {{Grin}}},\ }\bibfield  {title} {\bibinfo {title} {{Constraining ultralight
  axions with galaxy surveys}},\ }\href
  {https://doi.org/10.1088/1475-7516/2022/01/049} {\bibfield  {journal}
  {\bibinfo  {journal} {\jcap}\ }\textbf {\bibinfo {volume} {2022}},\ \bibinfo
  {eid} {049} (\bibinfo {year} {2022})},\ \Eprint
  {https://arxiv.org/abs/2104.07802} {arXiv:2104.07802 [astro-ph.CO]}
  \BibitemShut {NoStop}%
\bibitem [{\citenamefont {{Rogers}}\ and\ \citenamefont
  {{Peiris}}(2021)}]{2021PhRvL.126g1302R}%
  \BibitemOpen
  \bibfield  {author} {\bibinfo {author} {\bibfnamefont {K.~K.}\ \bibnamefont
  {{Rogers}}}\ and\ \bibinfo {author} {\bibfnamefont {H.~V.}\ \bibnamefont
  {{Peiris}}},\ }\bibfield  {title} {\bibinfo {title} {{Strong Bound on
  Canonical Ultralight Axion Dark Matter from the Lyman-Alpha Forest}},\ }\href
  {https://doi.org/10.1103/PhysRevLett.126.071302} {\bibfield  {journal}
  {\bibinfo  {journal} {\prl}\ }\textbf {\bibinfo {volume} {126}},\ \bibinfo
  {eid} {071302} (\bibinfo {year} {2021})},\ \Eprint
  {https://arxiv.org/abs/2007.12705} {arXiv:2007.12705 [astro-ph.CO]}
  \BibitemShut {NoStop}%
\bibitem [{\citenamefont {{Planck Collaboration}}\ \emph
  {et~al.}(2014)\citenamefont {{Planck Collaboration}}, \citenamefont {{Ade}},
  \citenamefont {{Aghanim}}, \citenamefont {{Armitage-Caplan}}, \citenamefont
  {{Arnaud}}, \citenamefont {{Ashdown}}, \citenamefont {{Atrio-Barandela}},
  \citenamefont {{Aumont}}, \citenamefont {{Baccigalupi}}, \citenamefont
  {{Banday}}, \citenamefont {{Barreiro}}, \citenamefont {{Bartlett}},
  \citenamefont {{Battaner}}, \citenamefont {{Benabed}}, \citenamefont
  {{Beno{\^\i}t}}, \citenamefont {{Benoit-L{\'e}vy}}, \citenamefont
  {{Bernard}}, \citenamefont {{Bersanelli}}, \citenamefont {{Bielewicz}},
  \citenamefont {{Bobin}}, \citenamefont {{Bock}}, \citenamefont {{Bonaldi}},
  \citenamefont {{Bonavera}}, \citenamefont {{Bond}}, \citenamefont
  {{Borrill}}, \citenamefont {{Bouchet}}, \citenamefont {{Boulanger}},
  \citenamefont {{Bridges}}, \citenamefont {{Bucher}}, \citenamefont
  {{Burigana}}, \citenamefont {{Butler}}, \citenamefont {{Calabrese}},
  \citenamefont {{Cardoso}}, \citenamefont {{Catalano}}, \citenamefont
  {{Challinor}}, \citenamefont {{Chamballu}}, \citenamefont {{Chiang}},
  \citenamefont {{Chiang}}, \citenamefont {{Christensen}}, \citenamefont
  {{Church}}, \citenamefont {{Clements}}, \citenamefont {{Colombi}},
  \citenamefont {{Colombo}}, \citenamefont {{Combet}}, \citenamefont
  {{Couchot}}, \citenamefont {{Coulais}}, \citenamefont {{Crill}},
  \citenamefont {{Curto}}, \citenamefont {{Cuttaia}}, \citenamefont {{Danese}},
  \citenamefont {{Davies}}, \citenamefont {{Davis}}, \citenamefont {{de
  Bernardis}}, \citenamefont {{de Rosa}}, \citenamefont {{de Zotti}},
  \citenamefont {{Delabrouille}}, \citenamefont {{Delouis}}, \citenamefont
  {{D{\'e}sert}}, \citenamefont {{Dickinson}}, \citenamefont {{Diego}},
  \citenamefont {{Dole}}, \citenamefont {{Donzelli}}, \citenamefont
  {{Dor{\'e}}}, \citenamefont {{Douspis}}, \citenamefont {{Dunkley}},
  \citenamefont {{Dupac}}, \citenamefont {{Efstathiou}}, \citenamefont
  {{Elsner}}, \citenamefont {{En{\ss}lin}}, \citenamefont {{Eriksen}},
  \citenamefont {{Finelli}}, \citenamefont {{Forni}}, \citenamefont
  {{Frailis}}, \citenamefont {{Fraisse}}, \citenamefont {{Franceschi}},
  \citenamefont {{Gaier}}, \citenamefont {{Galeotta}}, \citenamefont {{Galli}},
  \citenamefont {{Ganga}}, \citenamefont {{Giard}}, \citenamefont {{Giardino}},
  \citenamefont {{Giraud-H{\'e}raud}}, \citenamefont {{Gjerl{\o}w}},
  \citenamefont {{Gonz{\'a}lez-Nuevo}}, \citenamefont {{G{\'o}rski}},
  \citenamefont {{Gratton}}, \citenamefont {{Gregorio}}, \citenamefont
  {{Gruppuso}}, \citenamefont {{Gudmundsson}}, \citenamefont {{Hansen}},
  \citenamefont {{Hanson}}, \citenamefont {{Harrison}}, \citenamefont
  {{Helou}}, \citenamefont {{Henrot-Versill{\'e}}}, \citenamefont
  {{Hern{\'a}ndez-Monteagudo}}, \citenamefont {{Herranz}}, \citenamefont
  {{Hildebrandt}}, \citenamefont {{Hivon}}, \citenamefont {{Hobson}},
  \citenamefont {{Holmes}}, \citenamefont {{Hornstrup}}, \citenamefont
  {{Hovest}}, \citenamefont {{Huffenberger}}, \citenamefont {{Hurier}},
  \citenamefont {{Jaffe}}, \citenamefont {{Jaffe}}, \citenamefont {{Jewell}},
  \citenamefont {{Jones}}, \citenamefont {{Juvela}}, \citenamefont
  {{Keih{\"a}nen}}, \citenamefont {{Keskitalo}}, \citenamefont {{Kiiveri}},
  \citenamefont {{Kisner}}, \citenamefont {{Kneissl}}, \citenamefont
  {{Knoche}}, \citenamefont {{Knox}}, \citenamefont {{Kunz}}, \citenamefont
  {{Kurki-Suonio}}, \citenamefont {{Lagache}}, \citenamefont
  {{L{\"a}hteenm{\"a}ki}}, \citenamefont {{Lamarre}}, \citenamefont
  {{Lasenby}}, \citenamefont {{Lattanzi}}, \citenamefont {{Laureijs}},
  \citenamefont {{Lawrence}}, \citenamefont {{Le Jeune}}, \citenamefont
  {{Leach}}, \citenamefont {{Leahy}}, \citenamefont {{Leonardi}}, \citenamefont
  {{Le{\'o}n-Tavares}}, \citenamefont {{Lesgourgues}}, \citenamefont
  {{Liguori}}, \citenamefont {{Lilje}}, \citenamefont {{Linden-V{\o}rnle}},
  \citenamefont {{Lindholm}}, \citenamefont {{L{\'o}pez-Caniego}},
  \citenamefont {{Lubin}}, \citenamefont {{Mac{\'\i}as-P{\'e}rez}},
  \citenamefont {{Maffei}}, \citenamefont {{Maino}}, \citenamefont
  {{Mandolesi}}, \citenamefont {{Marinucci}}, \citenamefont {{Maris}},
  \citenamefont {{Marshall}}, \citenamefont {{Martin}}, \citenamefont
  {{Mart{\'\i}nez-Gonz{\'a}lez}}, \citenamefont {{Masi}}, \citenamefont
  {{Massardi}}, \citenamefont {{Matarrese}}, \citenamefont {{Matthai}},
  \citenamefont {{Mazzotta}}, \citenamefont {{Meinhold}}, \citenamefont
  {{Melchiorri}}, \citenamefont {{Mendes}}, \citenamefont {{Menegoni}},
  \citenamefont {{Mennella}}, \citenamefont {{Migliaccio}}, \citenamefont
  {{Millea}}, \citenamefont {{Mitra}}, \citenamefont {{Miville-Desch{\^e}nes}},
  \citenamefont {{Molinari}}, \citenamefont {{Moneti}}, \citenamefont
  {{Montier}}, \citenamefont {{Morgante}}, \citenamefont {{Mortlock}},
  \citenamefont {{Moss}}, \citenamefont {{Munshi}}, \citenamefont {{Murphy}},
  \citenamefont {{Naselsky}}, \citenamefont {{Nati}}, \citenamefont {{Natoli}},
  \citenamefont {{Netterfield}}, \citenamefont {{N{\o}rgaard-Nielsen}},
  \citenamefont {{Noviello}}, \citenamefont {{Novikov}}, \citenamefont
  {{Novikov}}, \citenamefont {{O'Dwyer}}, \citenamefont {{Orieux}},
  \citenamefont {{Osborne}}, \citenamefont {{Oxborrow}}, \citenamefont
  {{Paci}}, \citenamefont {{Pagano}}, \citenamefont {{Pajot}}, \citenamefont
  {{Paladini}}, \citenamefont {{Paoletti}}, \citenamefont {{Partridge}},
  \citenamefont {{Pasian}}, \citenamefont {{Patanchon}}, \citenamefont
  {{Paykari}}, \citenamefont {{Perdereau}}, \citenamefont {{Perotto}},
  \citenamefont {{Perrotta}}, \citenamefont {{Piacentini}}, \citenamefont
  {{Piat}}, \citenamefont {{Pierpaoli}}, \citenamefont {{Pietrobon}},
  \citenamefont {{Plaszczynski}}, \citenamefont {{Pointecouteau}},
  \citenamefont {{Polenta}}, \citenamefont {{Ponthieu}}, \citenamefont
  {{Popa}}, \citenamefont {{Poutanen}}, \citenamefont {{Pratt}}, \citenamefont
  {{Pr{\'e}zeau}}, \citenamefont {{Prunet}}, \citenamefont {{Puget}},
  \citenamefont {{Rachen}}, \citenamefont {{Rahlin}}, \citenamefont {{Rebolo}},
  \citenamefont {{Reinecke}}, \citenamefont {{Remazeilles}}, \citenamefont
  {{Renault}}, \citenamefont {{Ricciardi}}, \citenamefont {{Riller}},
  \citenamefont {{Ringeval}}, \citenamefont {{Ristorcelli}}, \citenamefont
  {{Rocha}}, \citenamefont {{Rosset}}, \citenamefont {{Roudier}}, \citenamefont
  {{Rowan-Robinson}}, \citenamefont {{Rubi{\~n}o-Mart{\'\i}n}}, \citenamefont
  {{Rusholme}}, \citenamefont {{Sandri}}, \citenamefont {{Sanselme}},
  \citenamefont {{Santos}}, \citenamefont {{Savini}}, \citenamefont {{Scott}},
  \citenamefont {{Seiffert}}, \citenamefont {{Shellard}}, \citenamefont
  {{Spencer}}, \citenamefont {{Starck}}, \citenamefont {{Stolyarov}},
  \citenamefont {{Stompor}}, \citenamefont {{Sudiwala}}, \citenamefont
  {{Sureau}}, \citenamefont {{Sutton}}, \citenamefont {{Suur-Uski}},
  \citenamefont {{Sygnet}}, \citenamefont {{Tauber}}, \citenamefont
  {{Tavagnacco}}, \citenamefont {{Terenzi}}, \citenamefont {{Toffolatti}},
  \citenamefont {{Tomasi}}, \citenamefont {{Tristram}}, \citenamefont
  {{Tucci}}, \citenamefont {{Tuovinen}}, \citenamefont {{T{\"u}rler}},
  \citenamefont {{Valenziano}}, \citenamefont {{Valiviita}}, \citenamefont
  {{Van Tent}}, \citenamefont {{Varis}}, \citenamefont {{Vielva}},
  \citenamefont {{Villa}}, \citenamefont {{Vittorio}}, \citenamefont {{Wade}},
  \citenamefont {{Wandelt}}, \citenamefont {{Wehus}}, \citenamefont {{White}},
  \citenamefont {{White}}, \citenamefont {{Yvon}}, \citenamefont {{Zacchei}},\
  and\ \citenamefont {{Zonca}}}]{2014A&A...571A..15P}%
  \BibitemOpen
  \bibfield  {author} {\bibinfo {author} {\bibnamefont {{Planck
  Collaboration}}}, \bibinfo {author} {\bibfnamefont {P.~A.~R.}\ \bibnamefont
  {{Ade}}}, \bibinfo {author} {\bibfnamefont {N.}~\bibnamefont {{Aghanim}}},
  \bibinfo {author} {\bibfnamefont {C.}~\bibnamefont {{Armitage-Caplan}}},
  \bibinfo {author} {\bibfnamefont {M.}~\bibnamefont {{Arnaud}}}, \bibinfo
  {author} {\bibfnamefont {M.}~\bibnamefont {{Ashdown}}}, \bibinfo {author}
  {\bibfnamefont {F.}~\bibnamefont {{Atrio-Barandela}}}, \bibinfo {author}
  {\bibfnamefont {J.}~\bibnamefont {{Aumont}}}, \bibinfo {author}
  {\bibfnamefont {C.}~\bibnamefont {{Baccigalupi}}}, \bibinfo {author}
  {\bibfnamefont {A.~J.}\ \bibnamefont {{Banday}}}, \bibinfo {author}
  {\bibfnamefont {R.~B.}\ \bibnamefont {{Barreiro}}}, \bibinfo {author}
  {\bibfnamefont {J.~G.}\ \bibnamefont {{Bartlett}}}, \bibinfo {author}
  {\bibfnamefont {E.}~\bibnamefont {{Battaner}}}, \bibinfo {author}
  {\bibfnamefont {K.}~\bibnamefont {{Benabed}}}, \bibinfo {author}
  {\bibfnamefont {A.}~\bibnamefont {{Beno{\^\i}t}}}, \bibinfo {author}
  {\bibfnamefont {A.}~\bibnamefont {{Benoit-L{\'e}vy}}}, \bibinfo {author}
  {\bibfnamefont {J.~P.}\ \bibnamefont {{Bernard}}}, \bibinfo {author}
  {\bibfnamefont {M.}~\bibnamefont {{Bersanelli}}}, \bibinfo {author}
  {\bibfnamefont {P.}~\bibnamefont {{Bielewicz}}}, \bibinfo {author}
  {\bibfnamefont {J.}~\bibnamefont {{Bobin}}}, \bibinfo {author} {\bibfnamefont
  {J.~J.}\ \bibnamefont {{Bock}}}, \bibinfo {author} {\bibfnamefont
  {A.}~\bibnamefont {{Bonaldi}}}, \bibinfo {author} {\bibfnamefont
  {L.}~\bibnamefont {{Bonavera}}}, \bibinfo {author} {\bibfnamefont {J.~R.}\
  \bibnamefont {{Bond}}}, \bibinfo {author} {\bibfnamefont {J.}~\bibnamefont
  {{Borrill}}}, \bibinfo {author} {\bibfnamefont {F.~R.}\ \bibnamefont
  {{Bouchet}}}, \bibinfo {author} {\bibfnamefont {F.}~\bibnamefont
  {{Boulanger}}}, \bibinfo {author} {\bibfnamefont {M.}~\bibnamefont
  {{Bridges}}}, \bibinfo {author} {\bibfnamefont {M.}~\bibnamefont {{Bucher}}},
  \bibinfo {author} {\bibfnamefont {C.}~\bibnamefont {{Burigana}}}, \bibinfo
  {author} {\bibfnamefont {R.~C.}\ \bibnamefont {{Butler}}}, \bibinfo {author}
  {\bibfnamefont {E.}~\bibnamefont {{Calabrese}}}, \bibinfo {author}
  {\bibfnamefont {J.~F.}\ \bibnamefont {{Cardoso}}}, \bibinfo {author}
  {\bibfnamefont {A.}~\bibnamefont {{Catalano}}}, \bibinfo {author}
  {\bibfnamefont {A.}~\bibnamefont {{Challinor}}}, \bibinfo {author}
  {\bibfnamefont {A.}~\bibnamefont {{Chamballu}}}, \bibinfo {author}
  {\bibfnamefont {H.~C.}\ \bibnamefont {{Chiang}}}, \bibinfo {author}
  {\bibfnamefont {L.~Y.}\ \bibnamefont {{Chiang}}}, \bibinfo {author}
  {\bibfnamefont {P.~R.}\ \bibnamefont {{Christensen}}}, \bibinfo {author}
  {\bibfnamefont {S.}~\bibnamefont {{Church}}}, \bibinfo {author}
  {\bibfnamefont {D.~L.}\ \bibnamefont {{Clements}}}, \bibinfo {author}
  {\bibfnamefont {S.}~\bibnamefont {{Colombi}}}, \bibinfo {author}
  {\bibfnamefont {L.~P.~L.}\ \bibnamefont {{Colombo}}}, \bibinfo {author}
  {\bibfnamefont {C.}~\bibnamefont {{Combet}}}, \bibinfo {author}
  {\bibfnamefont {F.}~\bibnamefont {{Couchot}}}, \bibinfo {author}
  {\bibfnamefont {A.}~\bibnamefont {{Coulais}}}, \bibinfo {author}
  {\bibfnamefont {B.~P.}\ \bibnamefont {{Crill}}}, \bibinfo {author}
  {\bibfnamefont {A.}~\bibnamefont {{Curto}}}, \bibinfo {author} {\bibfnamefont
  {F.}~\bibnamefont {{Cuttaia}}}, \bibinfo {author} {\bibfnamefont
  {L.}~\bibnamefont {{Danese}}}, \bibinfo {author} {\bibfnamefont {R.~D.}\
  \bibnamefont {{Davies}}}, \bibinfo {author} {\bibfnamefont {R.~J.}\
  \bibnamefont {{Davis}}}, \bibinfo {author} {\bibfnamefont {P.}~\bibnamefont
  {{de Bernardis}}}, \bibinfo {author} {\bibfnamefont {A.}~\bibnamefont {{de
  Rosa}}}, \bibinfo {author} {\bibfnamefont {G.}~\bibnamefont {{de Zotti}}},
  \bibinfo {author} {\bibfnamefont {J.}~\bibnamefont {{Delabrouille}}},
  \bibinfo {author} {\bibfnamefont {J.~M.}\ \bibnamefont {{Delouis}}}, \bibinfo
  {author} {\bibfnamefont {F.~X.}\ \bibnamefont {{D{\'e}sert}}}, \bibinfo
  {author} {\bibfnamefont {C.}~\bibnamefont {{Dickinson}}}, \bibinfo {author}
  {\bibfnamefont {J.~M.}\ \bibnamefont {{Diego}}}, \bibinfo {author}
  {\bibfnamefont {H.}~\bibnamefont {{Dole}}}, \bibinfo {author} {\bibfnamefont
  {S.}~\bibnamefont {{Donzelli}}}, \bibinfo {author} {\bibfnamefont
  {O.}~\bibnamefont {{Dor{\'e}}}}, \bibinfo {author} {\bibfnamefont
  {M.}~\bibnamefont {{Douspis}}}, \bibinfo {author} {\bibfnamefont
  {J.}~\bibnamefont {{Dunkley}}}, \bibinfo {author} {\bibfnamefont
  {X.}~\bibnamefont {{Dupac}}}, \bibinfo {author} {\bibfnamefont
  {G.}~\bibnamefont {{Efstathiou}}}, \bibinfo {author} {\bibfnamefont
  {F.}~\bibnamefont {{Elsner}}}, \bibinfo {author} {\bibfnamefont {T.~A.}\
  \bibnamefont {{En{\ss}lin}}}, \bibinfo {author} {\bibfnamefont {H.~K.}\
  \bibnamefont {{Eriksen}}}, \bibinfo {author} {\bibfnamefont {F.}~\bibnamefont
  {{Finelli}}}, \bibinfo {author} {\bibfnamefont {O.}~\bibnamefont {{Forni}}},
  \bibinfo {author} {\bibfnamefont {M.}~\bibnamefont {{Frailis}}}, \bibinfo
  {author} {\bibfnamefont {A.~A.}\ \bibnamefont {{Fraisse}}}, \bibinfo {author}
  {\bibfnamefont {E.}~\bibnamefont {{Franceschi}}}, \bibinfo {author}
  {\bibfnamefont {T.~C.}\ \bibnamefont {{Gaier}}}, \bibinfo {author}
  {\bibfnamefont {S.}~\bibnamefont {{Galeotta}}}, \bibinfo {author}
  {\bibfnamefont {S.}~\bibnamefont {{Galli}}}, \bibinfo {author} {\bibfnamefont
  {K.}~\bibnamefont {{Ganga}}}, \bibinfo {author} {\bibfnamefont
  {M.}~\bibnamefont {{Giard}}}, \bibinfo {author} {\bibfnamefont
  {G.}~\bibnamefont {{Giardino}}}, \bibinfo {author} {\bibfnamefont
  {Y.}~\bibnamefont {{Giraud-H{\'e}raud}}}, \bibinfo {author} {\bibfnamefont
  {E.}~\bibnamefont {{Gjerl{\o}w}}}, \bibinfo {author} {\bibfnamefont
  {J.}~\bibnamefont {{Gonz{\'a}lez-Nuevo}}}, \bibinfo {author} {\bibfnamefont
  {K.~M.}\ \bibnamefont {{G{\'o}rski}}}, \bibinfo {author} {\bibfnamefont
  {S.}~\bibnamefont {{Gratton}}}, \bibinfo {author} {\bibfnamefont
  {A.}~\bibnamefont {{Gregorio}}}, \bibinfo {author} {\bibfnamefont
  {A.}~\bibnamefont {{Gruppuso}}}, \bibinfo {author} {\bibfnamefont {J.~E.}\
  \bibnamefont {{Gudmundsson}}}, \bibinfo {author} {\bibfnamefont {F.~K.}\
  \bibnamefont {{Hansen}}}, \bibinfo {author} {\bibfnamefont {D.}~\bibnamefont
  {{Hanson}}}, \bibinfo {author} {\bibfnamefont {D.}~\bibnamefont
  {{Harrison}}}, \bibinfo {author} {\bibfnamefont {G.}~\bibnamefont {{Helou}}},
  \bibinfo {author} {\bibfnamefont {S.}~\bibnamefont {{Henrot-Versill{\'e}}}},
  \bibinfo {author} {\bibfnamefont {C.}~\bibnamefont
  {{Hern{\'a}ndez-Monteagudo}}}, \bibinfo {author} {\bibfnamefont
  {D.}~\bibnamefont {{Herranz}}}, \bibinfo {author} {\bibfnamefont {S.~R.}\
  \bibnamefont {{Hildebrandt}}}, \bibinfo {author} {\bibfnamefont
  {E.}~\bibnamefont {{Hivon}}}, \bibinfo {author} {\bibfnamefont
  {M.}~\bibnamefont {{Hobson}}}, \bibinfo {author} {\bibfnamefont {W.~A.}\
  \bibnamefont {{Holmes}}}, \bibinfo {author} {\bibfnamefont {A.}~\bibnamefont
  {{Hornstrup}}}, \bibinfo {author} {\bibfnamefont {W.}~\bibnamefont
  {{Hovest}}}, \bibinfo {author} {\bibfnamefont {K.~M.}\ \bibnamefont
  {{Huffenberger}}}, \bibinfo {author} {\bibfnamefont {G.}~\bibnamefont
  {{Hurier}}}, \bibinfo {author} {\bibfnamefont {A.~H.}\ \bibnamefont
  {{Jaffe}}}, \bibinfo {author} {\bibfnamefont {T.~R.}\ \bibnamefont
  {{Jaffe}}}, \bibinfo {author} {\bibfnamefont {J.}~\bibnamefont {{Jewell}}},
  \bibinfo {author} {\bibfnamefont {W.~C.}\ \bibnamefont {{Jones}}}, \bibinfo
  {author} {\bibfnamefont {M.}~\bibnamefont {{Juvela}}}, \bibinfo {author}
  {\bibfnamefont {E.}~\bibnamefont {{Keih{\"a}nen}}}, \bibinfo {author}
  {\bibfnamefont {R.}~\bibnamefont {{Keskitalo}}}, \bibinfo {author}
  {\bibfnamefont {K.}~\bibnamefont {{Kiiveri}}}, \bibinfo {author}
  {\bibfnamefont {T.~S.}\ \bibnamefont {{Kisner}}}, \bibinfo {author}
  {\bibfnamefont {R.}~\bibnamefont {{Kneissl}}}, \bibinfo {author}
  {\bibfnamefont {J.}~\bibnamefont {{Knoche}}}, \bibinfo {author}
  {\bibfnamefont {L.}~\bibnamefont {{Knox}}}, \bibinfo {author} {\bibfnamefont
  {M.}~\bibnamefont {{Kunz}}}, \bibinfo {author} {\bibfnamefont
  {H.}~\bibnamefont {{Kurki-Suonio}}}, \bibinfo {author} {\bibfnamefont
  {G.}~\bibnamefont {{Lagache}}}, \bibinfo {author} {\bibfnamefont
  {A.}~\bibnamefont {{L{\"a}hteenm{\"a}ki}}}, \bibinfo {author} {\bibfnamefont
  {J.~M.}\ \bibnamefont {{Lamarre}}}, \bibinfo {author} {\bibfnamefont
  {A.}~\bibnamefont {{Lasenby}}}, \bibinfo {author} {\bibfnamefont
  {M.}~\bibnamefont {{Lattanzi}}}, \bibinfo {author} {\bibfnamefont {R.~J.}\
  \bibnamefont {{Laureijs}}}, \bibinfo {author} {\bibfnamefont {C.~R.}\
  \bibnamefont {{Lawrence}}}, \bibinfo {author} {\bibfnamefont
  {M.}~\bibnamefont {{Le Jeune}}}, \bibinfo {author} {\bibfnamefont
  {S.}~\bibnamefont {{Leach}}}, \bibinfo {author} {\bibfnamefont {J.~P.}\
  \bibnamefont {{Leahy}}}, \bibinfo {author} {\bibfnamefont {R.}~\bibnamefont
  {{Leonardi}}}, \bibinfo {author} {\bibfnamefont {J.}~\bibnamefont
  {{Le{\'o}n-Tavares}}}, \bibinfo {author} {\bibfnamefont {J.}~\bibnamefont
  {{Lesgourgues}}}, \bibinfo {author} {\bibfnamefont {M.}~\bibnamefont
  {{Liguori}}}, \bibinfo {author} {\bibfnamefont {P.~B.}\ \bibnamefont
  {{Lilje}}}, \bibinfo {author} {\bibfnamefont {M.}~\bibnamefont
  {{Linden-V{\o}rnle}}}, \bibinfo {author} {\bibfnamefont {V.}~\bibnamefont
  {{Lindholm}}}, \bibinfo {author} {\bibfnamefont {M.}~\bibnamefont
  {{L{\'o}pez-Caniego}}}, \bibinfo {author} {\bibfnamefont {P.~M.}\
  \bibnamefont {{Lubin}}}, \bibinfo {author} {\bibfnamefont {J.~F.}\
  \bibnamefont {{Mac{\'\i}as-P{\'e}rez}}}, \bibinfo {author} {\bibfnamefont
  {B.}~\bibnamefont {{Maffei}}}, \bibinfo {author} {\bibfnamefont
  {D.}~\bibnamefont {{Maino}}}, \bibinfo {author} {\bibfnamefont
  {N.}~\bibnamefont {{Mandolesi}}}, \bibinfo {author} {\bibfnamefont
  {D.}~\bibnamefont {{Marinucci}}}, \bibinfo {author} {\bibfnamefont
  {M.}~\bibnamefont {{Maris}}}, \bibinfo {author} {\bibfnamefont {D.~J.}\
  \bibnamefont {{Marshall}}}, \bibinfo {author} {\bibfnamefont {P.~G.}\
  \bibnamefont {{Martin}}}, \bibinfo {author} {\bibfnamefont {E.}~\bibnamefont
  {{Mart{\'\i}nez-Gonz{\'a}lez}}}, \bibinfo {author} {\bibfnamefont
  {S.}~\bibnamefont {{Masi}}}, \bibinfo {author} {\bibfnamefont
  {M.}~\bibnamefont {{Massardi}}}, \bibinfo {author} {\bibfnamefont
  {S.}~\bibnamefont {{Matarrese}}}, \bibinfo {author} {\bibfnamefont
  {F.}~\bibnamefont {{Matthai}}}, \bibinfo {author} {\bibfnamefont
  {P.}~\bibnamefont {{Mazzotta}}}, \bibinfo {author} {\bibfnamefont {P.~R.}\
  \bibnamefont {{Meinhold}}}, \bibinfo {author} {\bibfnamefont
  {A.}~\bibnamefont {{Melchiorri}}}, \bibinfo {author} {\bibfnamefont
  {L.}~\bibnamefont {{Mendes}}}, \bibinfo {author} {\bibfnamefont
  {E.}~\bibnamefont {{Menegoni}}}, \bibinfo {author} {\bibfnamefont
  {A.}~\bibnamefont {{Mennella}}}, \bibinfo {author} {\bibfnamefont
  {M.}~\bibnamefont {{Migliaccio}}}, \bibinfo {author} {\bibfnamefont
  {M.}~\bibnamefont {{Millea}}}, \bibinfo {author} {\bibfnamefont
  {S.}~\bibnamefont {{Mitra}}}, \bibinfo {author} {\bibfnamefont {M.~A.}\
  \bibnamefont {{Miville-Desch{\^e}nes}}}, \bibinfo {author} {\bibfnamefont
  {D.}~\bibnamefont {{Molinari}}}, \bibinfo {author} {\bibfnamefont
  {A.}~\bibnamefont {{Moneti}}}, \bibinfo {author} {\bibfnamefont
  {L.}~\bibnamefont {{Montier}}}, \bibinfo {author} {\bibfnamefont
  {G.}~\bibnamefont {{Morgante}}}, \bibinfo {author} {\bibfnamefont
  {D.}~\bibnamefont {{Mortlock}}}, \bibinfo {author} {\bibfnamefont
  {A.}~\bibnamefont {{Moss}}}, \bibinfo {author} {\bibfnamefont
  {D.}~\bibnamefont {{Munshi}}}, \bibinfo {author} {\bibfnamefont {J.~A.}\
  \bibnamefont {{Murphy}}}, \bibinfo {author} {\bibfnamefont {P.}~\bibnamefont
  {{Naselsky}}}, \bibinfo {author} {\bibfnamefont {F.}~\bibnamefont {{Nati}}},
  \bibinfo {author} {\bibfnamefont {P.}~\bibnamefont {{Natoli}}}, \bibinfo
  {author} {\bibfnamefont {C.~B.}\ \bibnamefont {{Netterfield}}}, \bibinfo
  {author} {\bibfnamefont {H.~U.}\ \bibnamefont {{N{\o}rgaard-Nielsen}}},
  \bibinfo {author} {\bibfnamefont {F.}~\bibnamefont {{Noviello}}}, \bibinfo
  {author} {\bibfnamefont {D.}~\bibnamefont {{Novikov}}}, \bibinfo {author}
  {\bibfnamefont {I.}~\bibnamefont {{Novikov}}}, \bibinfo {author}
  {\bibfnamefont {I.~J.}\ \bibnamefont {{O'Dwyer}}}, \bibinfo {author}
  {\bibfnamefont {F.}~\bibnamefont {{Orieux}}}, \bibinfo {author}
  {\bibfnamefont {S.}~\bibnamefont {{Osborne}}}, \bibinfo {author}
  {\bibfnamefont {C.~A.}\ \bibnamefont {{Oxborrow}}}, \bibinfo {author}
  {\bibfnamefont {F.}~\bibnamefont {{Paci}}}, \bibinfo {author} {\bibfnamefont
  {L.}~\bibnamefont {{Pagano}}}, \bibinfo {author} {\bibfnamefont
  {F.}~\bibnamefont {{Pajot}}}, \bibinfo {author} {\bibfnamefont
  {R.}~\bibnamefont {{Paladini}}}, \bibinfo {author} {\bibfnamefont
  {D.}~\bibnamefont {{Paoletti}}}, \bibinfo {author} {\bibfnamefont
  {B.}~\bibnamefont {{Partridge}}}, \bibinfo {author} {\bibfnamefont
  {F.}~\bibnamefont {{Pasian}}}, \bibinfo {author} {\bibfnamefont
  {G.}~\bibnamefont {{Patanchon}}}, \bibinfo {author} {\bibfnamefont
  {P.}~\bibnamefont {{Paykari}}}, \bibinfo {author} {\bibfnamefont
  {O.}~\bibnamefont {{Perdereau}}}, \bibinfo {author} {\bibfnamefont
  {L.}~\bibnamefont {{Perotto}}}, \bibinfo {author} {\bibfnamefont
  {F.}~\bibnamefont {{Perrotta}}}, \bibinfo {author} {\bibfnamefont
  {F.}~\bibnamefont {{Piacentini}}}, \bibinfo {author} {\bibfnamefont
  {M.}~\bibnamefont {{Piat}}}, \bibinfo {author} {\bibfnamefont
  {E.}~\bibnamefont {{Pierpaoli}}}, \bibinfo {author} {\bibfnamefont
  {D.}~\bibnamefont {{Pietrobon}}}, \bibinfo {author} {\bibfnamefont
  {S.}~\bibnamefont {{Plaszczynski}}}, \bibinfo {author} {\bibfnamefont
  {E.}~\bibnamefont {{Pointecouteau}}}, \bibinfo {author} {\bibfnamefont
  {G.}~\bibnamefont {{Polenta}}}, \bibinfo {author} {\bibfnamefont
  {N.}~\bibnamefont {{Ponthieu}}}, \bibinfo {author} {\bibfnamefont
  {L.}~\bibnamefont {{Popa}}}, \bibinfo {author} {\bibfnamefont
  {T.}~\bibnamefont {{Poutanen}}}, \bibinfo {author} {\bibfnamefont {G.~W.}\
  \bibnamefont {{Pratt}}}, \bibinfo {author} {\bibfnamefont {G.}~\bibnamefont
  {{Pr{\'e}zeau}}}, \bibinfo {author} {\bibfnamefont {S.}~\bibnamefont
  {{Prunet}}}, \bibinfo {author} {\bibfnamefont {J.~L.}\ \bibnamefont
  {{Puget}}}, \bibinfo {author} {\bibfnamefont {J.~P.}\ \bibnamefont
  {{Rachen}}}, \bibinfo {author} {\bibfnamefont {A.}~\bibnamefont {{Rahlin}}},
  \bibinfo {author} {\bibfnamefont {R.}~\bibnamefont {{Rebolo}}}, \bibinfo
  {author} {\bibfnamefont {M.}~\bibnamefont {{Reinecke}}}, \bibinfo {author}
  {\bibfnamefont {M.}~\bibnamefont {{Remazeilles}}}, \bibinfo {author}
  {\bibfnamefont {C.}~\bibnamefont {{Renault}}}, \bibinfo {author}
  {\bibfnamefont {S.}~\bibnamefont {{Ricciardi}}}, \bibinfo {author}
  {\bibfnamefont {T.}~\bibnamefont {{Riller}}}, \bibinfo {author}
  {\bibfnamefont {C.}~\bibnamefont {{Ringeval}}}, \bibinfo {author}
  {\bibfnamefont {I.}~\bibnamefont {{Ristorcelli}}}, \bibinfo {author}
  {\bibfnamefont {G.}~\bibnamefont {{Rocha}}}, \bibinfo {author} {\bibfnamefont
  {C.}~\bibnamefont {{Rosset}}}, \bibinfo {author} {\bibfnamefont
  {G.}~\bibnamefont {{Roudier}}}, \bibinfo {author} {\bibfnamefont
  {M.}~\bibnamefont {{Rowan-Robinson}}}, \bibinfo {author} {\bibfnamefont
  {J.~A.}\ \bibnamefont {{Rubi{\~n}o-Mart{\'\i}n}}}, \bibinfo {author}
  {\bibfnamefont {B.}~\bibnamefont {{Rusholme}}}, \bibinfo {author}
  {\bibfnamefont {M.}~\bibnamefont {{Sandri}}}, \bibinfo {author}
  {\bibfnamefont {L.}~\bibnamefont {{Sanselme}}}, \bibinfo {author}
  {\bibfnamefont {D.}~\bibnamefont {{Santos}}}, \bibinfo {author}
  {\bibfnamefont {G.}~\bibnamefont {{Savini}}}, \bibinfo {author}
  {\bibfnamefont {D.}~\bibnamefont {{Scott}}}, \bibinfo {author} {\bibfnamefont
  {M.~D.}\ \bibnamefont {{Seiffert}}}, \bibinfo {author} {\bibfnamefont
  {E.~P.~S.}\ \bibnamefont {{Shellard}}}, \bibinfo {author} {\bibfnamefont
  {L.~D.}\ \bibnamefont {{Spencer}}}, \bibinfo {author} {\bibfnamefont {J.~L.}\
  \bibnamefont {{Starck}}}, \bibinfo {author} {\bibfnamefont {V.}~\bibnamefont
  {{Stolyarov}}}, \bibinfo {author} {\bibfnamefont {R.}~\bibnamefont
  {{Stompor}}}, \bibinfo {author} {\bibfnamefont {R.}~\bibnamefont
  {{Sudiwala}}}, \bibinfo {author} {\bibfnamefont {F.}~\bibnamefont
  {{Sureau}}}, \bibinfo {author} {\bibfnamefont {D.}~\bibnamefont {{Sutton}}},
  \bibinfo {author} {\bibfnamefont {A.~S.}\ \bibnamefont {{Suur-Uski}}},
  \bibinfo {author} {\bibfnamefont {J.~F.}\ \bibnamefont {{Sygnet}}}, \bibinfo
  {author} {\bibfnamefont {J.~A.}\ \bibnamefont {{Tauber}}}, \bibinfo {author}
  {\bibfnamefont {D.}~\bibnamefont {{Tavagnacco}}}, \bibinfo {author}
  {\bibfnamefont {L.}~\bibnamefont {{Terenzi}}}, \bibinfo {author}
  {\bibfnamefont {L.}~\bibnamefont {{Toffolatti}}}, \bibinfo {author}
  {\bibfnamefont {M.}~\bibnamefont {{Tomasi}}}, \bibinfo {author}
  {\bibfnamefont {M.}~\bibnamefont {{Tristram}}}, \bibinfo {author}
  {\bibfnamefont {M.}~\bibnamefont {{Tucci}}}, \bibinfo {author} {\bibfnamefont
  {J.}~\bibnamefont {{Tuovinen}}}, \bibinfo {author} {\bibfnamefont
  {M.}~\bibnamefont {{T{\"u}rler}}}, \bibinfo {author} {\bibfnamefont
  {L.}~\bibnamefont {{Valenziano}}}, \bibinfo {author} {\bibfnamefont
  {J.}~\bibnamefont {{Valiviita}}}, \bibinfo {author} {\bibfnamefont
  {B.}~\bibnamefont {{Van Tent}}}, \bibinfo {author} {\bibfnamefont
  {J.}~\bibnamefont {{Varis}}}, \bibinfo {author} {\bibfnamefont
  {P.}~\bibnamefont {{Vielva}}}, \bibinfo {author} {\bibfnamefont
  {F.}~\bibnamefont {{Villa}}}, \bibinfo {author} {\bibfnamefont
  {N.}~\bibnamefont {{Vittorio}}}, \bibinfo {author} {\bibfnamefont {L.~A.}\
  \bibnamefont {{Wade}}}, \bibinfo {author} {\bibfnamefont {B.~D.}\
  \bibnamefont {{Wandelt}}}, \bibinfo {author} {\bibfnamefont {I.~K.}\
  \bibnamefont {{Wehus}}}, \bibinfo {author} {\bibfnamefont {M.}~\bibnamefont
  {{White}}}, \bibinfo {author} {\bibfnamefont {S.~D.~M.}\ \bibnamefont
  {{White}}}, \bibinfo {author} {\bibfnamefont {D.}~\bibnamefont {{Yvon}}},
  \bibinfo {author} {\bibfnamefont {A.}~\bibnamefont {{Zacchei}}},\ and\
  \bibinfo {author} {\bibfnamefont {A.}~\bibnamefont {{Zonca}}},\ }\bibfield
  {title} {\bibinfo {title} {{Planck 2013 results. XV. CMB power spectra and
  likelihood}},\ }\href {https://doi.org/10.1051/0004-6361/201321573}
  {\bibfield  {journal} {\bibinfo  {journal} {\aap}\ }\textbf {\bibinfo
  {volume} {571}},\ \bibinfo {eid} {A15} (\bibinfo {year} {2014})},\ \Eprint
  {https://arxiv.org/abs/1303.5075} {arXiv:1303.5075 [astro-ph.CO]}
  \BibitemShut {NoStop}%
\bibitem [{\citenamefont {{Das}}\ \emph {et~al.}(2014)\citenamefont {{Das}},
  \citenamefont {{Louis}}, \citenamefont {{Nolta}}, \citenamefont {{Addison}},
  \citenamefont {{Battistelli}}, \citenamefont {{Bond}}, \citenamefont
  {{Calabrese}}, \citenamefont {{Crichton}}, \citenamefont {{Devlin}},
  \citenamefont {{Dicker}}, \citenamefont {{Dunkley}}, \citenamefont
  {{D{\"u}nner}}, \citenamefont {{Fowler}}, \citenamefont {{Gralla}},
  \citenamefont {{Hajian}}, \citenamefont {{Halpern}}, \citenamefont
  {{Hasselfield}}, \citenamefont {{Hilton}}, \citenamefont {{Hincks}},
  \citenamefont {{Hlozek}}, \citenamefont {{Huffenberger}}, \citenamefont
  {{Hughes}}, \citenamefont {{Irwin}}, \citenamefont {{Kosowsky}},
  \citenamefont {{Lupton}}, \citenamefont {{Marriage}}, \citenamefont
  {{Marsden}}, \citenamefont {{Menanteau}}, \citenamefont {{Moodley}},
  \citenamefont {{Niemack}}, \citenamefont {{Page}}, \citenamefont
  {{Partridge}}, \citenamefont {{Reese}}, \citenamefont {{Schmitt}},
  \citenamefont {{Sehgal}}, \citenamefont {{Sherwin}}, \citenamefont
  {{Sievers}}, \citenamefont {{Spergel}}, \citenamefont {{Staggs}},
  \citenamefont {{Swetz}}, \citenamefont {{Switzer}}, \citenamefont
  {{Thornton}}, \citenamefont {{Trac}},\ and\ \citenamefont
  {{Wollack}}}]{2014JCAP...04..014D}%
  \BibitemOpen
  \bibfield  {author} {\bibinfo {author} {\bibfnamefont {S.}~\bibnamefont
  {{Das}}}, \bibinfo {author} {\bibfnamefont {T.}~\bibnamefont {{Louis}}},
  \bibinfo {author} {\bibfnamefont {M.~R.}\ \bibnamefont {{Nolta}}}, \bibinfo
  {author} {\bibfnamefont {G.~E.}\ \bibnamefont {{Addison}}}, \bibinfo {author}
  {\bibfnamefont {E.~S.}\ \bibnamefont {{Battistelli}}}, \bibinfo {author}
  {\bibfnamefont {J.~R.}\ \bibnamefont {{Bond}}}, \bibinfo {author}
  {\bibfnamefont {E.}~\bibnamefont {{Calabrese}}}, \bibinfo {author}
  {\bibfnamefont {D.}~\bibnamefont {{Crichton}}}, \bibinfo {author}
  {\bibfnamefont {M.~J.}\ \bibnamefont {{Devlin}}}, \bibinfo {author}
  {\bibfnamefont {S.}~\bibnamefont {{Dicker}}}, \bibinfo {author}
  {\bibfnamefont {J.}~\bibnamefont {{Dunkley}}}, \bibinfo {author}
  {\bibfnamefont {R.}~\bibnamefont {{D{\"u}nner}}}, \bibinfo {author}
  {\bibfnamefont {J.~W.}\ \bibnamefont {{Fowler}}}, \bibinfo {author}
  {\bibfnamefont {M.}~\bibnamefont {{Gralla}}}, \bibinfo {author}
  {\bibfnamefont {A.}~\bibnamefont {{Hajian}}}, \bibinfo {author}
  {\bibfnamefont {M.}~\bibnamefont {{Halpern}}}, \bibinfo {author}
  {\bibfnamefont {M.}~\bibnamefont {{Hasselfield}}}, \bibinfo {author}
  {\bibfnamefont {M.}~\bibnamefont {{Hilton}}}, \bibinfo {author}
  {\bibfnamefont {A.~D.}\ \bibnamefont {{Hincks}}}, \bibinfo {author}
  {\bibfnamefont {R.}~\bibnamefont {{Hlozek}}}, \bibinfo {author}
  {\bibfnamefont {K.~M.}\ \bibnamefont {{Huffenberger}}}, \bibinfo {author}
  {\bibfnamefont {J.~P.}\ \bibnamefont {{Hughes}}}, \bibinfo {author}
  {\bibfnamefont {K.~D.}\ \bibnamefont {{Irwin}}}, \bibinfo {author}
  {\bibfnamefont {A.}~\bibnamefont {{Kosowsky}}}, \bibinfo {author}
  {\bibfnamefont {R.~H.}\ \bibnamefont {{Lupton}}}, \bibinfo {author}
  {\bibfnamefont {T.~A.}\ \bibnamefont {{Marriage}}}, \bibinfo {author}
  {\bibfnamefont {D.}~\bibnamefont {{Marsden}}}, \bibinfo {author}
  {\bibfnamefont {F.}~\bibnamefont {{Menanteau}}}, \bibinfo {author}
  {\bibfnamefont {K.}~\bibnamefont {{Moodley}}}, \bibinfo {author}
  {\bibfnamefont {M.~D.}\ \bibnamefont {{Niemack}}}, \bibinfo {author}
  {\bibfnamefont {L.~A.}\ \bibnamefont {{Page}}}, \bibinfo {author}
  {\bibfnamefont {B.}~\bibnamefont {{Partridge}}}, \bibinfo {author}
  {\bibfnamefont {E.~D.}\ \bibnamefont {{Reese}}}, \bibinfo {author}
  {\bibfnamefont {B.~L.}\ \bibnamefont {{Schmitt}}}, \bibinfo {author}
  {\bibfnamefont {N.}~\bibnamefont {{Sehgal}}}, \bibinfo {author}
  {\bibfnamefont {B.~D.}\ \bibnamefont {{Sherwin}}}, \bibinfo {author}
  {\bibfnamefont {J.~L.}\ \bibnamefont {{Sievers}}}, \bibinfo {author}
  {\bibfnamefont {D.~N.}\ \bibnamefont {{Spergel}}}, \bibinfo {author}
  {\bibfnamefont {S.~T.}\ \bibnamefont {{Staggs}}}, \bibinfo {author}
  {\bibfnamefont {D.~S.}\ \bibnamefont {{Swetz}}}, \bibinfo {author}
  {\bibfnamefont {E.~R.}\ \bibnamefont {{Switzer}}}, \bibinfo {author}
  {\bibfnamefont {R.}~\bibnamefont {{Thornton}}}, \bibinfo {author}
  {\bibfnamefont {H.}~\bibnamefont {{Trac}}},\ and\ \bibinfo {author}
  {\bibfnamefont {E.}~\bibnamefont {{Wollack}}},\ }\bibfield  {title} {\bibinfo
  {title} {{The Atacama Cosmology Telescope: temperature and gravitational
  lensing power spectrum measurements from three seasons of data}},\ }\href
  {https://doi.org/10.1088/1475-7516/2014/04/014} {\bibfield  {journal}
  {\bibinfo  {journal} {\jcap}\ }\textbf {\bibinfo {volume} {2014}},\ \bibinfo
  {eid} {014} (\bibinfo {year} {2014})},\ \Eprint
  {https://arxiv.org/abs/1301.1037} {arXiv:1301.1037 [astro-ph.CO]}
  \BibitemShut {NoStop}%
\bibitem [{\citenamefont {{Keisler}}\ \emph {et~al.}(2011)\citenamefont
  {{Keisler}}, \citenamefont {{Reichardt}}, \citenamefont {{Aird}},
  \citenamefont {{Benson}}, \citenamefont {{Bleem}}, \citenamefont
  {{Carlstrom}}, \citenamefont {{Chang}}, \citenamefont {{Cho}}, \citenamefont
  {{Crawford}}, \citenamefont {{Crites}}, \citenamefont {{de Haan}},
  \citenamefont {{Dobbs}}, \citenamefont {{Dudley}}, \citenamefont {{George}},
  \citenamefont {{Halverson}}, \citenamefont {{Holder}}, \citenamefont
  {{Holzapfel}}, \citenamefont {{Hoover}}, \citenamefont {{Hou}}, \citenamefont
  {{Hrubes}}, \citenamefont {{Joy}}, \citenamefont {{Knox}}, \citenamefont
  {{Lee}}, \citenamefont {{Leitch}}, \citenamefont {{Lueker}}, \citenamefont
  {{Luong-Van}}, \citenamefont {{McMahon}}, \citenamefont {{Mehl}},
  \citenamefont {{Meyer}}, \citenamefont {{Millea}}, \citenamefont {{Mohr}},
  \citenamefont {{Montroy}}, \citenamefont {{Natoli}}, \citenamefont {{Padin}},
  \citenamefont {{Plagge}}, \citenamefont {{Pryke}}, \citenamefont {{Ruhl}},
  \citenamefont {{Schaffer}}, \citenamefont {{Shaw}}, \citenamefont
  {{Shirokoff}}, \citenamefont {{Spieler}}, \citenamefont {{Staniszewski}},
  \citenamefont {{Stark}}, \citenamefont {{Story}}, \citenamefont {{van
  Engelen}}, \citenamefont {{Vanderlinde}}, \citenamefont {{Vieira}},
  \citenamefont {{Williamson}},\ and\ \citenamefont
  {{Zahn}}}]{2011ApJ...743...28K}%
  \BibitemOpen
  \bibfield  {author} {\bibinfo {author} {\bibfnamefont {R.}~\bibnamefont
  {{Keisler}}}, \bibinfo {author} {\bibfnamefont {C.~L.}\ \bibnamefont
  {{Reichardt}}}, \bibinfo {author} {\bibfnamefont {K.~A.}\ \bibnamefont
  {{Aird}}}, \bibinfo {author} {\bibfnamefont {B.~A.}\ \bibnamefont
  {{Benson}}}, \bibinfo {author} {\bibfnamefont {L.~E.}\ \bibnamefont
  {{Bleem}}}, \bibinfo {author} {\bibfnamefont {J.~E.}\ \bibnamefont
  {{Carlstrom}}}, \bibinfo {author} {\bibfnamefont {C.~L.}\ \bibnamefont
  {{Chang}}}, \bibinfo {author} {\bibfnamefont {H.~M.}\ \bibnamefont {{Cho}}},
  \bibinfo {author} {\bibfnamefont {T.~M.}\ \bibnamefont {{Crawford}}},
  \bibinfo {author} {\bibfnamefont {A.~T.}\ \bibnamefont {{Crites}}}, \bibinfo
  {author} {\bibfnamefont {T.}~\bibnamefont {{de Haan}}}, \bibinfo {author}
  {\bibfnamefont {M.~A.}\ \bibnamefont {{Dobbs}}}, \bibinfo {author}
  {\bibfnamefont {J.}~\bibnamefont {{Dudley}}}, \bibinfo {author}
  {\bibfnamefont {E.~M.}\ \bibnamefont {{George}}}, \bibinfo {author}
  {\bibfnamefont {N.~W.}\ \bibnamefont {{Halverson}}}, \bibinfo {author}
  {\bibfnamefont {G.~P.}\ \bibnamefont {{Holder}}}, \bibinfo {author}
  {\bibfnamefont {W.~L.}\ \bibnamefont {{Holzapfel}}}, \bibinfo {author}
  {\bibfnamefont {S.}~\bibnamefont {{Hoover}}}, \bibinfo {author}
  {\bibfnamefont {Z.}~\bibnamefont {{Hou}}}, \bibinfo {author} {\bibfnamefont
  {J.~D.}\ \bibnamefont {{Hrubes}}}, \bibinfo {author} {\bibfnamefont
  {M.}~\bibnamefont {{Joy}}}, \bibinfo {author} {\bibfnamefont
  {L.}~\bibnamefont {{Knox}}}, \bibinfo {author} {\bibfnamefont {A.~T.}\
  \bibnamefont {{Lee}}}, \bibinfo {author} {\bibfnamefont {E.~M.}\ \bibnamefont
  {{Leitch}}}, \bibinfo {author} {\bibfnamefont {M.}~\bibnamefont {{Lueker}}},
  \bibinfo {author} {\bibfnamefont {D.}~\bibnamefont {{Luong-Van}}}, \bibinfo
  {author} {\bibfnamefont {J.~J.}\ \bibnamefont {{McMahon}}}, \bibinfo {author}
  {\bibfnamefont {J.}~\bibnamefont {{Mehl}}}, \bibinfo {author} {\bibfnamefont
  {S.~S.}\ \bibnamefont {{Meyer}}}, \bibinfo {author} {\bibfnamefont
  {M.}~\bibnamefont {{Millea}}}, \bibinfo {author} {\bibfnamefont {J.~J.}\
  \bibnamefont {{Mohr}}}, \bibinfo {author} {\bibfnamefont {T.~E.}\
  \bibnamefont {{Montroy}}}, \bibinfo {author} {\bibfnamefont {T.}~\bibnamefont
  {{Natoli}}}, \bibinfo {author} {\bibfnamefont {S.}~\bibnamefont {{Padin}}},
  \bibinfo {author} {\bibfnamefont {T.}~\bibnamefont {{Plagge}}}, \bibinfo
  {author} {\bibfnamefont {C.}~\bibnamefont {{Pryke}}}, \bibinfo {author}
  {\bibfnamefont {J.~E.}\ \bibnamefont {{Ruhl}}}, \bibinfo {author}
  {\bibfnamefont {K.~K.}\ \bibnamefont {{Schaffer}}}, \bibinfo {author}
  {\bibfnamefont {L.}~\bibnamefont {{Shaw}}}, \bibinfo {author} {\bibfnamefont
  {E.}~\bibnamefont {{Shirokoff}}}, \bibinfo {author} {\bibfnamefont {H.~G.}\
  \bibnamefont {{Spieler}}}, \bibinfo {author} {\bibfnamefont {Z.}~\bibnamefont
  {{Staniszewski}}}, \bibinfo {author} {\bibfnamefont {A.~A.}\ \bibnamefont
  {{Stark}}}, \bibinfo {author} {\bibfnamefont {K.}~\bibnamefont {{Story}}},
  \bibinfo {author} {\bibfnamefont {A.}~\bibnamefont {{van Engelen}}}, \bibinfo
  {author} {\bibfnamefont {K.}~\bibnamefont {{Vanderlinde}}}, \bibinfo {author}
  {\bibfnamefont {J.~D.}\ \bibnamefont {{Vieira}}}, \bibinfo {author}
  {\bibfnamefont {R.}~\bibnamefont {{Williamson}}},\ and\ \bibinfo {author}
  {\bibfnamefont {O.}~\bibnamefont {{Zahn}}},\ }\bibfield  {title} {\bibinfo
  {title} {{A Measurement of the Damping Tail of the Cosmic Microwave
  Background Power Spectrum with the South Pole Telescope}},\ }\href
  {https://doi.org/10.1088/0004-637X/743/1/28} {\bibfield  {journal} {\bibinfo
  {journal} {\apj}\ }\textbf {\bibinfo {volume} {743}},\ \bibinfo {eid} {28}
  (\bibinfo {year} {2011})},\ \Eprint {https://arxiv.org/abs/1105.3182}
  {arXiv:1105.3182 [astro-ph.CO]} \BibitemShut {NoStop}%
\bibitem [{\citenamefont {{Blake}}\ \emph {et~al.}(2010)\citenamefont
  {{Blake}}, \citenamefont {{Brough}}, \citenamefont {{Colless}}, \citenamefont
  {{Couch}}, \citenamefont {{Croom}}, \citenamefont {{Davis}}, \citenamefont
  {{Drinkwater}}, \citenamefont {{Forster}}, \citenamefont {{Glazebrook}},
  \citenamefont {{Jelliffe}}, \citenamefont {{Jurek}}, \citenamefont {{Li}},
  \citenamefont {{Madore}}, \citenamefont {{Martin}}, \citenamefont
  {{Pimbblet}}, \citenamefont {{Poole}}, \citenamefont {{Pracy}}, \citenamefont
  {{Sharp}}, \citenamefont {{Wisnioski}}, \citenamefont {{Woods}},\ and\
  \citenamefont {{Wyder}}}]{2010MNRAS.406..803B}%
  \BibitemOpen
  \bibfield  {author} {\bibinfo {author} {\bibfnamefont {C.}~\bibnamefont
  {{Blake}}}, \bibinfo {author} {\bibfnamefont {S.}~\bibnamefont {{Brough}}},
  \bibinfo {author} {\bibfnamefont {M.}~\bibnamefont {{Colless}}}, \bibinfo
  {author} {\bibfnamefont {W.}~\bibnamefont {{Couch}}}, \bibinfo {author}
  {\bibfnamefont {S.}~\bibnamefont {{Croom}}}, \bibinfo {author} {\bibfnamefont
  {T.}~\bibnamefont {{Davis}}}, \bibinfo {author} {\bibfnamefont {M.~J.}\
  \bibnamefont {{Drinkwater}}}, \bibinfo {author} {\bibfnamefont
  {K.}~\bibnamefont {{Forster}}}, \bibinfo {author} {\bibfnamefont
  {K.}~\bibnamefont {{Glazebrook}}}, \bibinfo {author} {\bibfnamefont
  {B.}~\bibnamefont {{Jelliffe}}}, \bibinfo {author} {\bibfnamefont {R.~J.}\
  \bibnamefont {{Jurek}}}, \bibinfo {author} {\bibfnamefont {I.~H.}\
  \bibnamefont {{Li}}}, \bibinfo {author} {\bibfnamefont {B.}~\bibnamefont
  {{Madore}}}, \bibinfo {author} {\bibfnamefont {C.}~\bibnamefont {{Martin}}},
  \bibinfo {author} {\bibfnamefont {K.}~\bibnamefont {{Pimbblet}}}, \bibinfo
  {author} {\bibfnamefont {G.~B.}\ \bibnamefont {{Poole}}}, \bibinfo {author}
  {\bibfnamefont {M.}~\bibnamefont {{Pracy}}}, \bibinfo {author} {\bibfnamefont
  {R.}~\bibnamefont {{Sharp}}}, \bibinfo {author} {\bibfnamefont
  {E.}~\bibnamefont {{Wisnioski}}}, \bibinfo {author} {\bibfnamefont
  {D.}~\bibnamefont {{Woods}}},\ and\ \bibinfo {author} {\bibfnamefont
  {T.}~\bibnamefont {{Wyder}}},\ }\bibfield  {title} {\bibinfo {title} {{The
  WiggleZ Dark Energy Survey: the selection function and z = 0.6 galaxy power
  spectrum}},\ }\href {https://doi.org/10.1111/j.1365-2966.2010.16747.x}
  {\bibfield  {journal} {\bibinfo  {journal} {\mnras}\ }\textbf {\bibinfo
  {volume} {406}},\ \bibinfo {pages} {803} (\bibinfo {year} {2010})},\ \Eprint
  {https://arxiv.org/abs/1003.5721} {arXiv:1003.5721 [astro-ph.CO]}
  \BibitemShut {NoStop}%
\bibitem [{\citenamefont {{Hlo{\v{z}}ek}}\ \emph {et~al.}(2018)\citenamefont
  {{Hlo{\v{z}}ek}}, \citenamefont {{Marsh}},\ and\ \citenamefont
  {{Grin}}}]{2018MNRAS.476.3063H}%
  \BibitemOpen
  \bibfield  {author} {\bibinfo {author} {\bibfnamefont {R.}~\bibnamefont
  {{Hlo{\v{z}}ek}}}, \bibinfo {author} {\bibfnamefont {D.~J.~E.}\ \bibnamefont
  {{Marsh}}},\ and\ \bibinfo {author} {\bibfnamefont {D.}~\bibnamefont
  {{Grin}}},\ }\bibfield  {title} {\bibinfo {title} {{Using the full power of
  the cosmic microwave background to probe axion dark matter}},\ }\href
  {https://doi.org/10.1093/mnras/sty271} {\bibfield  {journal} {\bibinfo
  {journal} {\mnras}\ }\textbf {\bibinfo {volume} {476}},\ \bibinfo {pages}
  {3063} (\bibinfo {year} {2018})},\ \Eprint {https://arxiv.org/abs/1708.05681}
  {arXiv:1708.05681 [astro-ph.CO]} \BibitemShut {NoStop}%
\bibitem [{\citenamefont {{Dentler}}\ \emph {et~al.}(2022)\citenamefont
  {{Dentler}}, \citenamefont {{Marsh}}, \citenamefont {{Hlo{\v{z}}ek}},
  \citenamefont {{Lagu{\"e}}}, \citenamefont {{Rogers}},\ and\ \citenamefont
  {{Grin}}}]{2022MNRAS.515.5646D}%
  \BibitemOpen
  \bibfield  {author} {\bibinfo {author} {\bibfnamefont {M.}~\bibnamefont
  {{Dentler}}}, \bibinfo {author} {\bibfnamefont {D.~J.~E.}\ \bibnamefont
  {{Marsh}}}, \bibinfo {author} {\bibfnamefont {R.}~\bibnamefont
  {{Hlo{\v{z}}ek}}}, \bibinfo {author} {\bibfnamefont {A.}~\bibnamefont
  {{Lagu{\"e}}}}, \bibinfo {author} {\bibfnamefont {K.~K.}\ \bibnamefont
  {{Rogers}}},\ and\ \bibinfo {author} {\bibfnamefont {D.}~\bibnamefont
  {{Grin}}},\ }\bibfield  {title} {\bibinfo {title} {{Fuzzy dark matter and the
  Dark Energy Survey Year 1 data}},\ }\href
  {https://doi.org/10.1093/mnras/stac1946} {\bibfield  {journal} {\bibinfo
  {journal} {\mnras}\ }\textbf {\bibinfo {volume} {515}},\ \bibinfo {pages}
  {5646} (\bibinfo {year} {2022})},\ \Eprint {https://arxiv.org/abs/2111.01199}
  {arXiv:2111.01199 [astro-ph.CO]} \BibitemShut {NoStop}%
\bibitem [{\citenamefont {{Ipser}}\ and\ \citenamefont
  {{Sikivie}}(1983)}]{1983PhRvL..50..925I}%
  \BibitemOpen
  \bibfield  {author} {\bibinfo {author} {\bibfnamefont {J.}~\bibnamefont
  {{Ipser}}}\ and\ \bibinfo {author} {\bibfnamefont {P.}~\bibnamefont
  {{Sikivie}}},\ }\bibfield  {title} {\bibinfo {title} {{Can Galactic Halos Be
  Made of Axions?}},\ }\href {https://doi.org/10.1103/PhysRevLett.50.925}
  {\bibfield  {journal} {\bibinfo  {journal} {\prl}\ }\textbf {\bibinfo
  {volume} {50}},\ \bibinfo {pages} {925} (\bibinfo {year} {1983})}\BibitemShut
  {NoStop}%
\bibitem [{\citenamefont {{Eggemeier}}\ \emph {et~al.}(2024)\citenamefont
  {{Eggemeier}}, \citenamefont {{Krishnan Anilkumar}},\ and\ \citenamefont
  {{Dolag}}}]{2024arXiv240218221E}%
  \BibitemOpen
  \bibfield  {author} {\bibinfo {author} {\bibfnamefont {B.}~\bibnamefont
  {{Eggemeier}}}, \bibinfo {author} {\bibfnamefont {A.}~\bibnamefont {{Krishnan
  Anilkumar}}},\ and\ \bibinfo {author} {\bibfnamefont {K.}~\bibnamefont
  {{Dolag}}},\ }\bibfield  {title} {\bibinfo {title} {{Evidence for axion
  miniclusters with an increased central density}},\ }\href
  {https://doi.org/10.48550/arXiv.2402.18221} {\bibfield  {journal} {\bibinfo
  {journal} {arXiv e-prints}\ ,\ \bibinfo {eid} {arXiv:2402.18221}} (\bibinfo
  {year} {2024})},\ \Eprint {https://arxiv.org/abs/2402.18221}
  {arXiv:2402.18221 [astro-ph.CO]} \BibitemShut {NoStop}%
\bibitem [{\citenamefont {{Colpi}}\ \emph {et~al.}(1986)\citenamefont
  {{Colpi}}, \citenamefont {{Shapiro}},\ and\ \citenamefont
  {{Wasserman}}}]{1986PhRvL..57.2485C}%
  \BibitemOpen
  \bibfield  {author} {\bibinfo {author} {\bibfnamefont {M.}~\bibnamefont
  {{Colpi}}}, \bibinfo {author} {\bibfnamefont {S.~L.}\ \bibnamefont
  {{Shapiro}}},\ and\ \bibinfo {author} {\bibfnamefont {I.}~\bibnamefont
  {{Wasserman}}},\ }\bibfield  {title} {\bibinfo {title} {{Boson stars:
  Gravitational equilibria of self-interacting scalar fields}},\ }\href
  {https://doi.org/10.1103/PhysRevLett.57.2485} {\bibfield  {journal} {\bibinfo
   {journal} {\prl}\ }\textbf {\bibinfo {volume} {57}},\ \bibinfo {pages}
  {2485} (\bibinfo {year} {1986})}\BibitemShut {NoStop}%
\bibitem [{\citenamefont {{Bautista}}\ and\ \citenamefont
  {{Degollado}}(2024)}]{2024FrASS..1146820B}%
  \BibitemOpen
  \bibfield  {author} {\bibinfo {author} {\bibfnamefont {B.}~\bibnamefont
  {{Bautista}}}\ and\ \bibinfo {author} {\bibfnamefont {J.~C.}\ \bibnamefont
  {{Degollado}}},\ }\bibfield  {title} {\bibinfo {title} {{Static axion stars
  revisited}},\ }\href {https://doi.org/10.3389/fspas.2024.1346820} {\bibfield
  {journal} {\bibinfo  {journal} {Frontiers in Astronomy and Space Sciences}\
  }\textbf {\bibinfo {volume} {11}},\ \bibinfo {eid} {1346820} (\bibinfo {year}
  {2024})}\BibitemShut {NoStop}%
\bibitem [{\citenamefont {{Bauer}}\ \emph {et~al.}(2021)\citenamefont
  {{Bauer}}, \citenamefont {{Marsh}}, \citenamefont {{Hlo{\v{z}}ek}},
  \citenamefont {{Padmanabhan}},\ and\ \citenamefont
  {{Lagu{\"e}}}}]{2021MNRAS.500.3162B}%
  \BibitemOpen
  \bibfield  {author} {\bibinfo {author} {\bibfnamefont {J.~B.}\ \bibnamefont
  {{Bauer}}}, \bibinfo {author} {\bibfnamefont {D.~J.~E.}\ \bibnamefont
  {{Marsh}}}, \bibinfo {author} {\bibfnamefont {R.}~\bibnamefont
  {{Hlo{\v{z}}ek}}}, \bibinfo {author} {\bibfnamefont {H.}~\bibnamefont
  {{Padmanabhan}}},\ and\ \bibinfo {author} {\bibfnamefont {A.}~\bibnamefont
  {{Lagu{\"e}}}},\ }\bibfield  {title} {\bibinfo {title} {{Intensity mapping as
  a probe of axion dark matter}},\ }\href
  {https://doi.org/10.1093/mnras/staa3300} {\bibfield  {journal} {\bibinfo
  {journal} {\mnras}\ }\textbf {\bibinfo {volume} {500}},\ \bibinfo {pages}
  {3162} (\bibinfo {year} {2021})},\ \Eprint {https://arxiv.org/abs/2003.09655}
  {arXiv:2003.09655 [astro-ph.CO]} \BibitemShut {NoStop}%
\bibitem [{\citenamefont {{Furlanetto}}\ \emph {et~al.}(2006)\citenamefont
  {{Furlanetto}}, \citenamefont {{Oh}},\ and\ \citenamefont
  {{Briggs}}}]{2006PhR...433..181F}%
  \BibitemOpen
  \bibfield  {author} {\bibinfo {author} {\bibfnamefont {S.~R.}\ \bibnamefont
  {{Furlanetto}}}, \bibinfo {author} {\bibfnamefont {S.~P.}\ \bibnamefont
  {{Oh}}},\ and\ \bibinfo {author} {\bibfnamefont {F.~H.}\ \bibnamefont
  {{Briggs}}},\ }\bibfield  {title} {\bibinfo {title} {{Cosmology at low
  frequencies: The 21 cm transition and the high-redshift Universe}},\ }\href
  {https://doi.org/10.1016/j.physrep.2006.08.002} {\bibfield  {journal}
  {\bibinfo  {journal} {\physrep}\ }\textbf {\bibinfo {volume} {433}},\
  \bibinfo {pages} {181} (\bibinfo {year} {2006})},\ \Eprint
  {https://arxiv.org/abs/astro-ph/0608032} {arXiv:astro-ph/0608032 [astro-ph]}
  \BibitemShut {NoStop}%
\bibitem [{\citenamefont {{Switzer}}\ \emph {et~al.}(2013)\citenamefont
  {{Switzer}}, \citenamefont {{Masui}}, \citenamefont {{Bandura}},
  \citenamefont {{Calin}}, \citenamefont {{Chang}}, \citenamefont {{Chen}},
  \citenamefont {{Li}}, \citenamefont {{Liao}}, \citenamefont {{Natarajan}},
  \citenamefont {{Pen}}, \citenamefont {{Peterson}}, \citenamefont {{Shaw}},\
  and\ \citenamefont {{Voytek}}}]{2013MNRAS.434L..46S}%
  \BibitemOpen
  \bibfield  {author} {\bibinfo {author} {\bibfnamefont {E.~R.}\ \bibnamefont
  {{Switzer}}}, \bibinfo {author} {\bibfnamefont {K.~W.}\ \bibnamefont
  {{Masui}}}, \bibinfo {author} {\bibfnamefont {K.}~\bibnamefont {{Bandura}}},
  \bibinfo {author} {\bibfnamefont {L.~M.}\ \bibnamefont {{Calin}}}, \bibinfo
  {author} {\bibfnamefont {T.~C.}\ \bibnamefont {{Chang}}}, \bibinfo {author}
  {\bibfnamefont {X.~L.}\ \bibnamefont {{Chen}}}, \bibinfo {author}
  {\bibfnamefont {Y.~C.}\ \bibnamefont {{Li}}}, \bibinfo {author}
  {\bibfnamefont {Y.~W.}\ \bibnamefont {{Liao}}}, \bibinfo {author}
  {\bibfnamefont {A.}~\bibnamefont {{Natarajan}}}, \bibinfo {author}
  {\bibfnamefont {U.~L.}\ \bibnamefont {{Pen}}}, \bibinfo {author}
  {\bibfnamefont {J.~B.}\ \bibnamefont {{Peterson}}}, \bibinfo {author}
  {\bibfnamefont {J.~R.}\ \bibnamefont {{Shaw}}},\ and\ \bibinfo {author}
  {\bibfnamefont {T.~C.}\ \bibnamefont {{Voytek}}},\ }\bibfield  {title}
  {\bibinfo {title} {{Determination of z \raisebox{-0.5ex}\textasciitilde 0.8
  neutral hydrogen fluctuations using the 21cm intensity mapping
  autocorrelation.}},\ }\href {https://doi.org/10.1093/mnrasl/slt074}
  {\bibfield  {journal} {\bibinfo  {journal} {\mnras}\ }\textbf {\bibinfo
  {volume} {434}},\ \bibinfo {pages} {L46} (\bibinfo {year} {2013})},\ \Eprint
  {https://arxiv.org/abs/1304.3712} {arXiv:1304.3712 [astro-ph.CO]}
  \BibitemShut {NoStop}%
\bibitem [{\citenamefont {{Planck Collaboration}}\ \emph
  {et~al.}(2020)\citenamefont {{Planck Collaboration}}, \citenamefont
  {{Aghanim}}, \citenamefont {{Akrami}}, \citenamefont {{Ashdown}},
  \citenamefont {{Aumont}}, \citenamefont {{Baccigalupi}}, \citenamefont
  {{Ballardini}}, \citenamefont {{Banday}}, \citenamefont {{Barreiro}},
  \citenamefont {{Bartolo}}, \citenamefont {{Basak}}, \citenamefont {{Battye}},
  \citenamefont {{Benabed}}, \citenamefont {{Bernard}}, \citenamefont
  {{Bersanelli}}, \citenamefont {{Bielewicz}}, \citenamefont {{Bock}},
  \citenamefont {{Bond}}, \citenamefont {{Borrill}}, \citenamefont {{Bouchet}},
  \citenamefont {{Boulanger}}, \citenamefont {{Bucher}}, \citenamefont
  {{Burigana}}, \citenamefont {{Butler}}, \citenamefont {{Calabrese}},
  \citenamefont {{Cardoso}}, \citenamefont {{Carron}}, \citenamefont
  {{Challinor}}, \citenamefont {{Chiang}}, \citenamefont {{Chluba}},
  \citenamefont {{Colombo}}, \citenamefont {{Combet}}, \citenamefont
  {{Contreras}}, \citenamefont {{Crill}}, \citenamefont {{Cuttaia}},
  \citenamefont {{de Bernardis}}, \citenamefont {{de Zotti}}, \citenamefont
  {{Delabrouille}}, \citenamefont {{Delouis}}, \citenamefont {{Di Valentino}},
  \citenamefont {{Diego}}, \citenamefont {{Dor{\'e}}}, \citenamefont
  {{Douspis}}, \citenamefont {{Ducout}}, \citenamefont {{Dupac}}, \citenamefont
  {{Dusini}}, \citenamefont {{Efstathiou}}, \citenamefont {{Elsner}},
  \citenamefont {{En{\ss}lin}}, \citenamefont {{Eriksen}}, \citenamefont
  {{Fantaye}}, \citenamefont {{Farhang}}, \citenamefont {{Fergusson}},
  \citenamefont {{Fernandez-Cobos}}, \citenamefont {{Finelli}}, \citenamefont
  {{Forastieri}}, \citenamefont {{Frailis}}, \citenamefont {{Fraisse}},
  \citenamefont {{Franceschi}}, \citenamefont {{Frolov}}, \citenamefont
  {{Galeotta}}, \citenamefont {{Galli}}, \citenamefont {{Ganga}}, \citenamefont
  {{G{\'e}nova-Santos}}, \citenamefont {{Gerbino}}, \citenamefont {{Ghosh}},
  \citenamefont {{Gonz{\'a}lez-Nuevo}}, \citenamefont {{G{\'o}rski}},
  \citenamefont {{Gratton}}, \citenamefont {{Gruppuso}}, \citenamefont
  {{Gudmundsson}}, \citenamefont {{Hamann}}, \citenamefont {{Handley}},
  \citenamefont {{Hansen}}, \citenamefont {{Herranz}}, \citenamefont
  {{Hildebrandt}}, \citenamefont {{Hivon}}, \citenamefont {{Huang}},
  \citenamefont {{Jaffe}}, \citenamefont {{Jones}}, \citenamefont {{Karakci}},
  \citenamefont {{Keih{\"a}nen}}, \citenamefont {{Keskitalo}}, \citenamefont
  {{Kiiveri}}, \citenamefont {{Kim}}, \citenamefont {{Kisner}}, \citenamefont
  {{Knox}}, \citenamefont {{Krachmalnicoff}}, \citenamefont {{Kunz}},
  \citenamefont {{Kurki-Suonio}}, \citenamefont {{Lagache}}, \citenamefont
  {{Lamarre}}, \citenamefont {{Lasenby}}, \citenamefont {{Lattanzi}},
  \citenamefont {{Lawrence}}, \citenamefont {{Le Jeune}}, \citenamefont
  {{Lemos}}, \citenamefont {{Lesgourgues}}, \citenamefont {{Levrier}},
  \citenamefont {{Lewis}}, \citenamefont {{Liguori}}, \citenamefont {{Lilje}},
  \citenamefont {{Lilley}}, \citenamefont {{Lindholm}}, \citenamefont
  {{L{\'o}pez-Caniego}}, \citenamefont {{Lubin}}, \citenamefont {{Ma}},
  \citenamefont {{Mac{\'\i}as-P{\'e}rez}}, \citenamefont {{Maggio}},
  \citenamefont {{Maino}}, \citenamefont {{Mandolesi}}, \citenamefont
  {{Mangilli}}, \citenamefont {{Marcos-Caballero}}, \citenamefont {{Maris}},
  \citenamefont {{Martin}}, \citenamefont {{Martinelli}}, \citenamefont
  {{Mart{\'\i}nez-Gonz{\'a}lez}}, \citenamefont {{Matarrese}}, \citenamefont
  {{Mauri}}, \citenamefont {{McEwen}}, \citenamefont {{Meinhold}},
  \citenamefont {{Melchiorri}}, \citenamefont {{Mennella}}, \citenamefont
  {{Migliaccio}}, \citenamefont {{Millea}}, \citenamefont {{Mitra}},
  \citenamefont {{Miville-Desch{\^e}nes}}, \citenamefont {{Molinari}},
  \citenamefont {{Montier}}, \citenamefont {{Morgante}}, \citenamefont
  {{Moss}}, \citenamefont {{Natoli}}, \citenamefont {{N{\o}rgaard-Nielsen}},
  \citenamefont {{Pagano}}, \citenamefont {{Paoletti}}, \citenamefont
  {{Partridge}}, \citenamefont {{Patanchon}}, \citenamefont {{Peiris}},
  \citenamefont {{Perrotta}}, \citenamefont {{Pettorino}}, \citenamefont
  {{Piacentini}}, \citenamefont {{Polastri}}, \citenamefont {{Polenta}},
  \citenamefont {{Puget}}, \citenamefont {{Rachen}}, \citenamefont
  {{Reinecke}}, \citenamefont {{Remazeilles}}, \citenamefont {{Renzi}},
  \citenamefont {{Rocha}}, \citenamefont {{Rosset}}, \citenamefont {{Roudier}},
  \citenamefont {{Rubi{\~n}o-Mart{\'\i}n}}, \citenamefont {{Ruiz-Granados}},
  \citenamefont {{Salvati}}, \citenamefont {{Sandri}}, \citenamefont
  {{Savelainen}}, \citenamefont {{Scott}}, \citenamefont {{Shellard}},
  \citenamefont {{Sirignano}}, \citenamefont {{Sirri}}, \citenamefont
  {{Spencer}}, \citenamefont {{Sunyaev}}, \citenamefont {{Suur-Uski}},
  \citenamefont {{Tauber}}, \citenamefont {{Tavagnacco}}, \citenamefont
  {{Tenti}}, \citenamefont {{Toffolatti}}, \citenamefont {{Tomasi}},
  \citenamefont {{Trombetti}}, \citenamefont {{Valenziano}}, \citenamefont
  {{Valiviita}}, \citenamefont {{Van Tent}}, \citenamefont {{Vibert}},
  \citenamefont {{Vielva}}, \citenamefont {{Villa}}, \citenamefont
  {{Vittorio}}, \citenamefont {{Wandelt}}, \citenamefont {{Wehus}},
  \citenamefont {{White}}, \citenamefont {{White}}, \citenamefont {{Zacchei}},\
  and\ \citenamefont {{Zonca}}}]{2020A&A...641A...6P}%
  \BibitemOpen
  \bibfield  {author} {\bibinfo {author} {\bibnamefont {{Planck
  Collaboration}}}, \bibinfo {author} {\bibfnamefont {N.}~\bibnamefont
  {{Aghanim}}}, \bibinfo {author} {\bibfnamefont {Y.}~\bibnamefont {{Akrami}}},
  \bibinfo {author} {\bibfnamefont {M.}~\bibnamefont {{Ashdown}}}, \bibinfo
  {author} {\bibfnamefont {J.}~\bibnamefont {{Aumont}}}, \bibinfo {author}
  {\bibfnamefont {C.}~\bibnamefont {{Baccigalupi}}}, \bibinfo {author}
  {\bibfnamefont {M.}~\bibnamefont {{Ballardini}}}, \bibinfo {author}
  {\bibfnamefont {A.~J.}\ \bibnamefont {{Banday}}}, \bibinfo {author}
  {\bibfnamefont {R.~B.}\ \bibnamefont {{Barreiro}}}, \bibinfo {author}
  {\bibfnamefont {N.}~\bibnamefont {{Bartolo}}}, \bibinfo {author}
  {\bibfnamefont {S.}~\bibnamefont {{Basak}}}, \bibinfo {author} {\bibfnamefont
  {R.}~\bibnamefont {{Battye}}}, \bibinfo {author} {\bibfnamefont
  {K.}~\bibnamefont {{Benabed}}}, \bibinfo {author} {\bibfnamefont {J.~P.}\
  \bibnamefont {{Bernard}}}, \bibinfo {author} {\bibfnamefont {M.}~\bibnamefont
  {{Bersanelli}}}, \bibinfo {author} {\bibfnamefont {P.}~\bibnamefont
  {{Bielewicz}}}, \bibinfo {author} {\bibfnamefont {J.~J.}\ \bibnamefont
  {{Bock}}}, \bibinfo {author} {\bibfnamefont {J.~R.}\ \bibnamefont {{Bond}}},
  \bibinfo {author} {\bibfnamefont {J.}~\bibnamefont {{Borrill}}}, \bibinfo
  {author} {\bibfnamefont {F.~R.}\ \bibnamefont {{Bouchet}}}, \bibinfo {author}
  {\bibfnamefont {F.}~\bibnamefont {{Boulanger}}}, \bibinfo {author}
  {\bibfnamefont {M.}~\bibnamefont {{Bucher}}}, \bibinfo {author}
  {\bibfnamefont {C.}~\bibnamefont {{Burigana}}}, \bibinfo {author}
  {\bibfnamefont {R.~C.}\ \bibnamefont {{Butler}}}, \bibinfo {author}
  {\bibfnamefont {E.}~\bibnamefont {{Calabrese}}}, \bibinfo {author}
  {\bibfnamefont {J.~F.}\ \bibnamefont {{Cardoso}}}, \bibinfo {author}
  {\bibfnamefont {J.}~\bibnamefont {{Carron}}}, \bibinfo {author}
  {\bibfnamefont {A.}~\bibnamefont {{Challinor}}}, \bibinfo {author}
  {\bibfnamefont {H.~C.}\ \bibnamefont {{Chiang}}}, \bibinfo {author}
  {\bibfnamefont {J.}~\bibnamefont {{Chluba}}}, \bibinfo {author}
  {\bibfnamefont {L.~P.~L.}\ \bibnamefont {{Colombo}}}, \bibinfo {author}
  {\bibfnamefont {C.}~\bibnamefont {{Combet}}}, \bibinfo {author}
  {\bibfnamefont {D.}~\bibnamefont {{Contreras}}}, \bibinfo {author}
  {\bibfnamefont {B.~P.}\ \bibnamefont {{Crill}}}, \bibinfo {author}
  {\bibfnamefont {F.}~\bibnamefont {{Cuttaia}}}, \bibinfo {author}
  {\bibfnamefont {P.}~\bibnamefont {{de Bernardis}}}, \bibinfo {author}
  {\bibfnamefont {G.}~\bibnamefont {{de Zotti}}}, \bibinfo {author}
  {\bibfnamefont {J.}~\bibnamefont {{Delabrouille}}}, \bibinfo {author}
  {\bibfnamefont {J.~M.}\ \bibnamefont {{Delouis}}}, \bibinfo {author}
  {\bibfnamefont {E.}~\bibnamefont {{Di Valentino}}}, \bibinfo {author}
  {\bibfnamefont {J.~M.}\ \bibnamefont {{Diego}}}, \bibinfo {author}
  {\bibfnamefont {O.}~\bibnamefont {{Dor{\'e}}}}, \bibinfo {author}
  {\bibfnamefont {M.}~\bibnamefont {{Douspis}}}, \bibinfo {author}
  {\bibfnamefont {A.}~\bibnamefont {{Ducout}}}, \bibinfo {author}
  {\bibfnamefont {X.}~\bibnamefont {{Dupac}}}, \bibinfo {author} {\bibfnamefont
  {S.}~\bibnamefont {{Dusini}}}, \bibinfo {author} {\bibfnamefont
  {G.}~\bibnamefont {{Efstathiou}}}, \bibinfo {author} {\bibfnamefont
  {F.}~\bibnamefont {{Elsner}}}, \bibinfo {author} {\bibfnamefont {T.~A.}\
  \bibnamefont {{En{\ss}lin}}}, \bibinfo {author} {\bibfnamefont {H.~K.}\
  \bibnamefont {{Eriksen}}}, \bibinfo {author} {\bibfnamefont {Y.}~\bibnamefont
  {{Fantaye}}}, \bibinfo {author} {\bibfnamefont {M.}~\bibnamefont
  {{Farhang}}}, \bibinfo {author} {\bibfnamefont {J.}~\bibnamefont
  {{Fergusson}}}, \bibinfo {author} {\bibfnamefont {R.}~\bibnamefont
  {{Fernandez-Cobos}}}, \bibinfo {author} {\bibfnamefont {F.}~\bibnamefont
  {{Finelli}}}, \bibinfo {author} {\bibfnamefont {F.}~\bibnamefont
  {{Forastieri}}}, \bibinfo {author} {\bibfnamefont {M.}~\bibnamefont
  {{Frailis}}}, \bibinfo {author} {\bibfnamefont {A.~A.}\ \bibnamefont
  {{Fraisse}}}, \bibinfo {author} {\bibfnamefont {E.}~\bibnamefont
  {{Franceschi}}}, \bibinfo {author} {\bibfnamefont {A.}~\bibnamefont
  {{Frolov}}}, \bibinfo {author} {\bibfnamefont {S.}~\bibnamefont
  {{Galeotta}}}, \bibinfo {author} {\bibfnamefont {S.}~\bibnamefont {{Galli}}},
  \bibinfo {author} {\bibfnamefont {K.}~\bibnamefont {{Ganga}}}, \bibinfo
  {author} {\bibfnamefont {R.~T.}\ \bibnamefont {{G{\'e}nova-Santos}}},
  \bibinfo {author} {\bibfnamefont {M.}~\bibnamefont {{Gerbino}}}, \bibinfo
  {author} {\bibfnamefont {T.}~\bibnamefont {{Ghosh}}}, \bibinfo {author}
  {\bibfnamefont {J.}~\bibnamefont {{Gonz{\'a}lez-Nuevo}}}, \bibinfo {author}
  {\bibfnamefont {K.~M.}\ \bibnamefont {{G{\'o}rski}}}, \bibinfo {author}
  {\bibfnamefont {S.}~\bibnamefont {{Gratton}}}, \bibinfo {author}
  {\bibfnamefont {A.}~\bibnamefont {{Gruppuso}}}, \bibinfo {author}
  {\bibfnamefont {J.~E.}\ \bibnamefont {{Gudmundsson}}}, \bibinfo {author}
  {\bibfnamefont {J.}~\bibnamefont {{Hamann}}}, \bibinfo {author}
  {\bibfnamefont {W.}~\bibnamefont {{Handley}}}, \bibinfo {author}
  {\bibfnamefont {F.~K.}\ \bibnamefont {{Hansen}}}, \bibinfo {author}
  {\bibfnamefont {D.}~\bibnamefont {{Herranz}}}, \bibinfo {author}
  {\bibfnamefont {S.~R.}\ \bibnamefont {{Hildebrandt}}}, \bibinfo {author}
  {\bibfnamefont {E.}~\bibnamefont {{Hivon}}}, \bibinfo {author} {\bibfnamefont
  {Z.}~\bibnamefont {{Huang}}}, \bibinfo {author} {\bibfnamefont {A.~H.}\
  \bibnamefont {{Jaffe}}}, \bibinfo {author} {\bibfnamefont {W.~C.}\
  \bibnamefont {{Jones}}}, \bibinfo {author} {\bibfnamefont {A.}~\bibnamefont
  {{Karakci}}}, \bibinfo {author} {\bibfnamefont {E.}~\bibnamefont
  {{Keih{\"a}nen}}}, \bibinfo {author} {\bibfnamefont {R.}~\bibnamefont
  {{Keskitalo}}}, \bibinfo {author} {\bibfnamefont {K.}~\bibnamefont
  {{Kiiveri}}}, \bibinfo {author} {\bibfnamefont {J.}~\bibnamefont {{Kim}}},
  \bibinfo {author} {\bibfnamefont {T.~S.}\ \bibnamefont {{Kisner}}}, \bibinfo
  {author} {\bibfnamefont {L.}~\bibnamefont {{Knox}}}, \bibinfo {author}
  {\bibfnamefont {N.}~\bibnamefont {{Krachmalnicoff}}}, \bibinfo {author}
  {\bibfnamefont {M.}~\bibnamefont {{Kunz}}}, \bibinfo {author} {\bibfnamefont
  {H.}~\bibnamefont {{Kurki-Suonio}}}, \bibinfo {author} {\bibfnamefont
  {G.}~\bibnamefont {{Lagache}}}, \bibinfo {author} {\bibfnamefont {J.~M.}\
  \bibnamefont {{Lamarre}}}, \bibinfo {author} {\bibfnamefont {A.}~\bibnamefont
  {{Lasenby}}}, \bibinfo {author} {\bibfnamefont {M.}~\bibnamefont
  {{Lattanzi}}}, \bibinfo {author} {\bibfnamefont {C.~R.}\ \bibnamefont
  {{Lawrence}}}, \bibinfo {author} {\bibfnamefont {M.}~\bibnamefont {{Le
  Jeune}}}, \bibinfo {author} {\bibfnamefont {P.}~\bibnamefont {{Lemos}}},
  \bibinfo {author} {\bibfnamefont {J.}~\bibnamefont {{Lesgourgues}}}, \bibinfo
  {author} {\bibfnamefont {F.}~\bibnamefont {{Levrier}}}, \bibinfo {author}
  {\bibfnamefont {A.}~\bibnamefont {{Lewis}}}, \bibinfo {author} {\bibfnamefont
  {M.}~\bibnamefont {{Liguori}}}, \bibinfo {author} {\bibfnamefont {P.~B.}\
  \bibnamefont {{Lilje}}}, \bibinfo {author} {\bibfnamefont {M.}~\bibnamefont
  {{Lilley}}}, \bibinfo {author} {\bibfnamefont {V.}~\bibnamefont
  {{Lindholm}}}, \bibinfo {author} {\bibfnamefont {M.}~\bibnamefont
  {{L{\'o}pez-Caniego}}}, \bibinfo {author} {\bibfnamefont {P.~M.}\
  \bibnamefont {{Lubin}}}, \bibinfo {author} {\bibfnamefont {Y.~Z.}\
  \bibnamefont {{Ma}}}, \bibinfo {author} {\bibfnamefont {J.~F.}\ \bibnamefont
  {{Mac{\'\i}as-P{\'e}rez}}}, \bibinfo {author} {\bibfnamefont
  {G.}~\bibnamefont {{Maggio}}}, \bibinfo {author} {\bibfnamefont
  {D.}~\bibnamefont {{Maino}}}, \bibinfo {author} {\bibfnamefont
  {N.}~\bibnamefont {{Mandolesi}}}, \bibinfo {author} {\bibfnamefont
  {A.}~\bibnamefont {{Mangilli}}}, \bibinfo {author} {\bibfnamefont
  {A.}~\bibnamefont {{Marcos-Caballero}}}, \bibinfo {author} {\bibfnamefont
  {M.}~\bibnamefont {{Maris}}}, \bibinfo {author} {\bibfnamefont {P.~G.}\
  \bibnamefont {{Martin}}}, \bibinfo {author} {\bibfnamefont {M.}~\bibnamefont
  {{Martinelli}}}, \bibinfo {author} {\bibfnamefont {E.}~\bibnamefont
  {{Mart{\'\i}nez-Gonz{\'a}lez}}}, \bibinfo {author} {\bibfnamefont
  {S.}~\bibnamefont {{Matarrese}}}, \bibinfo {author} {\bibfnamefont
  {N.}~\bibnamefont {{Mauri}}}, \bibinfo {author} {\bibfnamefont {J.~D.}\
  \bibnamefont {{McEwen}}}, \bibinfo {author} {\bibfnamefont {P.~R.}\
  \bibnamefont {{Meinhold}}}, \bibinfo {author} {\bibfnamefont
  {A.}~\bibnamefont {{Melchiorri}}}, \bibinfo {author} {\bibfnamefont
  {A.}~\bibnamefont {{Mennella}}}, \bibinfo {author} {\bibfnamefont
  {M.}~\bibnamefont {{Migliaccio}}}, \bibinfo {author} {\bibfnamefont
  {M.}~\bibnamefont {{Millea}}}, \bibinfo {author} {\bibfnamefont
  {S.}~\bibnamefont {{Mitra}}}, \bibinfo {author} {\bibfnamefont {M.~A.}\
  \bibnamefont {{Miville-Desch{\^e}nes}}}, \bibinfo {author} {\bibfnamefont
  {D.}~\bibnamefont {{Molinari}}}, \bibinfo {author} {\bibfnamefont
  {L.}~\bibnamefont {{Montier}}}, \bibinfo {author} {\bibfnamefont
  {G.}~\bibnamefont {{Morgante}}}, \bibinfo {author} {\bibfnamefont
  {A.}~\bibnamefont {{Moss}}}, \bibinfo {author} {\bibfnamefont
  {P.}~\bibnamefont {{Natoli}}}, \bibinfo {author} {\bibfnamefont {H.~U.}\
  \bibnamefont {{N{\o}rgaard-Nielsen}}}, \bibinfo {author} {\bibfnamefont
  {L.}~\bibnamefont {{Pagano}}}, \bibinfo {author} {\bibfnamefont
  {D.}~\bibnamefont {{Paoletti}}}, \bibinfo {author} {\bibfnamefont
  {B.}~\bibnamefont {{Partridge}}}, \bibinfo {author} {\bibfnamefont
  {G.}~\bibnamefont {{Patanchon}}}, \bibinfo {author} {\bibfnamefont {H.~V.}\
  \bibnamefont {{Peiris}}}, \bibinfo {author} {\bibfnamefont {F.}~\bibnamefont
  {{Perrotta}}}, \bibinfo {author} {\bibfnamefont {V.}~\bibnamefont
  {{Pettorino}}}, \bibinfo {author} {\bibfnamefont {F.}~\bibnamefont
  {{Piacentini}}}, \bibinfo {author} {\bibfnamefont {L.}~\bibnamefont
  {{Polastri}}}, \bibinfo {author} {\bibfnamefont {G.}~\bibnamefont
  {{Polenta}}}, \bibinfo {author} {\bibfnamefont {J.~L.}\ \bibnamefont
  {{Puget}}}, \bibinfo {author} {\bibfnamefont {J.~P.}\ \bibnamefont
  {{Rachen}}}, \bibinfo {author} {\bibfnamefont {M.}~\bibnamefont
  {{Reinecke}}}, \bibinfo {author} {\bibfnamefont {M.}~\bibnamefont
  {{Remazeilles}}}, \bibinfo {author} {\bibfnamefont {A.}~\bibnamefont
  {{Renzi}}}, \bibinfo {author} {\bibfnamefont {G.}~\bibnamefont {{Rocha}}},
  \bibinfo {author} {\bibfnamefont {C.}~\bibnamefont {{Rosset}}}, \bibinfo
  {author} {\bibfnamefont {G.}~\bibnamefont {{Roudier}}}, \bibinfo {author}
  {\bibfnamefont {J.~A.}\ \bibnamefont {{Rubi{\~n}o-Mart{\'\i}n}}}, \bibinfo
  {author} {\bibfnamefont {B.}~\bibnamefont {{Ruiz-Granados}}}, \bibinfo
  {author} {\bibfnamefont {L.}~\bibnamefont {{Salvati}}}, \bibinfo {author}
  {\bibfnamefont {M.}~\bibnamefont {{Sandri}}}, \bibinfo {author}
  {\bibfnamefont {M.}~\bibnamefont {{Savelainen}}}, \bibinfo {author}
  {\bibfnamefont {D.}~\bibnamefont {{Scott}}}, \bibinfo {author} {\bibfnamefont
  {E.~P.~S.}\ \bibnamefont {{Shellard}}}, \bibinfo {author} {\bibfnamefont
  {C.}~\bibnamefont {{Sirignano}}}, \bibinfo {author} {\bibfnamefont
  {G.}~\bibnamefont {{Sirri}}}, \bibinfo {author} {\bibfnamefont {L.~D.}\
  \bibnamefont {{Spencer}}}, \bibinfo {author} {\bibfnamefont {R.}~\bibnamefont
  {{Sunyaev}}}, \bibinfo {author} {\bibfnamefont {A.~S.}\ \bibnamefont
  {{Suur-Uski}}}, \bibinfo {author} {\bibfnamefont {J.~A.}\ \bibnamefont
  {{Tauber}}}, \bibinfo {author} {\bibfnamefont {D.}~\bibnamefont
  {{Tavagnacco}}}, \bibinfo {author} {\bibfnamefont {M.}~\bibnamefont
  {{Tenti}}}, \bibinfo {author} {\bibfnamefont {L.}~\bibnamefont
  {{Toffolatti}}}, \bibinfo {author} {\bibfnamefont {M.}~\bibnamefont
  {{Tomasi}}}, \bibinfo {author} {\bibfnamefont {T.}~\bibnamefont
  {{Trombetti}}}, \bibinfo {author} {\bibfnamefont {L.}~\bibnamefont
  {{Valenziano}}}, \bibinfo {author} {\bibfnamefont {J.}~\bibnamefont
  {{Valiviita}}}, \bibinfo {author} {\bibfnamefont {B.}~\bibnamefont {{Van
  Tent}}}, \bibinfo {author} {\bibfnamefont {L.}~\bibnamefont {{Vibert}}},
  \bibinfo {author} {\bibfnamefont {P.}~\bibnamefont {{Vielva}}}, \bibinfo
  {author} {\bibfnamefont {F.}~\bibnamefont {{Villa}}}, \bibinfo {author}
  {\bibfnamefont {N.}~\bibnamefont {{Vittorio}}}, \bibinfo {author}
  {\bibfnamefont {B.~D.}\ \bibnamefont {{Wandelt}}}, \bibinfo {author}
  {\bibfnamefont {I.~K.}\ \bibnamefont {{Wehus}}}, \bibinfo {author}
  {\bibfnamefont {M.}~\bibnamefont {{White}}}, \bibinfo {author} {\bibfnamefont
  {S.~D.~M.}\ \bibnamefont {{White}}}, \bibinfo {author} {\bibfnamefont
  {A.}~\bibnamefont {{Zacchei}}},\ and\ \bibinfo {author} {\bibfnamefont
  {A.}~\bibnamefont {{Zonca}}},\ }\bibfield  {title} {\bibinfo {title} {{Planck
  2018 results. VI. Cosmological parameters}},\ }\href
  {https://doi.org/10.1051/0004-6361/201833910} {\bibfield  {journal} {\bibinfo
   {journal} {\aap}\ }\textbf {\bibinfo {volume} {641}},\ \bibinfo {eid} {A6}
  (\bibinfo {year} {2020})},\ \Eprint {https://arxiv.org/abs/1807.06209}
  {arXiv:1807.06209 [astro-ph.CO]} \BibitemShut {NoStop}%
\bibitem [{\citenamefont {{Abazajian}}\ \emph {et~al.}(2019)\citenamefont
  {{Abazajian}}, \citenamefont {{Addison}}, \citenamefont {{Adshead}},
  \citenamefont {{Ahmed}}, \citenamefont {{Allen}}, \citenamefont {{Alonso}},
  \citenamefont {{Alvarez}}, \citenamefont {{Anderson}}, \citenamefont
  {{Arnold}}, \citenamefont {{Baccigalupi}}, \citenamefont {{Bailey}},
  \citenamefont {{Barkats}}, \citenamefont {{Barron}}, \citenamefont {{Barry}},
  \citenamefont {{Bartlett}}, \citenamefont {{Basu Thakur}}, \citenamefont
  {{Battaglia}}, \citenamefont {{Baxter}}, \citenamefont {{Bean}},
  \citenamefont {{Bebek}}, \citenamefont {{Bender}}, \citenamefont {{Benson}},
  \citenamefont {{Berger}}, \citenamefont {{Bhimani}}, \citenamefont
  {{Bischoff}}, \citenamefont {{Bleem}}, \citenamefont {{Bocquet}},
  \citenamefont {{Boddy}}, \citenamefont {{Bonato}}, \citenamefont {{Bond}},
  \citenamefont {{Borrill}}, \citenamefont {{Bouchet}}, \citenamefont
  {{Brown}}, \citenamefont {{Bryan}}, \citenamefont {{Burkhart}}, \citenamefont
  {{Buza}}, \citenamefont {{Byrum}}, \citenamefont {{Calabrese}}, \citenamefont
  {{Calafut}}, \citenamefont {{Caldwell}}, \citenamefont {{Carlstrom}},
  \citenamefont {{Carron}}, \citenamefont {{Cecil}}, \citenamefont
  {{Challinor}}, \citenamefont {{Chang}}, \citenamefont {{Chinone}},
  \citenamefont {{Cho}}, \citenamefont {{Cooray}}, \citenamefont {{Crawford}},
  \citenamefont {{Crites}}, \citenamefont {{Cukierman}}, \citenamefont
  {{Cyr-Racine}}, \citenamefont {{de Haan}}, \citenamefont {{de Zotti}},
  \citenamefont {{Delabrouille}}, \citenamefont {{Demarteau}}, \citenamefont
  {{Devlin}}, \citenamefont {{Di Valentino}}, \citenamefont {{Dobbs}},
  \citenamefont {{Duff}}, \citenamefont {{Duivenvoorden}}, \citenamefont
  {{Dvorkin}}, \citenamefont {{Edwards}}, \citenamefont {{Eimer}},
  \citenamefont {{Errard}}, \citenamefont {{Essinger-Hileman}}, \citenamefont
  {{Fabbian}}, \citenamefont {{Feng}}, \citenamefont {{Ferraro}}, \citenamefont
  {{Filippini}}, \citenamefont {{Flauger}}, \citenamefont {{Flaugher}},
  \citenamefont {{Fraisse}}, \citenamefont {{Frolov}}, \citenamefont
  {{Galitzki}}, \citenamefont {{Galli}}, \citenamefont {{Ganga}}, \citenamefont
  {{Gerbino}}, \citenamefont {{Gilchriese}}, \citenamefont {{Gluscevic}},
  \citenamefont {{Green}}, \citenamefont {{Grin}}, \citenamefont {{Grohs}},
  \citenamefont {{Gualtieri}}, \citenamefont {{Guarino}}, \citenamefont
  {{Gudmundsson}}, \citenamefont {{Habib}}, \citenamefont {{Haller}},
  \citenamefont {{Halpern}}, \citenamefont {{Halverson}}, \citenamefont
  {{Hanany}}, \citenamefont {{Harrington}}, \citenamefont {{Hasegawa}},
  \citenamefont {{Hasselfield}}, \citenamefont {{Hazumi}}, \citenamefont
  {{Heitmann}}, \citenamefont {{Henderson}}, \citenamefont {{Henning}},
  \citenamefont {{Hill}}, \citenamefont {{Hlozek}}, \citenamefont {{Holder}},
  \citenamefont {{Holzapfel}}, \citenamefont {{Hubmayr}}, \citenamefont
  {{Huffenberger}}, \citenamefont {{Huffer}}, \citenamefont {{Hui}},
  \citenamefont {{Irwin}}, \citenamefont {{Johnson}}, \citenamefont
  {{Johnstone}}, \citenamefont {{Jones}}, \citenamefont {{Karkare}},
  \citenamefont {{Katayama}}, \citenamefont {{Kerby}}, \citenamefont
  {{Kernovsky}}, \citenamefont {{Keskitalo}}, \citenamefont {{Kisner}},
  \citenamefont {{Knox}}, \citenamefont {{Kosowsky}}, \citenamefont {{Kovac}},
  \citenamefont {{Kovetz}}, \citenamefont {{Kuhlmann}}, \citenamefont {{Kuo}},
  \citenamefont {{Kurita}}, \citenamefont {{Kusaka}}, \citenamefont
  {{Lahteenmaki}}, \citenamefont {{Lawrence}}, \citenamefont {{Lee}},
  \citenamefont {{Lewis}}, \citenamefont {{Li}}, \citenamefont {{Linder}},
  \citenamefont {{Loverde}}, \citenamefont {{Lowitz}}, \citenamefont
  {{Madhavacheril}}, \citenamefont {{Mantz}}, \citenamefont {{Matsuda}},
  \citenamefont {{Mauskopf}}, \citenamefont {{McMahon}}, \citenamefont
  {{McQuinn}}, \citenamefont {{Meerburg}}, \citenamefont {{Melin}},
  \citenamefont {{Meyers}}, \citenamefont {{Millea}}, \citenamefont {{Mohr}},
  \citenamefont {{Moncelsi}}, \citenamefont {{Mroczkowski}}, \citenamefont
  {{Mukherjee}}, \citenamefont {{M{\"u}nchmeyer}}, \citenamefont {{Nagai}},
  \citenamefont {{Nagy}}, \citenamefont {{Namikawa}}, \citenamefont {{Nati}},
  \citenamefont {{Natoli}}, \citenamefont {{Negrello}}, \citenamefont
  {{Newburgh}}, \citenamefont {{Niemack}}, \citenamefont {{Nishino}},
  \citenamefont {{Nordby}}, \citenamefont {{Novosad}}, \citenamefont
  {{O'Connor}}, \citenamefont {{Obied}}, \citenamefont {{Padin}}, \citenamefont
  {{Pandey}}, \citenamefont {{Partridge}}, \citenamefont {{Pierpaoli}},
  \citenamefont {{Pogosian}}, \citenamefont {{Pryke}}, \citenamefont
  {{Puglisi}}, \citenamefont {{Racine}}, \citenamefont {{Raghunathan}},
  \citenamefont {{Rahlin}}, \citenamefont {{Rajagopalan}}, \citenamefont
  {{Raveri}}, \citenamefont {{Reichanadter}}, \citenamefont {{Reichardt}},
  \citenamefont {{Remazeilles}}, \citenamefont {{Rocha}}, \citenamefont
  {{Roe}}, \citenamefont {{Roy}}, \citenamefont {{Ruhl}}, \citenamefont
  {{Salatino}}, \citenamefont {{Saliwanchik}}, \citenamefont {{Schaan}},
  \citenamefont {{Schillaci}}, \citenamefont {{Schmittfull}}, \citenamefont
  {{Scott}}, \citenamefont {{Sehgal}}, \citenamefont {{Shandera}},
  \citenamefont {{Sheehy}}, \citenamefont {{Sherwin}}, \citenamefont
  {{Shirokoff}}, \citenamefont {{Simon}}, \citenamefont {{Slosar}},
  \citenamefont {{Somerville}}, \citenamefont {{Spergel}}, \citenamefont
  {{Staggs}}, \citenamefont {{Stark}}, \citenamefont {{Stompor}}, \citenamefont
  {{Story}}, \citenamefont {{Stoughton}}, \citenamefont {{Suzuki}},
  \citenamefont {{Tajima}}, \citenamefont {{Teply}}, \citenamefont
  {{Thompson}}, \citenamefont {{Timbie}}, \citenamefont {{Tomasi}},
  \citenamefont {{Treu}}, \citenamefont {{Tristram}}, \citenamefont {{Tucker}},
  \citenamefont {{Umilt{\`a}}}, \citenamefont {{van Engelen}}, \citenamefont
  {{Vieira}}, \citenamefont {{Vieregg}}, \citenamefont {{Vogelsberger}},
  \citenamefont {{Wang}}, \citenamefont {{Watson}}, \citenamefont {{White}},
  \citenamefont {{Whitehorn}}, \citenamefont {{Wollack}}, \citenamefont {{Kimmy
  Wu}}, \citenamefont {{Xu}}, \citenamefont {{Yasini}}, \citenamefont {{Yeck}},
  \citenamefont {{Yoon}}, \citenamefont {{Young}},\ and\ \citenamefont
  {{Zonca}}}]{2019arXiv190704473A}%
  \BibitemOpen
  \bibfield  {author} {\bibinfo {author} {\bibfnamefont {K.}~\bibnamefont
  {{Abazajian}}}, \bibinfo {author} {\bibfnamefont {G.}~\bibnamefont
  {{Addison}}}, \bibinfo {author} {\bibfnamefont {P.}~\bibnamefont
  {{Adshead}}}, \bibinfo {author} {\bibfnamefont {Z.}~\bibnamefont {{Ahmed}}},
  \bibinfo {author} {\bibfnamefont {S.~W.}\ \bibnamefont {{Allen}}}, \bibinfo
  {author} {\bibfnamefont {D.}~\bibnamefont {{Alonso}}}, \bibinfo {author}
  {\bibfnamefont {M.}~\bibnamefont {{Alvarez}}}, \bibinfo {author}
  {\bibfnamefont {A.}~\bibnamefont {{Anderson}}}, \bibinfo {author}
  {\bibfnamefont {K.~S.}\ \bibnamefont {{Arnold}}}, \bibinfo {author}
  {\bibfnamefont {C.}~\bibnamefont {{Baccigalupi}}}, \bibinfo {author}
  {\bibfnamefont {K.}~\bibnamefont {{Bailey}}}, \bibinfo {author}
  {\bibfnamefont {D.}~\bibnamefont {{Barkats}}}, \bibinfo {author}
  {\bibfnamefont {D.}~\bibnamefont {{Barron}}}, \bibinfo {author}
  {\bibfnamefont {P.~S.}\ \bibnamefont {{Barry}}}, \bibinfo {author}
  {\bibfnamefont {J.~G.}\ \bibnamefont {{Bartlett}}}, \bibinfo {author}
  {\bibfnamefont {R.}~\bibnamefont {{Basu Thakur}}}, \bibinfo {author}
  {\bibfnamefont {N.}~\bibnamefont {{Battaglia}}}, \bibinfo {author}
  {\bibfnamefont {E.}~\bibnamefont {{Baxter}}}, \bibinfo {author}
  {\bibfnamefont {R.}~\bibnamefont {{Bean}}}, \bibinfo {author} {\bibfnamefont
  {C.}~\bibnamefont {{Bebek}}}, \bibinfo {author} {\bibfnamefont {A.~N.}\
  \bibnamefont {{Bender}}}, \bibinfo {author} {\bibfnamefont {B.~A.}\
  \bibnamefont {{Benson}}}, \bibinfo {author} {\bibfnamefont {E.}~\bibnamefont
  {{Berger}}}, \bibinfo {author} {\bibfnamefont {S.}~\bibnamefont {{Bhimani}}},
  \bibinfo {author} {\bibfnamefont {C.~A.}\ \bibnamefont {{Bischoff}}},
  \bibinfo {author} {\bibfnamefont {L.}~\bibnamefont {{Bleem}}}, \bibinfo
  {author} {\bibfnamefont {S.}~\bibnamefont {{Bocquet}}}, \bibinfo {author}
  {\bibfnamefont {K.}~\bibnamefont {{Boddy}}}, \bibinfo {author} {\bibfnamefont
  {M.}~\bibnamefont {{Bonato}}}, \bibinfo {author} {\bibfnamefont {J.~R.}\
  \bibnamefont {{Bond}}}, \bibinfo {author} {\bibfnamefont {J.}~\bibnamefont
  {{Borrill}}}, \bibinfo {author} {\bibfnamefont {F.~R.}\ \bibnamefont
  {{Bouchet}}}, \bibinfo {author} {\bibfnamefont {M.~L.}\ \bibnamefont
  {{Brown}}}, \bibinfo {author} {\bibfnamefont {S.}~\bibnamefont {{Bryan}}},
  \bibinfo {author} {\bibfnamefont {B.}~\bibnamefont {{Burkhart}}}, \bibinfo
  {author} {\bibfnamefont {V.}~\bibnamefont {{Buza}}}, \bibinfo {author}
  {\bibfnamefont {K.}~\bibnamefont {{Byrum}}}, \bibinfo {author} {\bibfnamefont
  {E.}~\bibnamefont {{Calabrese}}}, \bibinfo {author} {\bibfnamefont
  {V.}~\bibnamefont {{Calafut}}}, \bibinfo {author} {\bibfnamefont
  {R.}~\bibnamefont {{Caldwell}}}, \bibinfo {author} {\bibfnamefont {J.~E.}\
  \bibnamefont {{Carlstrom}}}, \bibinfo {author} {\bibfnamefont
  {J.}~\bibnamefont {{Carron}}}, \bibinfo {author} {\bibfnamefont
  {T.}~\bibnamefont {{Cecil}}}, \bibinfo {author} {\bibfnamefont
  {A.}~\bibnamefont {{Challinor}}}, \bibinfo {author} {\bibfnamefont {C.~L.}\
  \bibnamefont {{Chang}}}, \bibinfo {author} {\bibfnamefont {Y.}~\bibnamefont
  {{Chinone}}}, \bibinfo {author} {\bibfnamefont {H.-M.~S.}\ \bibnamefont
  {{Cho}}}, \bibinfo {author} {\bibfnamefont {A.}~\bibnamefont {{Cooray}}},
  \bibinfo {author} {\bibfnamefont {T.~M.}\ \bibnamefont {{Crawford}}},
  \bibinfo {author} {\bibfnamefont {A.}~\bibnamefont {{Crites}}}, \bibinfo
  {author} {\bibfnamefont {A.}~\bibnamefont {{Cukierman}}}, \bibinfo {author}
  {\bibfnamefont {F.-Y.}\ \bibnamefont {{Cyr-Racine}}}, \bibinfo {author}
  {\bibfnamefont {T.}~\bibnamefont {{de Haan}}}, \bibinfo {author}
  {\bibfnamefont {G.}~\bibnamefont {{de Zotti}}}, \bibinfo {author}
  {\bibfnamefont {J.}~\bibnamefont {{Delabrouille}}}, \bibinfo {author}
  {\bibfnamefont {M.}~\bibnamefont {{Demarteau}}}, \bibinfo {author}
  {\bibfnamefont {M.}~\bibnamefont {{Devlin}}}, \bibinfo {author}
  {\bibfnamefont {E.}~\bibnamefont {{Di Valentino}}}, \bibinfo {author}
  {\bibfnamefont {M.}~\bibnamefont {{Dobbs}}}, \bibinfo {author} {\bibfnamefont
  {S.}~\bibnamefont {{Duff}}}, \bibinfo {author} {\bibfnamefont
  {A.}~\bibnamefont {{Duivenvoorden}}}, \bibinfo {author} {\bibfnamefont
  {C.}~\bibnamefont {{Dvorkin}}}, \bibinfo {author} {\bibfnamefont
  {W.}~\bibnamefont {{Edwards}}}, \bibinfo {author} {\bibfnamefont
  {J.}~\bibnamefont {{Eimer}}}, \bibinfo {author} {\bibfnamefont
  {J.}~\bibnamefont {{Errard}}}, \bibinfo {author} {\bibfnamefont
  {T.}~\bibnamefont {{Essinger-Hileman}}}, \bibinfo {author} {\bibfnamefont
  {G.}~\bibnamefont {{Fabbian}}}, \bibinfo {author} {\bibfnamefont
  {C.}~\bibnamefont {{Feng}}}, \bibinfo {author} {\bibfnamefont
  {S.}~\bibnamefont {{Ferraro}}}, \bibinfo {author} {\bibfnamefont {J.~P.}\
  \bibnamefont {{Filippini}}}, \bibinfo {author} {\bibfnamefont
  {R.}~\bibnamefont {{Flauger}}}, \bibinfo {author} {\bibfnamefont
  {B.}~\bibnamefont {{Flaugher}}}, \bibinfo {author} {\bibfnamefont {A.~A.}\
  \bibnamefont {{Fraisse}}}, \bibinfo {author} {\bibfnamefont {A.}~\bibnamefont
  {{Frolov}}}, \bibinfo {author} {\bibfnamefont {N.}~\bibnamefont
  {{Galitzki}}}, \bibinfo {author} {\bibfnamefont {S.}~\bibnamefont {{Galli}}},
  \bibinfo {author} {\bibfnamefont {K.}~\bibnamefont {{Ganga}}}, \bibinfo
  {author} {\bibfnamefont {M.}~\bibnamefont {{Gerbino}}}, \bibinfo {author}
  {\bibfnamefont {M.}~\bibnamefont {{Gilchriese}}}, \bibinfo {author}
  {\bibfnamefont {V.}~\bibnamefont {{Gluscevic}}}, \bibinfo {author}
  {\bibfnamefont {D.}~\bibnamefont {{Green}}}, \bibinfo {author} {\bibfnamefont
  {D.}~\bibnamefont {{Grin}}}, \bibinfo {author} {\bibfnamefont
  {E.}~\bibnamefont {{Grohs}}}, \bibinfo {author} {\bibfnamefont
  {R.}~\bibnamefont {{Gualtieri}}}, \bibinfo {author} {\bibfnamefont
  {V.}~\bibnamefont {{Guarino}}}, \bibinfo {author} {\bibfnamefont {J.~E.}\
  \bibnamefont {{Gudmundsson}}}, \bibinfo {author} {\bibfnamefont
  {S.}~\bibnamefont {{Habib}}}, \bibinfo {author} {\bibfnamefont
  {G.}~\bibnamefont {{Haller}}}, \bibinfo {author} {\bibfnamefont
  {M.}~\bibnamefont {{Halpern}}}, \bibinfo {author} {\bibfnamefont {N.~W.}\
  \bibnamefont {{Halverson}}}, \bibinfo {author} {\bibfnamefont
  {S.}~\bibnamefont {{Hanany}}}, \bibinfo {author} {\bibfnamefont
  {K.}~\bibnamefont {{Harrington}}}, \bibinfo {author} {\bibfnamefont
  {M.}~\bibnamefont {{Hasegawa}}}, \bibinfo {author} {\bibfnamefont
  {M.}~\bibnamefont {{Hasselfield}}}, \bibinfo {author} {\bibfnamefont
  {M.}~\bibnamefont {{Hazumi}}}, \bibinfo {author} {\bibfnamefont
  {K.}~\bibnamefont {{Heitmann}}}, \bibinfo {author} {\bibfnamefont
  {S.}~\bibnamefont {{Henderson}}}, \bibinfo {author} {\bibfnamefont {J.~W.}\
  \bibnamefont {{Henning}}}, \bibinfo {author} {\bibfnamefont {J.~C.}\
  \bibnamefont {{Hill}}}, \bibinfo {author} {\bibfnamefont {R.}~\bibnamefont
  {{Hlozek}}}, \bibinfo {author} {\bibfnamefont {G.}~\bibnamefont {{Holder}}},
  \bibinfo {author} {\bibfnamefont {W.}~\bibnamefont {{Holzapfel}}}, \bibinfo
  {author} {\bibfnamefont {J.}~\bibnamefont {{Hubmayr}}}, \bibinfo {author}
  {\bibfnamefont {K.~M.}\ \bibnamefont {{Huffenberger}}}, \bibinfo {author}
  {\bibfnamefont {M.}~\bibnamefont {{Huffer}}}, \bibinfo {author}
  {\bibfnamefont {H.}~\bibnamefont {{Hui}}}, \bibinfo {author} {\bibfnamefont
  {K.}~\bibnamefont {{Irwin}}}, \bibinfo {author} {\bibfnamefont {B.~R.}\
  \bibnamefont {{Johnson}}}, \bibinfo {author} {\bibfnamefont {D.}~\bibnamefont
  {{Johnstone}}}, \bibinfo {author} {\bibfnamefont {W.~C.}\ \bibnamefont
  {{Jones}}}, \bibinfo {author} {\bibfnamefont {K.}~\bibnamefont {{Karkare}}},
  \bibinfo {author} {\bibfnamefont {N.}~\bibnamefont {{Katayama}}}, \bibinfo
  {author} {\bibfnamefont {J.}~\bibnamefont {{Kerby}}}, \bibinfo {author}
  {\bibfnamefont {S.}~\bibnamefont {{Kernovsky}}}, \bibinfo {author}
  {\bibfnamefont {R.}~\bibnamefont {{Keskitalo}}}, \bibinfo {author}
  {\bibfnamefont {T.}~\bibnamefont {{Kisner}}}, \bibinfo {author}
  {\bibfnamefont {L.}~\bibnamefont {{Knox}}}, \bibinfo {author} {\bibfnamefont
  {A.}~\bibnamefont {{Kosowsky}}}, \bibinfo {author} {\bibfnamefont
  {J.}~\bibnamefont {{Kovac}}}, \bibinfo {author} {\bibfnamefont {E.~D.}\
  \bibnamefont {{Kovetz}}}, \bibinfo {author} {\bibfnamefont {S.}~\bibnamefont
  {{Kuhlmann}}}, \bibinfo {author} {\bibfnamefont {C.-l.}\ \bibnamefont
  {{Kuo}}}, \bibinfo {author} {\bibfnamefont {N.}~\bibnamefont {{Kurita}}},
  \bibinfo {author} {\bibfnamefont {A.}~\bibnamefont {{Kusaka}}}, \bibinfo
  {author} {\bibfnamefont {A.}~\bibnamefont {{Lahteenmaki}}}, \bibinfo {author}
  {\bibfnamefont {C.~R.}\ \bibnamefont {{Lawrence}}}, \bibinfo {author}
  {\bibfnamefont {A.~T.}\ \bibnamefont {{Lee}}}, \bibinfo {author}
  {\bibfnamefont {A.}~\bibnamefont {{Lewis}}}, \bibinfo {author} {\bibfnamefont
  {D.}~\bibnamefont {{Li}}}, \bibinfo {author} {\bibfnamefont {E.}~\bibnamefont
  {{Linder}}}, \bibinfo {author} {\bibfnamefont {M.}~\bibnamefont {{Loverde}}},
  \bibinfo {author} {\bibfnamefont {A.}~\bibnamefont {{Lowitz}}}, \bibinfo
  {author} {\bibfnamefont {M.~S.}\ \bibnamefont {{Madhavacheril}}}, \bibinfo
  {author} {\bibfnamefont {A.}~\bibnamefont {{Mantz}}}, \bibinfo {author}
  {\bibfnamefont {F.}~\bibnamefont {{Matsuda}}}, \bibinfo {author}
  {\bibfnamefont {P.}~\bibnamefont {{Mauskopf}}}, \bibinfo {author}
  {\bibfnamefont {J.}~\bibnamefont {{McMahon}}}, \bibinfo {author}
  {\bibfnamefont {M.}~\bibnamefont {{McQuinn}}}, \bibinfo {author}
  {\bibfnamefont {P.~D.}\ \bibnamefont {{Meerburg}}}, \bibinfo {author}
  {\bibfnamefont {J.-B.}\ \bibnamefont {{Melin}}}, \bibinfo {author}
  {\bibfnamefont {J.}~\bibnamefont {{Meyers}}}, \bibinfo {author}
  {\bibfnamefont {M.}~\bibnamefont {{Millea}}}, \bibinfo {author}
  {\bibfnamefont {J.}~\bibnamefont {{Mohr}}}, \bibinfo {author} {\bibfnamefont
  {L.}~\bibnamefont {{Moncelsi}}}, \bibinfo {author} {\bibfnamefont
  {T.}~\bibnamefont {{Mroczkowski}}}, \bibinfo {author} {\bibfnamefont
  {S.}~\bibnamefont {{Mukherjee}}}, \bibinfo {author} {\bibfnamefont
  {M.}~\bibnamefont {{M{\"u}nchmeyer}}}, \bibinfo {author} {\bibfnamefont
  {D.}~\bibnamefont {{Nagai}}}, \bibinfo {author} {\bibfnamefont
  {J.}~\bibnamefont {{Nagy}}}, \bibinfo {author} {\bibfnamefont
  {T.}~\bibnamefont {{Namikawa}}}, \bibinfo {author} {\bibfnamefont
  {F.}~\bibnamefont {{Nati}}}, \bibinfo {author} {\bibfnamefont
  {T.}~\bibnamefont {{Natoli}}}, \bibinfo {author} {\bibfnamefont
  {M.}~\bibnamefont {{Negrello}}}, \bibinfo {author} {\bibfnamefont
  {L.}~\bibnamefont {{Newburgh}}}, \bibinfo {author} {\bibfnamefont {M.~D.}\
  \bibnamefont {{Niemack}}}, \bibinfo {author} {\bibfnamefont {H.}~\bibnamefont
  {{Nishino}}}, \bibinfo {author} {\bibfnamefont {M.}~\bibnamefont {{Nordby}}},
  \bibinfo {author} {\bibfnamefont {V.}~\bibnamefont {{Novosad}}}, \bibinfo
  {author} {\bibfnamefont {P.}~\bibnamefont {{O'Connor}}}, \bibinfo {author}
  {\bibfnamefont {G.}~\bibnamefont {{Obied}}}, \bibinfo {author} {\bibfnamefont
  {S.}~\bibnamefont {{Padin}}}, \bibinfo {author} {\bibfnamefont
  {S.}~\bibnamefont {{Pandey}}}, \bibinfo {author} {\bibfnamefont
  {B.}~\bibnamefont {{Partridge}}}, \bibinfo {author} {\bibfnamefont
  {E.}~\bibnamefont {{Pierpaoli}}}, \bibinfo {author} {\bibfnamefont
  {L.}~\bibnamefont {{Pogosian}}}, \bibinfo {author} {\bibfnamefont
  {C.}~\bibnamefont {{Pryke}}}, \bibinfo {author} {\bibfnamefont
  {G.}~\bibnamefont {{Puglisi}}}, \bibinfo {author} {\bibfnamefont
  {B.}~\bibnamefont {{Racine}}}, \bibinfo {author} {\bibfnamefont
  {S.}~\bibnamefont {{Raghunathan}}}, \bibinfo {author} {\bibfnamefont
  {A.}~\bibnamefont {{Rahlin}}}, \bibinfo {author} {\bibfnamefont
  {S.}~\bibnamefont {{Rajagopalan}}}, \bibinfo {author} {\bibfnamefont
  {M.}~\bibnamefont {{Raveri}}}, \bibinfo {author} {\bibfnamefont
  {M.}~\bibnamefont {{Reichanadter}}}, \bibinfo {author} {\bibfnamefont
  {C.~L.}\ \bibnamefont {{Reichardt}}}, \bibinfo {author} {\bibfnamefont
  {M.}~\bibnamefont {{Remazeilles}}}, \bibinfo {author} {\bibfnamefont
  {G.}~\bibnamefont {{Rocha}}}, \bibinfo {author} {\bibfnamefont {N.~A.}\
  \bibnamefont {{Roe}}}, \bibinfo {author} {\bibfnamefont {A.}~\bibnamefont
  {{Roy}}}, \bibinfo {author} {\bibfnamefont {J.}~\bibnamefont {{Ruhl}}},
  \bibinfo {author} {\bibfnamefont {M.}~\bibnamefont {{Salatino}}}, \bibinfo
  {author} {\bibfnamefont {B.}~\bibnamefont {{Saliwanchik}}}, \bibinfo {author}
  {\bibfnamefont {E.}~\bibnamefont {{Schaan}}}, \bibinfo {author}
  {\bibfnamefont {A.}~\bibnamefont {{Schillaci}}}, \bibinfo {author}
  {\bibfnamefont {M.~M.}\ \bibnamefont {{Schmittfull}}}, \bibinfo {author}
  {\bibfnamefont {D.}~\bibnamefont {{Scott}}}, \bibinfo {author} {\bibfnamefont
  {N.}~\bibnamefont {{Sehgal}}}, \bibinfo {author} {\bibfnamefont
  {S.}~\bibnamefont {{Shandera}}}, \bibinfo {author} {\bibfnamefont
  {C.}~\bibnamefont {{Sheehy}}}, \bibinfo {author} {\bibfnamefont {B.~D.}\
  \bibnamefont {{Sherwin}}}, \bibinfo {author} {\bibfnamefont {E.}~\bibnamefont
  {{Shirokoff}}}, \bibinfo {author} {\bibfnamefont {S.~M.}\ \bibnamefont
  {{Simon}}}, \bibinfo {author} {\bibfnamefont {A.}~\bibnamefont {{Slosar}}},
  \bibinfo {author} {\bibfnamefont {R.}~\bibnamefont {{Somerville}}}, \bibinfo
  {author} {\bibfnamefont {D.}~\bibnamefont {{Spergel}}}, \bibinfo {author}
  {\bibfnamefont {S.~T.}\ \bibnamefont {{Staggs}}}, \bibinfo {author}
  {\bibfnamefont {A.}~\bibnamefont {{Stark}}}, \bibinfo {author} {\bibfnamefont
  {R.}~\bibnamefont {{Stompor}}}, \bibinfo {author} {\bibfnamefont {K.~T.}\
  \bibnamefont {{Story}}}, \bibinfo {author} {\bibfnamefont {C.}~\bibnamefont
  {{Stoughton}}}, \bibinfo {author} {\bibfnamefont {A.}~\bibnamefont
  {{Suzuki}}}, \bibinfo {author} {\bibfnamefont {O.}~\bibnamefont {{Tajima}}},
  \bibinfo {author} {\bibfnamefont {G.~P.}\ \bibnamefont {{Teply}}}, \bibinfo
  {author} {\bibfnamefont {K.}~\bibnamefont {{Thompson}}}, \bibinfo {author}
  {\bibfnamefont {P.}~\bibnamefont {{Timbie}}}, \bibinfo {author}
  {\bibfnamefont {M.}~\bibnamefont {{Tomasi}}}, \bibinfo {author}
  {\bibfnamefont {J.~I.}\ \bibnamefont {{Treu}}}, \bibinfo {author}
  {\bibfnamefont {M.}~\bibnamefont {{Tristram}}}, \bibinfo {author}
  {\bibfnamefont {G.}~\bibnamefont {{Tucker}}}, \bibinfo {author}
  {\bibfnamefont {C.}~\bibnamefont {{Umilt{\`a}}}}, \bibinfo {author}
  {\bibfnamefont {A.}~\bibnamefont {{van Engelen}}}, \bibinfo {author}
  {\bibfnamefont {J.~D.}\ \bibnamefont {{Vieira}}}, \bibinfo {author}
  {\bibfnamefont {A.~G.}\ \bibnamefont {{Vieregg}}}, \bibinfo {author}
  {\bibfnamefont {M.}~\bibnamefont {{Vogelsberger}}}, \bibinfo {author}
  {\bibfnamefont {G.}~\bibnamefont {{Wang}}}, \bibinfo {author} {\bibfnamefont
  {S.}~\bibnamefont {{Watson}}}, \bibinfo {author} {\bibfnamefont
  {M.}~\bibnamefont {{White}}}, \bibinfo {author} {\bibfnamefont
  {N.}~\bibnamefont {{Whitehorn}}}, \bibinfo {author} {\bibfnamefont {E.~J.}\
  \bibnamefont {{Wollack}}}, \bibinfo {author} {\bibfnamefont {W.~L.}\
  \bibnamefont {{Kimmy Wu}}}, \bibinfo {author} {\bibfnamefont
  {Z.}~\bibnamefont {{Xu}}}, \bibinfo {author} {\bibfnamefont {S.}~\bibnamefont
  {{Yasini}}}, \bibinfo {author} {\bibfnamefont {J.}~\bibnamefont {{Yeck}}},
  \bibinfo {author} {\bibfnamefont {K.~W.}\ \bibnamefont {{Yoon}}}, \bibinfo
  {author} {\bibfnamefont {E.}~\bibnamefont {{Young}}},\ and\ \bibinfo {author}
  {\bibfnamefont {A.}~\bibnamefont {{Zonca}}},\ }\bibfield  {title} {\bibinfo
  {title} {{CMB-S4 Science Case, Reference Design, and Project Plan}},\ }\href
  {https://doi.org/10.48550/arXiv.1907.04473} {\bibfield  {journal} {\bibinfo
  {journal} {arXiv e-prints}\ ,\ \bibinfo {eid} {arXiv:1907.04473}} (\bibinfo
  {year} {2019})},\ \Eprint {https://arxiv.org/abs/1907.04473}
  {arXiv:1907.04473 [astro-ph.IM]} \BibitemShut {NoStop}%
\bibitem [{\citenamefont {{Cooray}}\ and\ \citenamefont
  {{Sheth}}(2002)}]{2002PhR...372....1C}%
  \BibitemOpen
  \bibfield  {author} {\bibinfo {author} {\bibfnamefont {A.}~\bibnamefont
  {{Cooray}}}\ and\ \bibinfo {author} {\bibfnamefont {R.}~\bibnamefont
  {{Sheth}}},\ }\bibfield  {title} {\bibinfo {title} {{Halo models of large
  scale structure}},\ }\href {https://doi.org/10.1016/S0370-1573(02)00276-4}
  {\bibfield  {journal} {\bibinfo  {journal} {\physrep}\ }\textbf {\bibinfo
  {volume} {372}},\ \bibinfo {pages} {1} (\bibinfo {year} {2002})},\ \Eprint
  {https://arxiv.org/abs/astro-ph/0206508} {arXiv:astro-ph/0206508 [astro-ph]}
  \BibitemShut {NoStop}%
\bibitem [{\citenamefont {{Limber}}(1953)}]{1953ApJ...117..134L}%
  \BibitemOpen
  \bibfield  {author} {\bibinfo {author} {\bibfnamefont {D.~N.}\ \bibnamefont
  {{Limber}}},\ }\bibfield  {title} {\bibinfo {title} {{The Analysis of Counts
  of the Extragalactic Nebulae in Terms of a Fluctuating Density Field.}},\
  }\href {https://doi.org/10.1086/145672} {\bibfield  {journal} {\bibinfo
  {journal} {\apj}\ }\textbf {\bibinfo {volume} {117}},\ \bibinfo {pages} {134}
  (\bibinfo {year} {1953})}\BibitemShut {NoStop}%
\bibitem [{\citenamefont {{Fang}}\ \emph {et~al.}(2020)\citenamefont {{Fang}},
  \citenamefont {{Krause}}, \citenamefont {{Eifler}},\ and\ \citenamefont
  {{MacCrann}}}]{2020JCAP...05..010F}%
  \BibitemOpen
  \bibfield  {author} {\bibinfo {author} {\bibfnamefont {X.}~\bibnamefont
  {{Fang}}}, \bibinfo {author} {\bibfnamefont {E.}~\bibnamefont {{Krause}}},
  \bibinfo {author} {\bibfnamefont {T.}~\bibnamefont {{Eifler}}},\ and\
  \bibinfo {author} {\bibfnamefont {N.}~\bibnamefont {{MacCrann}}},\ }\bibfield
   {title} {\bibinfo {title} {{Beyond Limber: efficient computation of angular
  power spectra for galaxy clustering and weak lensing}},\ }\href
  {https://doi.org/10.1088/1475-7516/2020/05/010} {\bibfield  {journal}
  {\bibinfo  {journal} {\jcap}\ }\textbf {\bibinfo {volume} {2020}},\ \bibinfo
  {eid} {010} (\bibinfo {year} {2020})},\ \Eprint
  {https://arxiv.org/abs/1911.11947} {arXiv:1911.11947 [astro-ph.CO]}
  \BibitemShut {NoStop}%
\bibitem [{\citenamefont {{LoVerde}}\ and\ \citenamefont
  {{Afshordi}}(2008)}]{2008PhRvD..78l3506L}%
  \BibitemOpen
  \bibfield  {author} {\bibinfo {author} {\bibfnamefont {M.}~\bibnamefont
  {{LoVerde}}}\ and\ \bibinfo {author} {\bibfnamefont {N.}~\bibnamefont
  {{Afshordi}}},\ }\bibfield  {title} {\bibinfo {title} {{Extended Limber
  approximation}},\ }\href {https://doi.org/10.1103/PhysRevD.78.123506}
  {\bibfield  {journal} {\bibinfo  {journal} {\prd}\ }\textbf {\bibinfo
  {volume} {78}},\ \bibinfo {eid} {123506} (\bibinfo {year} {2008})},\ \Eprint
  {https://arxiv.org/abs/0809.5112} {arXiv:0809.5112 [astro-ph]} \BibitemShut
  {NoStop}%
\bibitem [{\citenamefont {{Lesgourgues}}\ and\ \citenamefont
  {{Pastor}}(2012)}]{2012arXiv1212.6154L}%
  \BibitemOpen
  \bibfield  {author} {\bibinfo {author} {\bibfnamefont {J.}~\bibnamefont
  {{Lesgourgues}}}\ and\ \bibinfo {author} {\bibfnamefont {S.}~\bibnamefont
  {{Pastor}}},\ }\bibfield  {title} {\bibinfo {title} {{Neutrino mass from
  Cosmology}},\ }\href {https://doi.org/10.48550/arXiv.1212.6154} {\bibfield
  {journal} {\bibinfo  {journal} {arXiv e-prints}\ ,\ \bibinfo {eid}
  {arXiv:1212.6154}} (\bibinfo {year} {2012})},\ \Eprint
  {https://arxiv.org/abs/1212.6154} {arXiv:1212.6154 [hep-ph]} \BibitemShut
  {NoStop}%
\bibitem [{\citenamefont {{Bull}}\ \emph {et~al.}(2015)\citenamefont {{Bull}},
  \citenamefont {{Ferreira}}, \citenamefont {{Patel}},\ and\ \citenamefont
  {{Santos}}}]{2015ApJ...803...21B}%
  \BibitemOpen
  \bibfield  {author} {\bibinfo {author} {\bibfnamefont {P.}~\bibnamefont
  {{Bull}}}, \bibinfo {author} {\bibfnamefont {P.~G.}\ \bibnamefont
  {{Ferreira}}}, \bibinfo {author} {\bibfnamefont {P.}~\bibnamefont
  {{Patel}}},\ and\ \bibinfo {author} {\bibfnamefont {M.~G.}\ \bibnamefont
  {{Santos}}},\ }\bibfield  {title} {\bibinfo {title} {{Late-time Cosmology
  with 21 cm Intensity Mapping Experiments}},\ }\href
  {https://doi.org/10.1088/0004-637X/803/1/21} {\bibfield  {journal} {\bibinfo
  {journal} {\apj}\ }\textbf {\bibinfo {volume} {803}},\ \bibinfo {eid} {21}
  (\bibinfo {year} {2015})},\ \Eprint {https://arxiv.org/abs/1405.1452}
  {arXiv:1405.1452 [astro-ph.CO]} \BibitemShut {NoStop}%
\bibitem [{\citenamefont {{Xiao}}\ \emph {et~al.}(2022)\citenamefont {{Xiao}},
  \citenamefont {{Costa}},\ and\ \citenamefont {{Wang}}}]{2022MNRAS.510.1495X}%
  \BibitemOpen
  \bibfield  {author} {\bibinfo {author} {\bibfnamefont {L.}~\bibnamefont
  {{Xiao}}}, \bibinfo {author} {\bibfnamefont {A.~A.}\ \bibnamefont
  {{Costa}}},\ and\ \bibinfo {author} {\bibfnamefont {B.}~\bibnamefont
  {{Wang}}},\ }\bibfield  {title} {\bibinfo {title} {{Forecasts on interacting
  dark energy from the 21-cm angular power spectrum with BINGO and SKA
  observations}},\ }\href {https://doi.org/10.1093/mnras/stab3256} {\bibfield
  {journal} {\bibinfo  {journal} {\mnras}\ }\textbf {\bibinfo {volume} {510}},\
  \bibinfo {pages} {1495} (\bibinfo {year} {2022})},\ \Eprint
  {https://arxiv.org/abs/2103.01796} {arXiv:2103.01796 [astro-ph.CO]}
  \BibitemShut {NoStop}%
\bibitem [{\citenamefont {{Seiffert}}\ \emph {et~al.}(2002)\citenamefont
  {{Seiffert}}, \citenamefont {{Mennella}}, \citenamefont {{Burigana}},
  \citenamefont {{Mandolesi}}, \citenamefont {{Bersanelli}}, \citenamefont
  {{Meinhold}},\ and\ \citenamefont {{Lubin}}}]{2002A&A...391.1185S}%
  \BibitemOpen
  \bibfield  {author} {\bibinfo {author} {\bibfnamefont {M.}~\bibnamefont
  {{Seiffert}}}, \bibinfo {author} {\bibfnamefont {A.}~\bibnamefont
  {{Mennella}}}, \bibinfo {author} {\bibfnamefont {C.}~\bibnamefont
  {{Burigana}}}, \bibinfo {author} {\bibfnamefont {N.}~\bibnamefont
  {{Mandolesi}}}, \bibinfo {author} {\bibfnamefont {M.}~\bibnamefont
  {{Bersanelli}}}, \bibinfo {author} {\bibfnamefont {P.}~\bibnamefont
  {{Meinhold}}},\ and\ \bibinfo {author} {\bibfnamefont {P.}~\bibnamefont
  {{Lubin}}},\ }\bibfield  {title} {\bibinfo {title} {{1/f noise and other
  systematic effects in the Planck-LFI radiometers}},\ }\href
  {https://doi.org/10.1051/0004-6361:20020880} {\bibfield  {journal} {\bibinfo
  {journal} {\aap}\ }\textbf {\bibinfo {volume} {391}},\ \bibinfo {pages}
  {1185} (\bibinfo {year} {2002})},\ \Eprint
  {https://arxiv.org/abs/astro-ph/0206093} {arXiv:astro-ph/0206093 [astro-ph]}
  \BibitemShut {NoStop}%
\bibitem [{\citenamefont {{Harper}}\ \emph {et~al.}(2018)\citenamefont
  {{Harper}}, \citenamefont {{Dickinson}}, \citenamefont {{Battye}},
  \citenamefont {{Roychowdhury}}, \citenamefont {{Browne}}, \citenamefont
  {{Ma}}, \citenamefont {{Olivari}},\ and\ \citenamefont
  {{Chen}}}]{2018MNRAS.478.2416H}%
  \BibitemOpen
  \bibfield  {author} {\bibinfo {author} {\bibfnamefont {S.~E.}\ \bibnamefont
  {{Harper}}}, \bibinfo {author} {\bibfnamefont {C.}~\bibnamefont
  {{Dickinson}}}, \bibinfo {author} {\bibfnamefont {R.~A.}\ \bibnamefont
  {{Battye}}}, \bibinfo {author} {\bibfnamefont {S.}~\bibnamefont
  {{Roychowdhury}}}, \bibinfo {author} {\bibfnamefont {I.~W.~A.}\ \bibnamefont
  {{Browne}}}, \bibinfo {author} {\bibfnamefont {Y.~Z.}\ \bibnamefont {{Ma}}},
  \bibinfo {author} {\bibfnamefont {L.~C.}\ \bibnamefont {{Olivari}}},\ and\
  \bibinfo {author} {\bibfnamefont {T.}~\bibnamefont {{Chen}}},\ }\bibfield
  {title} {\bibinfo {title} {{Impact of simulated 1/f noise for HI intensity
  mapping experiments}},\ }\href {https://doi.org/10.1093/mnras/sty1238}
  {\bibfield  {journal} {\bibinfo  {journal} {\mnras}\ }\textbf {\bibinfo
  {volume} {478}},\ \bibinfo {pages} {2416} (\bibinfo {year} {2018})},\ \Eprint
  {https://arxiv.org/abs/1711.07843} {arXiv:1711.07843 [astro-ph.CO]}
  \BibitemShut {NoStop}%
\bibitem [{\citenamefont {{Bennett}}\ \emph {et~al.}(2003)\citenamefont
  {{Bennett}}, \citenamefont {{Hill}}, \citenamefont {{Hinshaw}}, \citenamefont
  {{Nolta}}, \citenamefont {{Odegard}}, \citenamefont {{Page}}, \citenamefont
  {{Spergel}}, \citenamefont {{Weiland}}, \citenamefont {{Wright}},
  \citenamefont {{Halpern}}, \citenamefont {{Jarosik}}, \citenamefont
  {{Kogut}}, \citenamefont {{Limon}}, \citenamefont {{Meyer}}, \citenamefont
  {{Tucker}},\ and\ \citenamefont {{Wollack}}}]{2003ApJS..148...97B}%
  \BibitemOpen
  \bibfield  {author} {\bibinfo {author} {\bibfnamefont {C.~L.}\ \bibnamefont
  {{Bennett}}}, \bibinfo {author} {\bibfnamefont {R.~S.}\ \bibnamefont
  {{Hill}}}, \bibinfo {author} {\bibfnamefont {G.}~\bibnamefont {{Hinshaw}}},
  \bibinfo {author} {\bibfnamefont {M.~R.}\ \bibnamefont {{Nolta}}}, \bibinfo
  {author} {\bibfnamefont {N.}~\bibnamefont {{Odegard}}}, \bibinfo {author}
  {\bibfnamefont {L.}~\bibnamefont {{Page}}}, \bibinfo {author} {\bibfnamefont
  {D.~N.}\ \bibnamefont {{Spergel}}}, \bibinfo {author} {\bibfnamefont {J.~L.}\
  \bibnamefont {{Weiland}}}, \bibinfo {author} {\bibfnamefont {E.~L.}\
  \bibnamefont {{Wright}}}, \bibinfo {author} {\bibfnamefont {M.}~\bibnamefont
  {{Halpern}}}, \bibinfo {author} {\bibfnamefont {N.}~\bibnamefont
  {{Jarosik}}}, \bibinfo {author} {\bibfnamefont {A.}~\bibnamefont {{Kogut}}},
  \bibinfo {author} {\bibfnamefont {M.}~\bibnamefont {{Limon}}}, \bibinfo
  {author} {\bibfnamefont {S.~S.}\ \bibnamefont {{Meyer}}}, \bibinfo {author}
  {\bibfnamefont {G.~S.}\ \bibnamefont {{Tucker}}},\ and\ \bibinfo {author}
  {\bibfnamefont {E.}~\bibnamefont {{Wollack}}},\ }\bibfield  {title} {\bibinfo
  {title} {{First-Year Wilkinson Microwave Anisotropy Probe (WMAP)
  Observations: Foreground Emission}},\ }\href {https://doi.org/10.1086/377252}
  {\bibfield  {journal} {\bibinfo  {journal} {\apjs}\ }\textbf {\bibinfo
  {volume} {148}},\ \bibinfo {pages} {97} (\bibinfo {year} {2003})},\ \Eprint
  {https://arxiv.org/abs/astro-ph/0302208} {arXiv:astro-ph/0302208 [astro-ph]}
  \BibitemShut {NoStop}%
\bibitem [{\citenamefont {{Delabrouille}}\ \emph {et~al.}(2009)\citenamefont
  {{Delabrouille}}, \citenamefont {{Cardoso}}, \citenamefont {{Le Jeune}},
  \citenamefont {{Betoule}}, \citenamefont {{Fay}},\ and\ \citenamefont
  {{Guilloux}}}]{2009A&A...493..835D}%
  \BibitemOpen
  \bibfield  {author} {\bibinfo {author} {\bibfnamefont {J.}~\bibnamefont
  {{Delabrouille}}}, \bibinfo {author} {\bibfnamefont {J.~F.}\ \bibnamefont
  {{Cardoso}}}, \bibinfo {author} {\bibfnamefont {M.}~\bibnamefont {{Le
  Jeune}}}, \bibinfo {author} {\bibfnamefont {M.}~\bibnamefont {{Betoule}}},
  \bibinfo {author} {\bibfnamefont {G.}~\bibnamefont {{Fay}}},\ and\ \bibinfo
  {author} {\bibfnamefont {F.}~\bibnamefont {{Guilloux}}},\ }\bibfield  {title}
  {\bibinfo {title} {{A full sky, low foreground, high resolution CMB map from
  WMAP}},\ }\href {https://doi.org/10.1051/0004-6361:200810514} {\bibfield
  {journal} {\bibinfo  {journal} {\aap}\ }\textbf {\bibinfo {volume} {493}},\
  \bibinfo {pages} {835} (\bibinfo {year} {2009})},\ \Eprint
  {https://arxiv.org/abs/0807.0773} {arXiv:0807.0773 [astro-ph]} \BibitemShut
  {NoStop}%
\bibitem [{\citenamefont {{Liu}}\ \emph
  {et~al.}(2014{\natexlab{a}})\citenamefont {{Liu}}, \citenamefont
  {{Parsons}},\ and\ \citenamefont {{Trott}}}]{2014PhRvD..90b3018L}%
  \BibitemOpen
  \bibfield  {author} {\bibinfo {author} {\bibfnamefont {A.}~\bibnamefont
  {{Liu}}}, \bibinfo {author} {\bibfnamefont {A.~R.}\ \bibnamefont
  {{Parsons}}},\ and\ \bibinfo {author} {\bibfnamefont {C.~M.}\ \bibnamefont
  {{Trott}}},\ }\bibfield  {title} {\bibinfo {title} {{Epoch of reionization
  window. I. Mathematical formalism}},\ }\href
  {https://doi.org/10.1103/PhysRevD.90.023018} {\bibfield  {journal} {\bibinfo
  {journal} {\prd}\ }\textbf {\bibinfo {volume} {90}},\ \bibinfo {eid} {023018}
  (\bibinfo {year} {2014}{\natexlab{a}})},\ \Eprint
  {https://arxiv.org/abs/1404.2596} {arXiv:1404.2596 [astro-ph.CO]}
  \BibitemShut {NoStop}%
\bibitem [{\citenamefont {{Liu}}\ \emph
  {et~al.}(2014{\natexlab{b}})\citenamefont {{Liu}}, \citenamefont
  {{Parsons}},\ and\ \citenamefont {{Trott}}}]{2014PhRvD..90b3019L}%
  \BibitemOpen
  \bibfield  {author} {\bibinfo {author} {\bibfnamefont {A.}~\bibnamefont
  {{Liu}}}, \bibinfo {author} {\bibfnamefont {A.~R.}\ \bibnamefont
  {{Parsons}}},\ and\ \bibinfo {author} {\bibfnamefont {C.~M.}\ \bibnamefont
  {{Trott}}},\ }\bibfield  {title} {\bibinfo {title} {{Epoch of reionization
  window. II. Statistical methods for foreground wedge reduction}},\ }\href
  {https://doi.org/10.1103/PhysRevD.90.023019} {\bibfield  {journal} {\bibinfo
  {journal} {\prd}\ }\textbf {\bibinfo {volume} {90}},\ \bibinfo {eid} {023019}
  (\bibinfo {year} {2014}{\natexlab{b}})},\ \Eprint
  {https://arxiv.org/abs/1404.4372} {arXiv:1404.4372 [astro-ph.CO]}
  \BibitemShut {NoStop}%
\bibitem [{\citenamefont {{Feng}}\ and\ \citenamefont
  {{Abdalla}}(2023)}]{2023arXiv230814777F}%
  \BibitemOpen
  \bibfield  {author} {\bibinfo {author} {\bibfnamefont {C.}~\bibnamefont
  {{Feng}}}\ and\ \bibinfo {author} {\bibfnamefont {F.~B.}\ \bibnamefont
  {{Abdalla}}},\ }\bibfield  {title} {\bibinfo {title} {{Power-spectrum space
  decomposition of frequency tomographic data for intensity mapping
  experiments}},\ }\href {https://doi.org/10.48550/arXiv.2308.14777} {\bibfield
   {journal} {\bibinfo  {journal} {arXiv e-prints}\ ,\ \bibinfo {eid}
  {arXiv:2308.14777}} (\bibinfo {year} {2023})},\ \Eprint
  {https://arxiv.org/abs/2308.14777} {arXiv:2308.14777 [astro-ph.IM]}
  \BibitemShut {NoStop}%
\bibitem [{\citenamefont {{Sellentin}}\ \emph {et~al.}(2014)\citenamefont
  {{Sellentin}}, \citenamefont {{Quartin}},\ and\ \citenamefont
  {{Amendola}}}]{2014MNRAS.441.1831S}%
  \BibitemOpen
  \bibfield  {author} {\bibinfo {author} {\bibfnamefont {E.}~\bibnamefont
  {{Sellentin}}}, \bibinfo {author} {\bibfnamefont {M.}~\bibnamefont
  {{Quartin}}},\ and\ \bibinfo {author} {\bibfnamefont {L.}~\bibnamefont
  {{Amendola}}},\ }\bibfield  {title} {\bibinfo {title} {{Breaking the spell of
  Gaussianity: forecasting with higher order Fisher matrices}},\ }\href
  {https://doi.org/10.1093/mnras/stu689} {\bibfield  {journal} {\bibinfo
  {journal} {\mnras}\ }\textbf {\bibinfo {volume} {441}},\ \bibinfo {pages}
  {1831} (\bibinfo {year} {2014})},\ \Eprint {https://arxiv.org/abs/1401.6892}
  {arXiv:1401.6892 [astro-ph.CO]} \BibitemShut {NoStop}%
\bibitem [{\citenamefont {{Perotto}}\ \emph {et~al.}(2006)\citenamefont
  {{Perotto}}, \citenamefont {{Lesgourgues}}, \citenamefont {{Hannestad}},
  \citenamefont {{Tu}},\ and\ \citenamefont {{Y Y
  Wong}}}]{2006JCAP...10..013P}%
  \BibitemOpen
  \bibfield  {author} {\bibinfo {author} {\bibfnamefont {L.}~\bibnamefont
  {{Perotto}}}, \bibinfo {author} {\bibfnamefont {J.}~\bibnamefont
  {{Lesgourgues}}}, \bibinfo {author} {\bibfnamefont {S.}~\bibnamefont
  {{Hannestad}}}, \bibinfo {author} {\bibfnamefont {H.}~\bibnamefont {{Tu}}},\
  and\ \bibinfo {author} {\bibfnamefont {Y.}~\bibnamefont {{Y Y Wong}}},\
  }\bibfield  {title} {\bibinfo {title} {{Probing cosmological parameters with
  the CMB: forecasts from Monte Carlo simulations}},\ }\href
  {https://doi.org/10.1088/1475-7516/2006/10/013} {\bibfield  {journal}
  {\bibinfo  {journal} {\jcap}\ }\textbf {\bibinfo {volume} {2006}},\ \bibinfo
  {eid} {013} (\bibinfo {year} {2006})},\ \Eprint
  {https://arxiv.org/abs/astro-ph/0606227} {arXiv:astro-ph/0606227 [astro-ph]}
  \BibitemShut {NoStop}%
\bibitem [{\citenamefont {{Chevallier}}\ and\ \citenamefont
  {{Polarski}}(2001)}]{2001IJMPD..10..213C}%
  \BibitemOpen
  \bibfield  {author} {\bibinfo {author} {\bibfnamefont {M.}~\bibnamefont
  {{Chevallier}}}\ and\ \bibinfo {author} {\bibfnamefont {D.}~\bibnamefont
  {{Polarski}}},\ }\bibfield  {title} {\bibinfo {title} {{Accelerating
  Universes with Scaling Dark Matter}},\ }\href
  {https://doi.org/10.1142/S0218271801000822} {\bibfield  {journal} {\bibinfo
  {journal} {International Journal of Modern Physics D}\ }\textbf {\bibinfo
  {volume} {10}},\ \bibinfo {pages} {213} (\bibinfo {year} {2001})},\ \Eprint
  {https://arxiv.org/abs/gr-qc/0009008} {arXiv:gr-qc/0009008 [gr-qc]}
  \BibitemShut {NoStop}%
\bibitem [{\citenamefont {{Linder}}(2003)}]{2003PhRvL..90i1301L}%
  \BibitemOpen
  \bibfield  {author} {\bibinfo {author} {\bibfnamefont {E.~V.}\ \bibnamefont
  {{Linder}}},\ }\bibfield  {title} {\bibinfo {title} {{Exploring the Expansion
  History of the Universe}},\ }\href
  {https://doi.org/10.1103/PhysRevLett.90.091301} {\bibfield  {journal}
  {\bibinfo  {journal} {\prl}\ }\textbf {\bibinfo {volume} {90}},\ \bibinfo
  {eid} {091301} (\bibinfo {year} {2003})},\ \Eprint
  {https://arxiv.org/abs/astro-ph/0208512} {arXiv:astro-ph/0208512 [astro-ph]}
  \BibitemShut {NoStop}%
\bibitem [{\citenamefont {{G{\'o}rski}}\ \emph {et~al.}(2005)\citenamefont
  {{G{\'o}rski}}, \citenamefont {{Hivon}}, \citenamefont {{Banday}},
  \citenamefont {{Wandelt}}, \citenamefont {{Hansen}}, \citenamefont
  {{Reinecke}},\ and\ \citenamefont {{Bartelmann}}}]{2005ApJ...622..759G}%
  \BibitemOpen
  \bibfield  {author} {\bibinfo {author} {\bibfnamefont {K.~M.}\ \bibnamefont
  {{G{\'o}rski}}}, \bibinfo {author} {\bibfnamefont {E.}~\bibnamefont
  {{Hivon}}}, \bibinfo {author} {\bibfnamefont {A.~J.}\ \bibnamefont
  {{Banday}}}, \bibinfo {author} {\bibfnamefont {B.~D.}\ \bibnamefont
  {{Wandelt}}}, \bibinfo {author} {\bibfnamefont {F.~K.}\ \bibnamefont
  {{Hansen}}}, \bibinfo {author} {\bibfnamefont {M.}~\bibnamefont
  {{Reinecke}}},\ and\ \bibinfo {author} {\bibfnamefont {M.}~\bibnamefont
  {{Bartelmann}}},\ }\bibfield  {title} {\bibinfo {title} {{HEALPix: A
  Framework for High-Resolution Discretization and Fast Analysis of Data
  Distributed on the Sphere}},\ }\href {https://doi.org/10.1086/427976}
  {\bibfield  {journal} {\bibinfo  {journal} {\apj}\ }\textbf {\bibinfo
  {volume} {622}},\ \bibinfo {pages} {759} (\bibinfo {year} {2005})},\ \Eprint
  {https://arxiv.org/abs/astro-ph/0409513} {arXiv:astro-ph/0409513 [astro-ph]}
  \BibitemShut {NoStop}%
\bibitem [{\citenamefont {{Foreman-Mackey}}\ \emph {et~al.}(2013)\citenamefont
  {{Foreman-Mackey}}, \citenamefont {{Hogg}}, \citenamefont {{Lang}},\ and\
  \citenamefont {{Goodman}}}]{2013PASP..125..306F}%
  \BibitemOpen
  \bibfield  {author} {\bibinfo {author} {\bibfnamefont {D.}~\bibnamefont
  {{Foreman-Mackey}}}, \bibinfo {author} {\bibfnamefont {D.~W.}\ \bibnamefont
  {{Hogg}}}, \bibinfo {author} {\bibfnamefont {D.}~\bibnamefont {{Lang}}},\
  and\ \bibinfo {author} {\bibfnamefont {J.}~\bibnamefont {{Goodman}}},\
  }\bibfield  {title} {\bibinfo {title} {{emcee: The MCMC Hammer}},\ }\href
  {https://doi.org/10.1086/670067} {\bibfield  {journal} {\bibinfo  {journal}
  {\pasp}\ }\textbf {\bibinfo {volume} {125}},\ \bibinfo {pages} {306}
  (\bibinfo {year} {2013})},\ \Eprint {https://arxiv.org/abs/1202.3665}
  {arXiv:1202.3665 [astro-ph.IM]} \BibitemShut {NoStop}%
\end{thebibliography}%
\end{document}